\documentclass[twocolumn]{aastex631}

\shorttitle{Evolutionary Period Changes for 52 CVs}
\shortauthors{Schaefer}
\graphicspath{{./}{figures/}}

\begin{document}
\title{Evolutionary Period Changes For 52 Cataclysmic Variables, and the Failure For the Most Fundamental Prediction of the Magnetic Braking Model}

\author[0000-0002-2659-8763]{Bradley E. Schaefer}
\affiliation{Department of Physics and Astronomy,
Louisiana State University,
Baton Rouge, LA 70803, USA}



\begin{abstract}

The evolution of Cataclysmic Variables (CVs) is driven by period-changes ($\dot{P}$), for which the long-venerable consensus is the Magnetic Braking Model (MBM).  The MBM has its only distinctive assumption being a power-law `recipe' describing the angular momentum loss (AML) in the binary, producing a single unique evolutionary track with $\dot{P}$ as a function of the orbital period.  This required prediction can be used to test the most-fundamental assumption of MBM, but it has never been tested previously.  In this paper, I collect $\dot{P}$ measures for 52 CVs of all types.  First, 44 per cent of the CVs have {\it positive}-$\dot{P}$, with such being impossible in MBM.  Second, even amongst the CVs with negative-$\dot{P}$, their $\dot{P}$ measures are always more-negative than required by MBM, with an average deviation of 110$\times$.  Third, three CVs have large chaotic variations in $\dot{P}$ that are impossible for MBM, proving that some unknown mechanism exists and is operating that dominates for these systems.  Fourth, the MBM does not account for the long-term effects on evolution arising from the large sudden period decreases seen across many nova events, with this unaccounted effect dominating for the majority of nova systems and changing the sign of the overall evolutionary $\dot{P}$.  Fifth, three recurrent novae are observed to suddenly change $\dot{P}$ by an order-of-magnitude across a nova event, with this being impossible in the MBM.  In all, the required MBM $\dot{P}$ predictions all fail for my 52 CVs, usually by orders-of-magnitude, so the MBM AML-recipe is wrong by orders-of-magnitude.

\end{abstract}


\section{Introduction}

Cataclysmic Variables (CVs) are close interacting binaries systems with a donor star (or companion star) losing mass through Roche lobe overflow (RLOF) on to a white dwarf (WD).  The forefront and most important research topic for CVs is their long-term evolution, where they must start out at contact with an orbital period ($P$) that is long, then evolving to a short $P$.  CV evolution is driven by the various physical mechanisms that make for angular momentum loss (AML) of the binary and the steady long-term period changes $\dot{P}$.

The long-time consensus model for CV evolution is called the Magnetic Braking Model (MBM).  The originating papers include Rappaport, Verbunt, \& Joss (1983) and Patterson (1984), while the model has been perfected by Knigge, Baraffe, \& Patterson (2011, K2011).  The one key ingredient to the MBM is its `recipe' for AML, with this included on top of standard physical mechanisms and stellar properties.  So MBM is distinguished and defined only by its AML `recipe'.  K2011 presume an empirical power-law with the change in the orbital angular momentum to be $\dot{J}_{\rm MBM}$$\propto$$M_{\rm comp}$$R_{\rm comp}^3$$P^{-3}$, which, for the mass ($M_{\rm comp}$) and radius ($R_{\rm comp}$) of the companion star on the lower part of the main sequence and the usual Roche lobe geometry, comes to $\dot{J}_{\rm MBM}$$\propto$$P^1$.  So the MBM is just the surmise that the AML behaves as $\dot{J}_{\rm MBM}$$\propto$$P$, then with standard physics and stellar properties producing a single unique track for binary evolution.  Knigge emphatically states `Theoretically, all CVs with initially unevolved donors are expected to quickly join on to a unique evolution track, whose properties are determined solely by the mechanism for AML from the system.'

The core assumption of MBM is the one power-law `recipe' of $\dot{J}_{\rm MBM}$$\propto$$P$.  But $\dot{J}$ cannot be measured.  Rather, we can measure $\dot{P}$, and this is simply related as $\dot{P}_{\rm MBM}/P = 3 \dot{J}_{\rm MBM}/J$.  So MBM has its most-fundamental prediction that the observed $\dot{P}$ follows the unique track of K2011 as a function of $P$.  Now we have a required and exacting prediction that can be tested for the most fundamental assumption of MBM.

Many CVs have had reported $\dot{P}$ measures, but the majority of the older values are poor because of short observing intervals, are corrupted by various data and analysis errors, and are subjected to bad analyses such as claimed sinewaves fitting jitter and noise.  Most critical, for the purposes of this paper, no one has considered $\dot{P}$ values for more than one CV and looking at the big picture of testing the MBM.  Schaefer (2023) is the preliminary study for this paper, measuring $\dot{P}$ for 8 classical novae (CNe) and 6 recurrent novae (RNe).  The many measured $\dot{P}$ values were strongly contradictory to the required predictions of the MBM.  In particular, 8 out of 20 $\dot{P}$ measures have {\it positive} values, with such being impossible by MBM.  In particular, the large changes in $\dot{P}$ from inter-eruption interval to the next inter-eruption interval for U Sco, T CrB, and CI Aql (having nova events that are identical spectroscopically and photometrically) is impossible within MBM.  In particular, even the negative-$\dot{P}$ systems have most of them with period-changes up to 10,000$\times$ larger in size than required by MBM.  The conclusion is that nova CVs almost always have $\dot{P}$ and $\dot{J}$ orders-of-magnitude different from the required MBM predictions.  Thus, the preliminary work for two classes of CVs shows that the MBM is catastrophically in error because its one core assumption for AML is greatly wrong for almost all systems.

For such a sweeping conclusion, at odds with such a venerable consensus model, we need further tests.  For example, perhaps the MBM failures are due to the nova events themselves, so we should test MBM for CVs that do not have any proximate nova eruptions.  The plan for this paper is to measure, collect, and improve well-measured $\dot{P}$ values for many CVs of all types and use these to test the most-fundamental prediction of the MBM.

\section{$\dot{P}$ MEASURES FOR 52 CATACLYSMIC VARIABLES}

I collect times of eclipses and photometric minima for 52 CVs, with the sources being many of my own telescope observations, from many papers in the literature, and from light curves in many public sources.  For observations from 1890--1989, I have measured by-eye large numbers of magnitudes from the archival Harvard plates, and supplemented with the light curves from the Digital Access to a Sky Century $@$ Harvard (DASCH)\footnote{\url{http://dasch.rc.fas.harvard.edu/lightcurve.php}} programme, with J. Grindlay (Harvard) as Principle Investigator.  For  CCD light curves over the last two decades, the American Association of Variable Star Observers (AAVSO) International Database\footnote{\url{https://www.aavso.org/data-download}} archives a vast quantity of professional-quality magnitudes for most of the CVs.  The Zwicky Transient Factory\footnote{\url{https://irsa.ipac.caltech.edu/cgi-bin/Gator/nph-scan?projshort=ZTF}} (ZTF) produces light curves in three bands from 2018--2023 for many CVs.  The amazing light curves from the {\it TESS} satellite from 2018 to present are publicly available from the Mikulski Archive for Space Telescopes (MAST)\footnote{\url{https://mast.stsci.edu/portal/Mashup/Clients/Mast/Portal.html}}.  Extensive collections of published eclipse times from widely-scatter bulletins and such are presented in Bob Nelson's Database of Eclipsing Binary $O-C$ Files\footnote{\url{https://www.aavso.org/bob-nelsons-o-c-files}} and in the Czech $O-C$ Gateway\footnote{\url{http://var2.astro.cz/ocgate/index.php}}.

My particular standard methods, procedures, and techniques for measuring eclipse/minimum times, creating consistent $O-C$ curves, and measuring the period changes are presented in prior papers of mine, see Schaefer (2011; 2020a; 2020b; 2023).  A technical point that needs explicit statement in this paper is that the $\dot{P}$ values are dimensionless (or equivalently with units of seconds-per-second or days-per-day) with the quadratic coefficient for the $O-C$ parabola equalling $\frac{1}{2}P\dot{P}$.  For some of the stars, a high timing accuracy is needed with linearity across decades, so the eclipse/minimum times in the native heliocentric Julian Day (HJD) system have to be converted to barycentric Julian Day (BJD, or more specifically, BJD$_{\rm TDB}$) through the applet\footnote{\url{https://astroutils.astronomy.osu.edu/time/hjd2bjd.html}} provided by Eastman, Siverd, \& Gaudi (2010)

In Section 2, 17 individual systems are discussed in subsections, in order of $P$.  These are followed by subsections with discussions for measuring $\dot{P}$ for 5 AM CVn stars, 3 Compact Binary Supersoft Sources (CBSSs), 13 Novalike systems of the SW Sex class, 8 CNe, and 6 RNe.  These $\dot{P}$ values are collected in Table 1 and Fig. 1.

\startlongtable
\begin{deluxetable*}{llllllc}
\tablenum{1}
\tablecaption{$\dot{P}$ measures for 52 CVs}
\tablehead{
\colhead{CV} & \colhead{Class} & \colhead{Primary} & \colhead{Companion} & \colhead{$P$} &
\colhead{$\Delta Y$} & \colhead{$\dot{P}$}  \\
\colhead{} & \colhead{} & \colhead{} & \colhead{} & \colhead{(days)} &
\colhead{(years)} & \colhead{(10$^{-12}$)}  }
\startdata
HM Cnc	&	AM CVn	&	WD	&	WD	&	0.00372	&	2000--2022	&	-36.57	$\pm$	0.01	\\
V407 Vul	&	AM CVn	&	WD	&	WD	&	0.00660	&	1993--2003	&	3.17	$\pm$	0.10	\\
ES Cet	&	AM CVn	&	WD	&	WD	&	0.00717	&	2003--2017	&	3.18	$\pm$	0.11	\\
AM CVn	&	AM CVn	&	WD	&	WD	&	0.0119	&	1992--2019	&	-27	$\pm$	3	\\
YZ LMi	&	AM CVn	&	WD	&	WD	&	0.0197	&	2006--2012	&	0.39	$\pm$	0.05	\\
WZ Sge	&	DN	&	WD	&	MS	&	0.0567	&	1961--2017	&	-0.102	$\pm$	0.014	\\
V2051 Oph	&	DN	&	WD	&	MS	&	0.0624	&	1924--2023	&	0.44	$\pm$	0.20	\\
OY Car	&	DN	&	WD	&	MS	&	0.0631	&	1979--2023	&	6-segments			\\
EX Hya	&	DN	&	WD	&	MS	&	0.0682	&	1962--2023	&	-0.72	$\pm$	0.05	\\
HT Cas	&	DN	&	WD	&	MS	&	0.0736	&	1978--2022	&	-2.85	$\pm$	0.18	\\
Z Cha	&	DN	&	WD	&	MS	&	0.0745	&	1969--2021	&	4-segments			\\
T Pyx	&	RN	&	WD	&	MS	&	0.0762	&	1986--2011	&	649	$\pm$	7	\\
T Pyx	&	RN	&	WD	&	MS	&	0.0762	&	2011--2022	&	367	$\pm$	27	\\
DV UMa	&	DN	&	WD	&	MS	&	0.0859	&	1987--2021	&	0.28	$\pm$	0.08	\\
IM Nor	&	RN	&	WD	&	MS	&	0.103	&	2003--2021	&	25.1	$\pm$	0.7	\\
V482 Cam	&	N-like	&	WD	&	MS	&	0.134	&	2001--2020	&	-4.5	$\pm$	1.5	\\
SW Sex	&	N-like	&	WD	&	MS	&	0.135	&	1980--2023	&	3-curves			\\
DW UMa	&	N-like	&	WD	&	MS	&	0.137	&	1984--2023	&	2.49	$\pm$	0.73	\\
V1315 Aql	&	N-like	&	WD	&	MS	&	0.140	&	1984--2022	&	9.74	$\pm$	0.43	\\
TT Tri	&	N-like	&	WD	&	MS	&	0.140	&	2002--2022	&	-60.2	$\pm$	4.3	\\
V1500 Cyg	&	Nova	&	WD	&	MS	&	0.140	&	1978--2022	&	-27	$\pm$	10	\\
RR Pic	&	Nova	&	WD	&	MS	&	0.145	&	1945--2021	&	95.8	$\pm$	3.4	\\
PX And	&	N-like	&	WD	&	MS	&	0.146	&	1990--2020	&	-1.37	$\pm$	1.37	\\
V1024 Cep	&	N-like	&	WD	&	MS	&	0.149	&	2000--2021	&	25.6	$\pm$	6.7	\\
HS 0220	&	N-like	&	WD	&	MS	&	0.149	&	2002--2022	&	-18.8	$\pm$	2.7	\\
BP Lyn	&	N-like	&	WD	&	MS	&	0.153	&	1989--2021	&	19.6	$\pm$	3.9	\\
BH Lyn	&	N-like	&	WD	&	MS	&	0.156	&	1989--2021	&	-7.073	$\pm$	0.039	\\
IP Peg	&	DN	&	WD	&	MS	&	0.158	&	1984--2022	&	-24.2	$\pm$	1.8	\\
LX Ser	&	N-like	&	WD	&	MS	&	0.158	&	1980--2022	&	5.2	$\pm$	0.3	\\
UU Aqr	&	N-like	&	WD	&	MS	&	0.164	&	1985--2022	&	-20.8	$\pm$	1.1	\\
V1776 Cyg	&	N-like	&	WD	&	MS	&	0.165	&	1987--2021	&	-6.1	$\pm$	2.4	\\
U Gem	&	DN	&	WD	&	MS	&	0.177	&	1876--2021	&	4.8	$\pm$	0.7	\\
DQ Her	&	Nova	&	WD	&	MS	&	0.194	&	1954--2022	&	0.13	$\pm$	0.12	\\
UX UMa	&	N-like	&	WD	&	MS	&	0.197	&	1914--2023	&	-2	$\pm$	1	\\
T Aur	&	Nova	&	WD	&	MS	&	0.204	&	1954--2021	&	-5.4	$\pm$	2.4	\\
V617 Sgr	&	CBSS	&	WD	&	MS	&	0.207	&	1987--2022	&	460	$\pm$	8	\\
EX Dra	&	DN	&	WD	&	MS	&	0.210	&	1991--2022	&	30.5	$\pm$	3.2	\\
HR Del	&	Nova	&	WD	&	MS	&	0.214	&	1967--2022	&	163	$\pm$	37	\\
RW Tri	&	N-like	&	WD	&	MS	&	0.232	&	1891--2022	&	-4.6	$\pm$	0.8	\\
1RXS J0644	&	N-like	&	WD	&	MS	&	0.269	&	2005--2021	&	-89	$\pm$	15	\\
EM Cyg	&	N-like	&	WD	&	MS	&	0.291	&	1962--2023	&	6.6	$\pm$	3.4	\\
AC Cnc	&	DN	&	WD	&	MS	&	0.300	&	1952--2023	&	-22.9	$\pm$	3.4	\\
V363 Aur	&	N-like	&	WD	&	MS	&	0.321	&	1980--2023	&	-189.9	$\pm$	1.2	\\
BT Mon	&	Nova	&	WD	&	MS	&	0.334	&	1941--2020	&	-68	$\pm$	3	\\
QZ Aur	&	Nova	&	WD	&	MS	&	0.358	&	1990--2021	&	-39	$\pm$	14	\\
WX Cen	&	CBSS	&	WD	&	MS	&	0.417	&	1901--2021	&	-1200	$\pm$	100	\\
V Sge	&	V Sge	&	WD	&	MS	&	0.514	&	1903--2019	&	-506	$\pm$	6	\\
CI Aql	&	RN	&	WD	&	MS	&	0.618	&	1991--2001	&	-470	$\pm$	600	\\
CI Aql	&	RN	&	WD	&	MS	&	0.618	&	2001--2022	&	-516	$\pm$	45	\\
QR And	&	CBSS	&	WD	&	MS	&	0.660	&	1941--2022	&	1130	$\pm$	40	\\
U Sco	&	RN	&	WD	&	subG	&	1.23	&	1987--1999	&	-3200	$\pm$	1900	\\
U Sco	&	RN	&	WD	&	subG	&	1.23	&	1999--2010	&	-1100	$\pm$	1100	\\
U Sco	&	RN	&	WD	&	subG	&	1.23	&	2010--2016	&	-21100	$\pm$	3200	\\
U Sco	&	RN	&	WD	&	subG	&	1.23	&	2016--2022	&	-8800	$\pm$	2900	\\
V394 CrA	&	RN	&	WD	&	subG	&	1.52	&	1989--2021	&	-500	$\pm$	900	\\
V1017 Sgr	&	Nova	&	WD	&	subG	&	5.79	&	1919--2019	&	9500	$\pm$	2900	\\
T CrB	&	RN	&	WD	&	RG	&	227	&	1867--1946	&	1750000	$\pm$	4500000	\\
T CrB	&	RN	&	WD	&	RG	&	227	&	1947--2022	&	-8900000	$\pm$	1600000	\\\enddata
\end{deluxetable*}

\begin{figure}
\epsscale{1.2}
\plotone{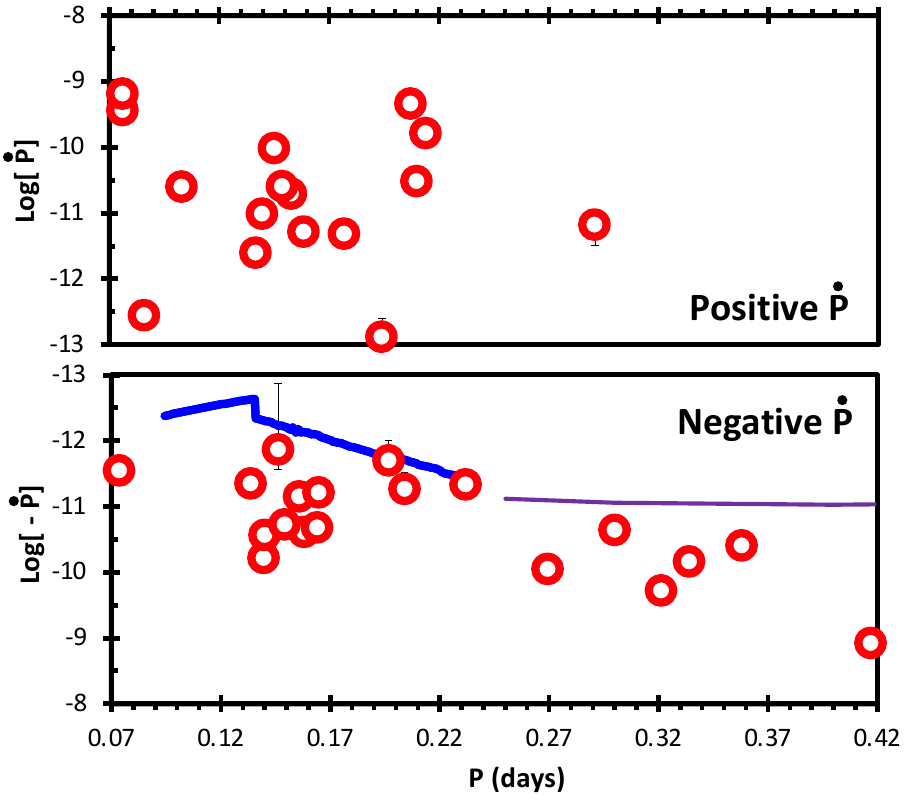}
\caption{$\dot{P}$ versus $P$.   This plot shows the distribution of $\dot{P}$ for a closeup for periods from 0.07 to 0.42 days, with this covering CVs with main sequence companion stars, as the core of the systems that MBM is designed to model.  The MBM prediction of K2011 is depicted with the blue curve, while the purple curve shows $\dot{P}_{\rm MBM}$ for longer period systems.  Each CV in this range is displayed as a red circle with a white center.  This plot is broken into two panels, with the upper panel for positive-$\dot{P}$ systems and the lower panel for negative-$\dot{P}$ systems, due to the awkwardness of plotting values that range over four magnitudes in size, both positive and negative.  One point of this plot is that 44 per cent of the systems have {\it positive} $\dot{P}$, with such being impossible in MBM.  Another point is that even the negative-$\dot{P}$ systems are all below the MBM predictions, by an average of 110$\times$.  That is to say, the MBM period change, and hence the angular momentum loss rate, are always in error, either with the wrong-sign or two orders-of-magnitude too small.  So the primary result of this paper is that the first test of the most-fundamental prediction of MBM shows that MBM is always failing by orders-of-magnitude in its angular momentum loss recipe. }
\end{figure}

\subsection{WZ Sge; Prototype Dwarf Nova with $P$=0.0567 days}

{\bf WZ Sge} is the prototype of a class of dwarf nova (DN), characterized by rare, large-amplitude, and long-duration super-outbursts.  WZ Sge is a very fast eclipsing binary, with a period of 0.0567 days (81.6 minutes), making it a prototype of the period-bounce systems.  

Patterson et al. (2018) has performed a wonderful and complete $O-C$ study of WZ Sge.  They collected 167 eclipse times 1961--2017, the majority taken by the co-authors.  They constructed an $O-C$ curve, which I have adapted into my Fig. 2.  Their fitted parabola has $\dot{P}$=($-$1.02$\pm$0.14)$\times$10$^{-13}$, and this represents the general evolution over the last six decades.  Patterson notes that there are significant-but-small systematic deviations from the best-fitting parabola, in particular the `wiggle' from 2002--2015 where the $O-C$ curve has a downward excursion with eclipses coming roughly 0.0002 days (17 seconds) early centered on 2008.

The Patterson paper presents four high-level results from the WZ Sge $O-C$ curve:  {\bf (1)} Patterson considered whether the $O-C$ had a sinewave component, even running a Fourier transform on the residuals, but correctly concluded that the `wiggles' are {\it not} caused by some periodic signal.  This is refreshing to see the wisdom of not jumping on the band-wagon of `discovering' a new planet.  {\bf (2)} Patterson pointed to a mechanism that can easily make for wiggles in the $O-C$ curve, and that is movement of the hot spot, where the accretion stream impacts the accretion disc.  The hot spot must move around, both azimuthally and radially, as the accretion disc and accretion rate change over the seasons.  For WZ Sge, the measured times of mid-eclipse are for the eclipse of the hot spot (not the WD), so movement of the hot spot will make for shifts in the time of the eclipse.  This is a reasonable and inevitable mechanism that can easily explain the bumps and wiggles in the $O-C$ curves for WZ Sge and many other CVs.  $O-C$ bumps from this mechanism must be small (when compared to $P$), they can be on most any time-scale, and they must average out to some constant phase of mid-eclipse.  So in the long-run, these bumps can only provide a jittering noise on top of the overall evolutionary parabola.  {\bf (3)} The {\it negative} $\dot{P}$ shows that WZ Sge has not yet gone past the bounce at the minimum period.  {\bf (4)} Patterson compares the $O-C$ diagrams for WZ Sge, other short-period DNe, and detached post-common-envelope binaries, highlighting that they all show the same nonperiodic bumps and wiggles with time-scales of 5--30 years.  Patterson concludes `The commonality of this wander suggests an origin in the one ingredient present in all three classes: the low-mass secondary stars.'

I can extend Patterson's third point, presenting a new challenge to standard theory.  Standard theory for the period-bouncers is that the minimum $P$ is 82 minutes and that any system near this bounce must have very small $P$ changes as the period bounces, for $|\dot{P}|$$<$2$\times$10$^{-14}$ (K2011).  WZ Sge has a $\dot{P}$ that is 5$\times$ larger, meaning that the period still is far above the minimum period.  When combined with the period of 81.6 minutes already being significantly below the standard minimum period of 82 minutes, this means that WZ Sge has an operational bounce period greatly smaller than the accepted minimum period.

\begin{figure}
\epsscale{1.2}
\plotone{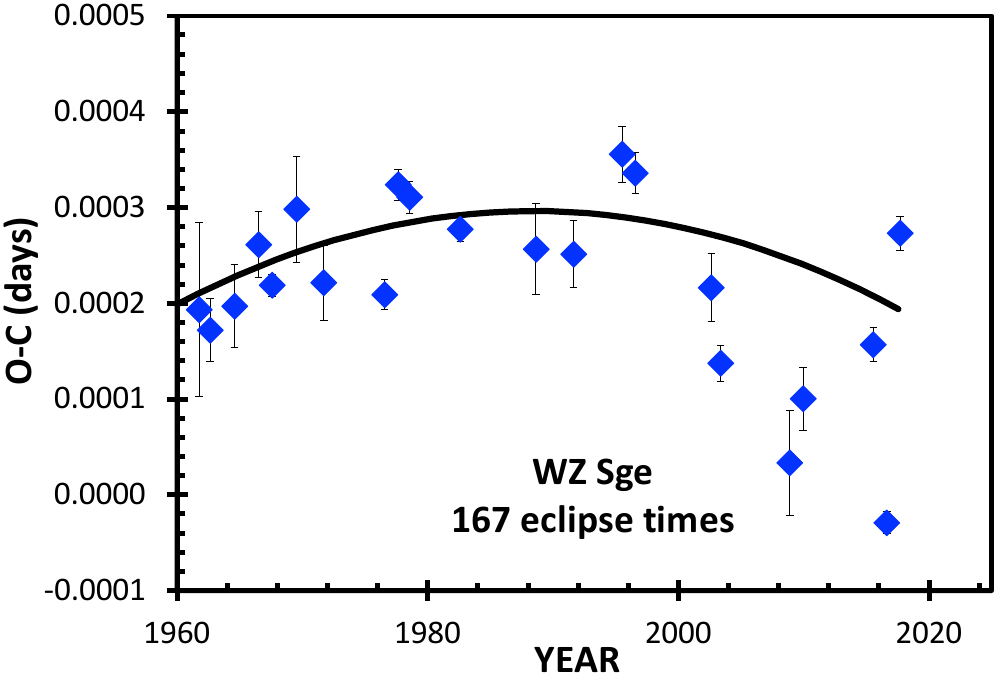}
\caption{$O-C$ curve for WZ Sge.  This $O-C$ plot is adopted from Patterson et al. (2018), using only their quoted times of mid-eclipse (blue diamonds), their ephemeris (with epoch BJD 2437547.72863 and period 0.056687847 days), and their best-fitting parabola (the black curve with $\dot{P}$=($-$1.02$\pm$0.14)$\times$10$^{-13}$).  My only change is in the display with their 167 eclipse times binned together to form yearly averages.  The parabola provides a reasonable description of the overall period change over six decades.  But the $O-C$ curve has significant `wiggles' or bumps as deviations from the parabola.  Patterson reasonably argues that these bumps are not real period changes, but rather caused by secular movements of the position of the hot-spot, as expected for the usual variations in the accretion rate.  This mechanism can provide small $O-C$ bumps for many CVs, for which RW Tri, HT Cas, and UX UMa might provide clear examples.  Such moving-hot-spot bumps can only provide small noise above and below the basic evolutionary parabola.  If the $O-C$ curve contains only one or two substantial bumps, the evolutionary $\dot{P}$ can be hidden, or an unwise analyst might claim a sinewave periodicity. }
\end{figure}

\begin{deluxetable}{llrl}
\tablenum{2}
\tablecaption{117 V2051 Oph eclipse times (full table in machine readable formation in the electronic article)}
\tablehead{
\colhead{Eclipse Time (BJD)} & \colhead{Year} & \colhead{$O-C$} &
\colhead{Source}  }
\startdata
2444787.32187	$\pm$	0.00010	&	1981.50	&	0.00015	&	Warner et al.	\\
...	&	...	&	...	&	...	\\
2450956.50478	$\pm$	0.00014	&	1998.39	&	-0.00049	&	AAVSO	\\
2452386.47803	$\pm$	0.00004	&	2002.30	&	0.00019	&	AAVSO	\\
2452408.45264	$\pm$	0.00002	&	2002.36	&	0.00020	&	AAVSO	\\
2452427.61790	$\pm$	0.00006	&	2002.42	&	0.00011	&	AAVSO	\\
2452427.61805	$\pm$	0.00010	&	2002.42	&	0.00026	&	AAVSO	\\
2452431.30117	$\pm$	0.00004	&	2002.43	&	0.00013	&	AAVSO	\\
2452751.49310	$\pm$	0.00005	&	2003.30	&	-0.00043	&	AAVSO	\\
2452751.49353	$\pm$	0.00006	&	2003.30	&	0.00000	&	AAVSO	\\
2454705.92261	$\pm$	0.00005	&	2008.65	&	0.00006	&	AAVSO	\\
2454707.04615	$\pm$	0.00002	&	2008.66	&	-0.00010	&	AAVSO	\\
2454707.98266	$\pm$	0.00011	&	2008.66	&	0.00000	&	AAVSO	\\
2455383.07771	$\pm$	0.00016	&	2010.51	&	0.00017	&	AAVSO	\\
2456117.66598	$\pm$	0.00009	&	2012.52	&	-0.00019	&	AAVSO	\\
2457179.87638	$\pm$	0.00009	&	2015.43	&	0.00017	&	AAVSO	\\
2457183.80919	$\pm$	0.00006	&	2015.44	&	0.00002	&	AAVSO	\\
2457187.74241	$\pm$	0.00010	&	2015.45	&	0.00029	&	AAVSO	\\
2457188.67847	$\pm$	0.00009	&	2015.45	&	-0.00007	&	AAVSO	\\
2457191.79978	$\pm$	0.00015	&	2015.46	&	-0.00015	&	AAVSO	\\
2458294.58841	$\pm$	0.00019	&	2018.48	&	0.00033	&	AAVSO	\\
2458299.58249	$\pm$	0.00010	&	2018.49	&	0.00019	&	AAVSO	\\
2458299.58269	$\pm$	0.00006	&	2018.49	&	0.00039	&	AAVSO	\\
2458303.01610	$\pm$	0.00007	&	2018.50	&	0.00026	&	ZTF 2018	\\
2458666.03395	$\pm$	0.00008	&	2019.50	&	0.00010	&	ZTF 2019	\\
2459383.01808	$\pm$	0.00016	&	2021.46	&	0.00027	&	ZTF 2021	\\
2459747.03492	$\pm$	0.00024	&	2022.46	&	0.00025	&	ZTF 2022	\\
\enddata
\tablecomments{{\bf (1)} The calculated $O-C$ values are in units of days for a linear ephemeris with epoch BJD 2452751.49353 and period 0.06242786 days.  
{\bf (2)}  Sources in the full table include references Beuermann \& Pakull (1984), Cook \& Brunt (1983), Echevarria \& Alvarez (1993), Hollander, Kraakman, \& van Paradijs (1993), Warner \& Cropper (1983), Warner \& O'Donoghue (1987), Watts \& Watson (1986), and Watts et al. (1986).}
\end{deluxetable}

\subsection{V2051 Oph; Dwarf Novae with $P$=0.0624 days}

\begin{figure}
\epsscale{1.16}
\plotone{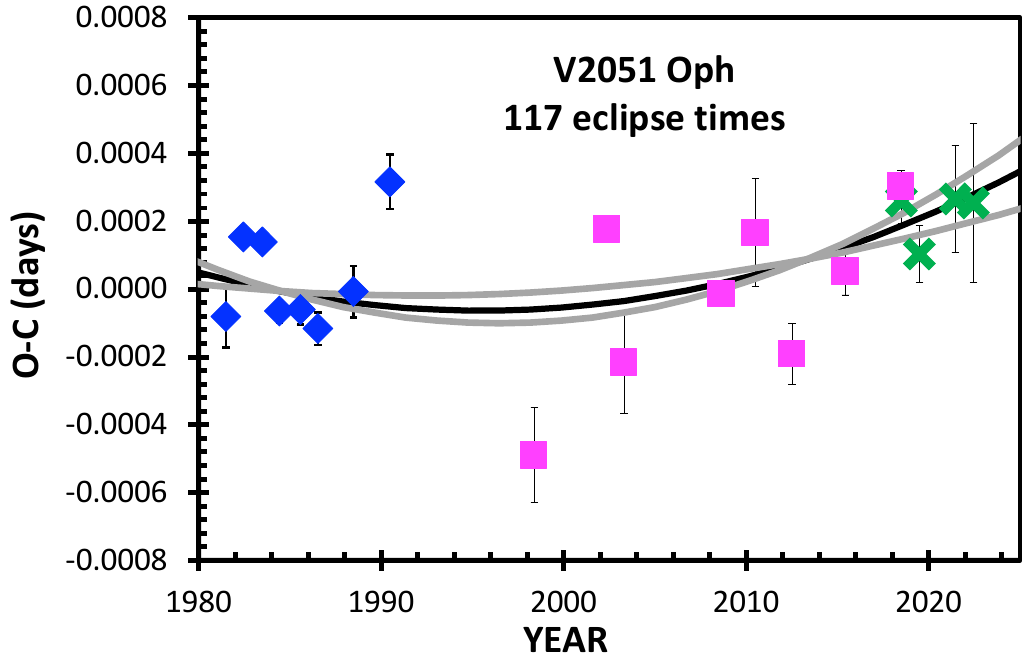}
\caption{$O-C$ curve for V2051 Oph.  This curve contains the blue diamonds for yearly-averages of all useable measures taken from the literature, with the magenta squares for AAVSO yearly-average times, and with the green $\times$ symbols for the ZTF seasonal-average times.  The $O-C$ curve shows no obvious period changes, with a near-zero $\dot{P}$ with some scatter.  The scatter is relatively small at $\pm$13 seconds, and consistent with the usual timing jitter arising from flickering in the light curve.  A formal fit for a parabola gives $\dot{P}$=($+$4.4$\pm$2.0)$\times$10$^{-13}$. }
\end{figure}

{\bf V2051 Oph} is an ordinary dwarf novae, with eclipses that are 1.5 mag deep with a sharp triangular shape with total duration of 0.10 in phase, all for a period of 0.0624 days (89 minutes).  Many times of mid-eclipse have been reported in the literature for the years 1979--2015.  Period changes in the $O-C$ diagrams have been reported to be sinewaves with periods of 5.5, 6.87, 11, and 21.6 years superposed on top of a parabola, but the wide range of claimed periods indicates that all the claims are dubious and there are major problems in the data or analysis.  Reported $\dot{P}$ values of $+$1.27$\pm$0.77, $+$1.04$\pm$0.26, and $-$16.2$\pm$2.2 (all in units of 10$^{-12}$) come from Echevarria \& Alvarez (1993), Baptista et al. (2003), and Qian et al. (2015) respectively, with this large range of positive and negative signs further pointing to major problems with the data or analysis.

The problem is with the definition of the time of minimum.  The long traditional definition is to report the time of deepest minimum, with this reliably marking the instant of conjunction.  This minimum is variously found by fitting a parabola to the deepest parts of the eclipse, or by fitting an appropriate light curve template to the observed light curve.  However,  Baptista et al. (2003) and Qian et al. (2015) both have adopted alternate definitions for the minimum as being the time halfway from the ingress to the egress.  Their two definitions differ, with the Qian times of ingress and egress corresponding to high up in the eclipse, for which the asymmetrical eclipse profile makes for a large offset from the time of minimum light.  These alternate definitions might be adequate for a symmetrical eclipse light curve, but such calculated times are far from minimum light for a light curve with substantial and variable asymmetries, like for V2051 Oph.  Such definitions are poor for the case of V2051 Oph, because they measure anisotropies in the disc, rather than the position of the WD.  This poor definition makes for a systematic offset between times reported by the traditional definition, as seen in the top panel of fig. 2 of Baptista et al. (2003), even to the extent of changing the {\it shape} of the $O-C$ curve and creating apparent sinewaves.  The times reported by Qian are systematically offset by over 2 minutes from those of Baptista and everyone else, with this being the only cause of their claimed large negative $\dot{P}$.  So, the times reported by Baptista and Qian have no utility for measuring the long-term evolution.

The solution to see the long-term evolution is to only use the consistent times from the common definition (available only for 1981--1990), and to extend the $O-C$ curve by a factor of over 4$\times$.  For this, I have collected light curves and times of minimum light from the literature with the traditional definition (1981--1990), from the AAVSO (1998-2018), and from ZTF (2018--2022).

I have a total of 117 times of minimum light, stretching from 1981--2022, all using the same consistent traditional definition of minimum light (see Table 2).  My new times are given explicitly in the Table, while most of the 92 times from the literature (and their citations) appear only in the Supplementary Material.   With this, I construct an $O-C$ curve in Fig. 3.  The $O-C$ curve appears with no blatant curvature and no sinewave modulation.  There is substantial scatter about the best-fitting curve, more than the scatter expected from the quoted measurement errors, with this undoubtedly due to the usual flickering causing small random shifts in the measured times of minimum light.  Quantitatively, this additional variance has an RMS of 0.00018 days, and this must be added in quadrature to the measurement errors so as to get the total uncertainty in each time for use in chi-square fits.  I fit a parabola and derive $\dot{P}$=($+$4.4$\pm$2.0)$\times$10$^{-13}$.

\subsection{OY Car; Dwarf Nova with $P$=0.0631 days}

\begin{deluxetable}{llrl}
\tablenum{3}
\tablecaption{142 eclipse times for OY Car (full table in machine readable formation in the electronic article)}
\tablehead{
\colhead{Eclipse Time (BJD)} & \colhead{Year} & \colhead{$O-C$} &
\colhead{Source}  }
\startdata
2444042.535643	$\pm$	0.000100	&	1979.46	&	-0.00059	&	Vogt et al.	\\
2444044.492403	$\pm$	0.000100	&	1979.46	&	-0.00057	&	Vogt et al.	\\
...			&	...	&	...	&	...	\\
2448594.879420	$\pm$	0.000047	&	1991.92	&	0.00003	&	Horne et al.	\\
2450162.550603	$\pm$	0.000029	&	1996.22	&	0.00029	&	K. Horne	\\
2458583.006601	$\pm$	0.000009	&	2019.27	&	0.00063	&	TESS 10	\\
2458603.015909	$\pm$	0.000021	&	2019.33	&	0.00061	&	TESS 11	\\
2459324.046357	$\pm$	0.000012	&	2021.30	&	0.00090	&	TESS 37	\\
2459347.022421	$\pm$	0.000004	&	2021.36	&	0.00095	&	TESS 38	\\
2460028.034446	$\pm$	0.000013	&	2023.23	&	0.00148	&	TESS 63	\\
2460055.050201	$\pm$	0.000009	&	2023.30	&	0.00149	&	TESS 64	\\
2460081.056047	$\pm$	0.000010	&	2023.37	&	0.00152	&	TESS 65	\\
\enddata	
\tablecomments{{\bf (1)}~~The $O-C$ values, in days, with the linear ephemeris of Pilar\u{c}ik  et al. (2018) for epoch BJD 2443993.554404 and period 0.06312091014 days.  {\bf (2)}~~Sources in the full table include references Cook (1985a), Greenhill et al. (2006), Horne et al. (1994), Horne (2023), Pilarcik et al. (2018), Pratt et al. (1999), Schoembs et al. (1987), Schoembs \& Hartmann (1983), Vogt et al. (1981), and Wood et al. (1989).}
\end{deluxetable}

\begin{figure}
\epsscale{1.16}
\plotone{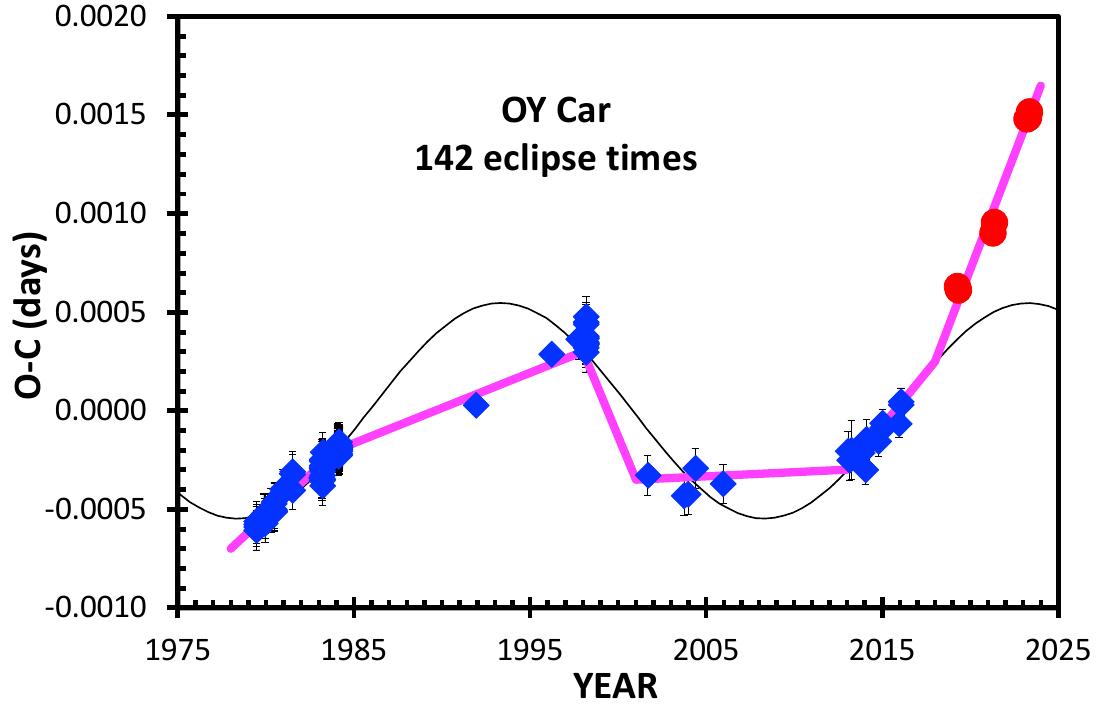}
\caption{$O-C$ curve for OY Car.  This plot has 142 eclipse times, with 133 coming from the literature plus 2 from light curves provided by K. Horne (blue diamonds), while 7 eclipse times come from {\it TESS} (overlapping red circles) with very high accuracy.  The general run of the $O-C$ curve cannot be reasonably described by any parabola.  And the previous conjectural sinusoidal shape (the thin black curve) is now completely refuted because the {\it TESS} data deviate greatly from its predictions for 1991 and 2019--2023.  Rather, the observed $O-C$ curve can be accurately described with a 6-segment broken line (the thick magenta curve).  Such a curve has no known physical mechanism or theoretical explanation.  Importantly, the period changes for this unknown mechanism are dominating over any evolutionary parabola (even with bumps and wiggles superposed), with the implication that the MBM prescription is greatly in error for this one system.}
\end{figure}

{\bf OY Car} is an ordinary dwarf nova of the SU UMa class.  Eclipses were discovered in 1979, with the very short period of 0.0631 days.  Collections of eclipse timings soon revealed a non-linear $O-C$ curve, with first one bump up, and then one bump down.  Various workers fit sinewaves to the $O-C$ curve, and such can always be fitted to data with one up-bump followed by one down-bump.  These claims can be tested with recent eclipse timings from {\it TESS}. 

For the times from the literature, I have compiled 133 eclipse times reported in the literature, as listed in Table 3.  Among published eclipse times, I have not used the four outliers from Greenhill et al. (2006), as these outliers are exactly the times that are contradicted by their quoted $O-C$ in their table and plot.  Further, the first quoted eclipse times in both Vogt et al. (1981) and Pratt et al. (1999), plus all the times in Nicholson (2009) cannot be used, because OY Car was in eruption at the times.  Finally, the eclipse times in Han et al. (2015) cannot be used, as they adopted a definition of the time of mid-eclipse that was different from all other workers, leading to an 80-second difference in the $O-C$.  Many sources do not quote any sort of error, and for these I have adopted an error bar of 0.0001 days, with this being typical for the RMS scatter in the $O-C$ within the various observing runs.

{\it TESS} has good nearly-gap-free photometry in 7 Sectors, covering the years 2019, 2021, and 2023.  The two Sectors in 2021 have 20-second exposure times, while the remaining five Sectors in 2019 and 2023 have 120-second exposures.  The 20-second light curves adequately show the steep photometric steps associated with the ingress and egress of the WD, while the 120-second light curves have a smeared version that retains the basic structure.  With these time resolutions being substantially poorer than for the high-speed photometry for all the literature eclipse times for individual eclipses, I cannot usefully measure individual eclipse times in the {\it TESS} light curves.  So I have fitted a realistic template to the folded {\it TESS} light curve for each Sector individually, only for orbital phases from $-$0.04 to $+$0.04.  The template adopted was the average eclipse light curve quantified in Vogt et al. (1981), with this providing a good representation of the observed light curves.  This template has the zero-phase corresponding to the time halfway between the WD ingress and egress, so as to match the definitions used by all other papers with OY Car eclipse times.  The folded light curves contain up to 420 individual eclipses (with the elimination of all in-eruption data), so the final eclipse time representative of the Sector has the timing jitter from flickering reduced by a factor of $\sqrt{420}$, or 20$\times$.

For the exquisite eclipse time with the {\it Hubble Space Telescope} in the ultraviolet and optical (Horne et al. 1994), K. Horne (University of St. Andrews) has passed along their original timing data, with the ingress and egress steps sharply defined.  Prof. Horne has also passed along a light curve showing two sequential eclipses on 1996 March 20, taken with the Danish 1.5-m telescope at the European Southern Observatory with 26-s time resolution.  I have fitted the folded light curve to the Vogt template, then converted the eclipse time into BJD.  Both eclipse times from Horne's light curves are listed in Table 3.

Fig. 4 shows my resultant $O-C$ curve derived from the 141 eclipse times.  This $O-C$ curve is far from any line or parabola.  I cannot suggest any steady $\dot{P}$ that describes the general run over 44-years, so I have no measure of the evolutionary period change.

The upward bump in the 1990s has inspired claims of a periodic sinusoidal modulation in the $O-C$ curve, variously with cycles of 35.5$\pm$3.5 (Greenhill et al. 2006), 35 (Han et al. 2015), and 28.0$\pm$1.1 years (Pilar\u{c}ik et al. 2018).  The Pilar\u{c}ik et al. (2018) claim is displayed as the thin black curve in Fig. 4.  These claims were ill-advised because the data only covered $\sim$1 cycle, and such can never be significant, or even suggestive.  Further, their $O-C$ curve as-used had measures at only four epochs (three epochs for the two earliest sinewave claims), and all datasets with only three-or-four epochs must fit a sinusoid+line, so any measurement error or intrinsic variability for any cause could result in a claimed sinusoidal component, hence making the enterprise dubious.  And the claim of Han et al. (2015) was fumbled because their data set has a systematic error of 0.0009 days.  A more direct refutation of the three claims is simply that they are disproven by my seven new {\it TESS} eclipse times.

Rather, the $O-C$ curve can be described by six segments of a broken line (see the magenta curve in the figure).  The period changes at the breaks range from $-$0.78 to $+$0.70 parts-per million.

\subsection{EX Hya; Dwarf Nova with $P$=0.0682 days}

\begin{deluxetable}{llrl}
\tablenum{4}
\tablecaption{695 eclipse times for EX Hya (full table in machine readable formation in the electronic article)}
\tablehead{
\colhead{Eclipse Time (BJD)} & \colhead{Year} & \colhead{$O-C$} &
\colhead{Source}  }
\startdata
2437699.94170	$\pm$	0.00030	&	1962.09	&	0.00027	&	Vogt et al.	\\
...			&	...	&	...	&	...	\\
2458564.75882	$\pm$	0.00043	&	2019.22	&	0.00019	&	AAVSO (DKS)	\\
2458584.068703	$\pm$	0.000039	&	2019.27	&	-0.00010	&	TESS 10	\\
2458590.61877	$\pm$	0.00044	&	2019.29	&	-0.00048	&	AAVSO (DKS)	\\
2458590.68779	$\pm$	0.00045	&	2019.29	&	0.00030	&	AAVSO (DKS)	\\
2458597.71524	$\pm$	0.00044	&	2019.31	&	-0.00033	&	AAVSO (DKS)	\\
2458630.60415	$\pm$	0.00039	&	2019.40	&	-0.00013	&	AAVSO (DKS)	\\
2458633.60618	$\pm$	0.00032	&	2019.41	&	-0.00039	&	AAVSO (DKS)	\\
2459321.062495	$\pm$	0.000023	&	2021.29	&	-0.00004	&	TESS 37	\\
2460053.006960	$\pm$	0.000017	&	2023.29	&	0.00000	&	TESS 64	\\
\enddata	
\tablecomments{{\bf (1)}~~The calculated $O-C$ values are in units of days for a linear ephemeris with epoch BJD 2460053.00696 and period 0.0682338422 days.  {\bf (2)}~~Sources in the full table include references Bond \& Freeth (1988), Gilliland (1982), Hellier \& Sproats (1992), Honkova et al. (2013), Hill \& Watson (1984), Hill et al. (1986), Jablonski \& Busko (1985), Mumford (1976a), Sterken \& Vogt (1995), Vogt et al. (1980), Warner (1972), and Watson \& Rayner (1974).}
\end{deluxetable}

\begin{figure}
\epsscale{1.16}
\plotone{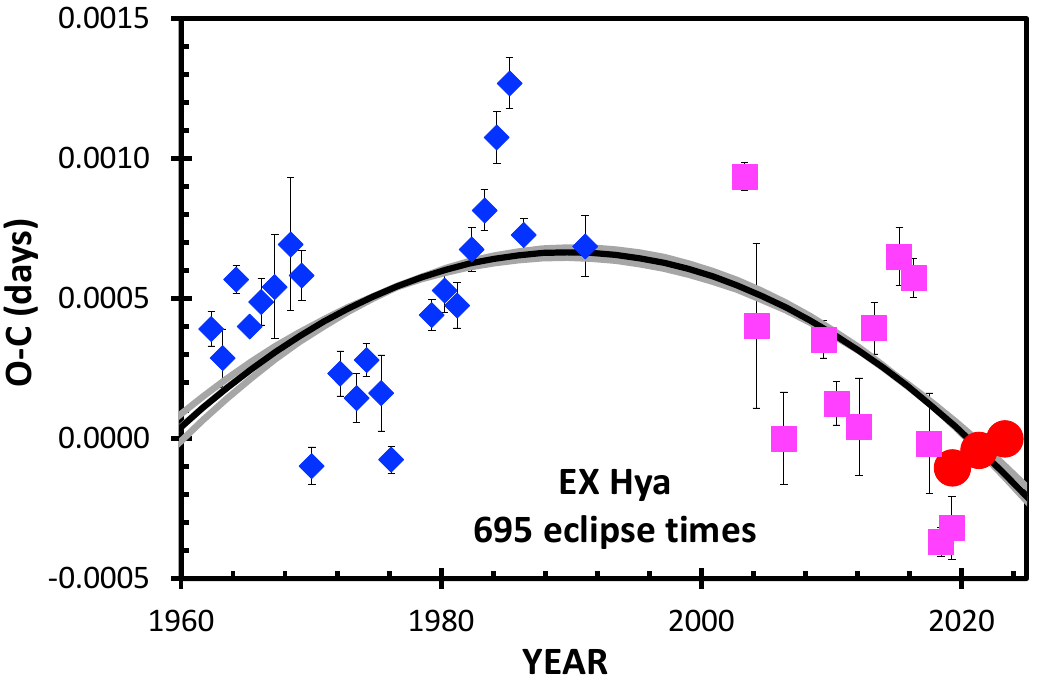}
\caption{$O-C$ curve for EX Hya.  This curve contains the yearly-averages of all useable measures taken from the literature (blue diamonds), the AAVSO yearly-average times (magenta squares), and the average times for each {\it TESS} Sector (red circles).  A formal fit for a parabola gives $\dot{P}$ equal to ($-$7.2$\pm$0.5)$\times$10$^{-13}$, as shown by the black curve, with the gray curves representing the 1-sigma uncertainty.  The measured $O-C$ curve is a simple parabola representing the long-term evolution) with `bumps' that represent real-but-small period changes on time-scales of 1--10 years.}
\end{figure}

{\bf EX Hya} is a bright nearby dwarf nova of the DQ Her class, which is an Intermediate Polar with a WD spin period of 67 minutes.  EX Hya has eclipses showing $P$=0.0682 days, but the fast flickering and the spin-modulation frequently camouflages the eclipses, while the jitter makes for substantial intrinsic variations in eclipse times.  

Many eclipse times have been published for 1962--1991, and I have collected 350 times from the literature.  To fill the gap after 1991, I have collected 342 eclipse times from 2003--2019 with light curves in the AAVSO database.  Further, I have fitted an average eclipse template to the {\it TESS} light curve from Sectors 10 (in year 2019), 37 (in 2021), and 64 (in 2023), with these light curves having 20-second time resolution and 355 eclipses each.  The result is 695 times of minimum 1962--2023 for the eclipses (see Table 4).

Fig. 5 is my $O-C$ curve, with yearly averages plotted to avoid crowding, plus the 3 {\it TESS} values.  The figure shows that there is substantial scatter larger than expected for the quoted error bars, and I attribute this to the inevitable timing jitter from ordinary flickering in the light curve.  Quantitatively, the reduced chi-square is near unity with a jitter noise of 0.0010 days added in quadrature with the measurement errors reported in Table 4.  With this, I have fitted a parabola to all 695 eclipse times, and I get $\dot{P}$=($-$7.2$\pm$0.5)$\times$10$^{-13}$.

The residuals from the parabola fit appear to have systematic structure, with high values around the 1960s and the mid-1980s, while low values appear around the 1970s.  This is the structure that has encouraged various prior claims for a periodic term in the $O-C$ curve, with periods proposed around 20 years.  But this sinewave has not continued in subsequent decades, so the periodicity is not real.  In the end, we are left with a good parabola for the evolution, with superposed bumps of unknown origin.

\subsection{HT Cas; Dwarf Nova with $P$=0.0736 days}

{\bf HT Cas} is a normal faint DN with superhumps discovered in 1978.  It shows deep eclipses in optical and X-ray light with a period of 0.0736 days and duration of 14 minutes.

Nine publications have reported eclipse times, as summarized in Borges et al. (2008).  Borges constructed an $O-C$ curve (essentially the blue diamonds in Fig. 6) and fitted a parabola, with $\dot{P}$=($-$3.72$\pm$0.19)$\times$10$^{-11}$, and a sinewave with cycle length of 36$\pm$4 years.  The sinewave component is ill-advised because a model with a parabola and sinewave is meaningless with only 6 effective epochs of eclipse, because the interval of observations is only 0.8$\times$ the cycle length so no sinewave can possibly be significant, and because the curvature from the parabola and the alleged short-sinewave will always be confounded together.

I have extended the $O-C$ curve up to 2022 by adding in eclipse times from light curves reported in the AAVSO and ZTF databases.  The AAVSO International Database has 23440 $CV$ magnitudes from 2012--2023, all measured with CCDs, mostly in time series covering many separate eclipses.  The time resolution was such that each eclipse was usually recorded at depth for only 4--6 times (with this being much poorer than the large-telescope measures from the literature), while the overall accuracy of the eclipse time is made good by having of-order one-hundred eclipses each year.  ZTF has 1685 $zg$ and $zr$ magnitudes from 2018.3--2023.2.  On two nights in late 2018, ZTF recorded long and fast times series that cover four eclipses, with reasonably good minimum times.  Aside from these two nights, the five observing seasons from 2018--2022 produce good folded light curves with many eclipses, with each eclipse represented by one magnitude each.  All the AAVSO and ZTF minimum times were converted to BJD and presented in Table 5.  The quoted error bars are measurement errors from chi-square fits, which, in this case substantially under-estimate the real total error, partly due to timing jitter from flickering and partly due to ordinary problems with minimum times by folding a sparse light curve.  The real total error of the AAVSO and ZTF times is likely $\sim$0.0003 days, as shown by the scatter seen in the $O-C$ curve.

I have constructed an $O-C$ curve with the 82 times from AAVSO, ZTF, and the sources in Borges et al. (2008), see Fig. 6.  The curve has a concave-down shape, for a decreasing $P$.  The best-fitting parabola (the black curve in the figure) has $\dot{P}$=($-$2.85$\pm$0.18)$\times$10$^{-12}$.  The shape of the $O-C$ curve is a good parabola, but the eclipse times around 1995 are systematically late by 17 seconds or so.  The claimed 36$\pm$4 year cycle is denied by the data from the last decade.

\begin{deluxetable}{llrl}
\tablenum{5}
\tablecaption{New eclipse times for HT Cas}
\tablehead{
\colhead{Eclipse Time (BJD)} & \colhead{Year} & \colhead{$O-C$} &
\colhead{Source}  }
\startdata
2456250.023333	$\pm$	0.000018	&	2012.883	&	-0.00132	&	AAVSO	\\
2456568.032226	$\pm$	0.000023	&	2013.754	&	-0.00105	&	AAVSO	\\
2458071.024323	$\pm$	0.000013	&	2017.869	&	-0.00107	&	AAVSO	\\
2458359.058409	$\pm$	0.000040	&	2018.657	&	-0.00153	&	ZTF	\\
2458459.071050	$\pm$	0.000029	&	2018.931	&	-0.00120	&	AAVSO	\\
2458474.684179	$\pm$	0.000050	&	2018.974	&	-0.00149	&	ZTF	\\
2458474.757869	$\pm$	0.000050	&	2018.974	&	-0.00150	&	ZTF	\\
2458475.715269	$\pm$	0.000050	&	2018.977	&	-0.00159	&	ZTF	\\
2458475.788829	$\pm$	0.000050	&	2018.977	&	-0.00153	&	ZTF	\\
2458552.013743	$\pm$	0.000039	&	2019.186	&	-0.00146	&	AAVSO	\\
2458773.029173	$\pm$	0.000056	&	2019.791	&	-0.00136	&	ZTF	\\
2459099.065257	$\pm$	0.000032	&	2020.683	&	-0.00144	&	ZTF	\\
2459463.030010	$\pm$	0.000090	&	2021.680	&	-0.00117	&	ZTF	\\
2459823.017462	$\pm$	0.000080	&	2022.665	&	-0.00124	&	ZTF	\\
\enddata	
\tablecomments{The $O-C$ values, in days, with the linear ephemeris of Borges et al. (2008) with epoch BJD 2443727.93804 and period 0.0736472029 days.}
\end{deluxetable}

\begin{figure}
\epsscale{1.16}
\plotone{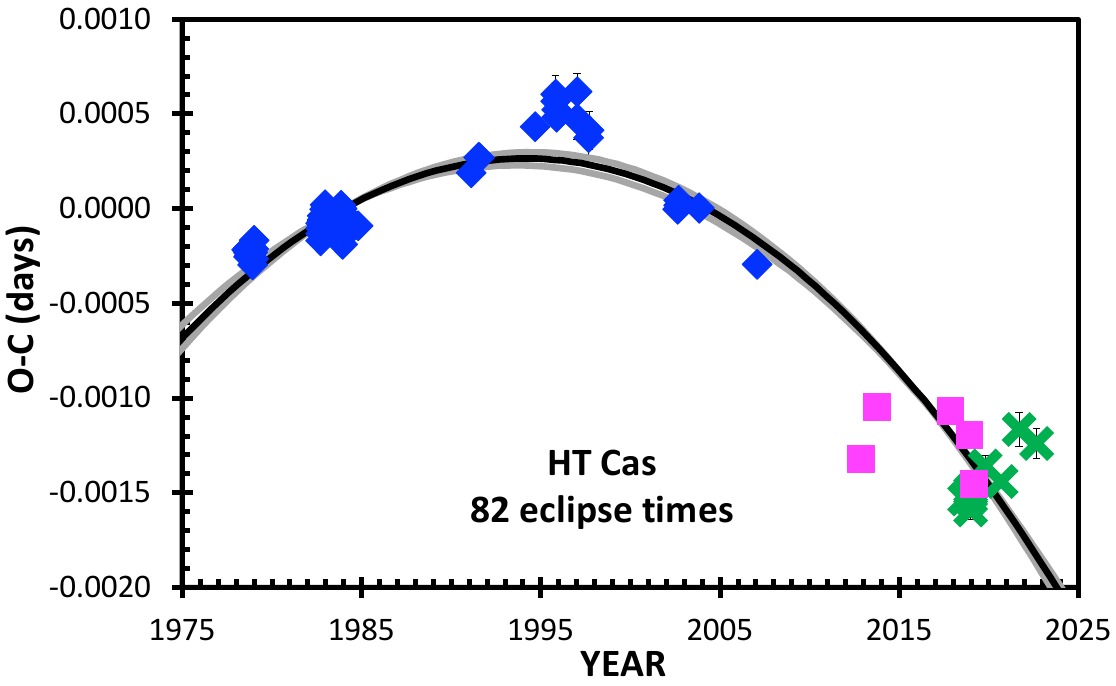}
\caption{$O-C$ curve for HT Cas.  This $O-C$ curve is created with eclipse times from the literature (see Borges et al. 2008) in the blue diamonds, from the AAVSO light curves in the magenta squares, and from the ZTF light curves in the green $\times$ symbols.  The best-fitting parabola is displayed as the black curve, while the one-sigma extremes are shown as gray curves.  We see a good parabola with $\dot{P}$=($-$2.85$\pm$0.18)$\times$10$^{-12}$, which appears to have a small possibly-significant bump around 1995.}
\end{figure}

\subsection{Z Cha; Dwarf Nova with $P$=0.0745 days}

{\bf Z Cha} is an ordinary DN of the SU UMa subclass, which means that it shows occasional  supermaxima and superhumps.  Z Cha is also an eclipsing system, with orbital period 0.0745 days, and deep eclipses.  The eclipse profiles have substantial variability from time-to-time and with the observing band, showing sharp ingress and egress edges for the WD and the bright spot.  This makes it hard to measure any stable phase, say of the conjunction, because the position of the bright spot moves around the accretion disc both radially and azimuthally, while the eclipse profile often has small and variable asymmetries around the bottom (e.g. Cook 1985b).  So single eclipse times have an intrinsic jitter, which is substantially larger than most formal measurement errors.

Eclipse times have been reported in many papers in the literature since early 1969.  For purposes of creating an $O-C$ curve, care must be taken to only use times for when Z Cha is in quiescence, because there are systematic time offsets that vary throughout the eruptions and supermaxima.  Previous compilations of Z Cha eclipse times have all included only roughly-half of the n-available times, included times from eruptions, and variously averaged together groups of times.  I have made a new compilation of eclipse times from the literature that includes all published times for when Z Cha is in quiescence.  Some of the papers use an alternative definition of bisecting the eclipse profile at half the eclipse depth or bisecting some perceived ingress/egress steps.  For the case of Z Cha, the difference between the bisector times and the minimum times is small compared to the overall variability, so I have chosen to make no attempts at correcting these times to the minimum times, and I just use the quoted times as reported.  The measures of Dai, Qian, \& Fern\'{a}ndez Laj\'{u}s (2009) likely should be deprecated as outliers because they greatly disagree with other contemporaneous measures as well as the general trend line for 2008, and because their scatter and reported measurement error bars are greatly larger than for all other datasets.    Most of the papers report times in HJD, which I have corrected to BJD.  However, some of the papers report their times in units of HJED and BJDD, which have only negligible differences from BJD, so I include these times as reported with no further correction.  Most of the published eclipse times have no quoted one-sigma error bar, so, following Dai et al. (2009), I adopt a default error bar of $\pm$0.0001 day.  In all, I have 117 eclipse times from the literature from 1969--2016 (see Table 6).

With Z Cha being near the south ecliptic pole, {\it TESS} has targeted light curves for 12 Sectors to date, each with 120 second time resolution.  This time resolution is greatly poorer than from the high-speed photometry usually used in the literature.  This time resolution creates problems for measuring the eclipse time.  The ingress/egress steps cannot be recognized with useable accuracy, and the lowest minimum (when both the WD and the bright spot are eclipsed) lasts only $\sim$400 seconds, so the {\it TESS} light curves have only 3 or 4 fluxes of relevance for measuring the time of each eclipse.  I have experimented with four methods to determine {\it TESS} eclipse times, as in Schaefer (2021).  The best method (i.e., that which returns the smallest RMS scatter in the $O-C$) uses a parabola fit to the three lowest points.  For the three-point-parabola, the average RMS for the $O-C$ is 0.0054 in phase (0.00040 days), which will be the uncertainty arising from the measurement error plus the intrinsic variation in the times of minimum.  After masking out the DN eruptions, each {\it TESS} Sector has 83--319 eclipse times, for an average of 231.  From {\it TESS}, I have derived 12 eclipse times from 2018--2021.

The AAVSO International Database contains many long time series of CCD photometry since 2007.  These runs usually have no filter (to improve the SNR) and have integration times from 60--200 seconds.  These integration times are greatly longer than for the high-speed photometry of the literature, so the eclipse profile is somewhat smeared in time, resulting in a poor determination of the minimum time for any one eclipse.  But over one observing season, the database has up to 200 eclipses measured, and this can make for a fairly good accuracy for a seasonal eclipse time.  The real timing accuracy is larger than the formal measurement errors, as can be seen by making small differences in the fitting range for the folded light curve, and as seen by differing definitions for measuring the times of minimum light.  The total error for the AAVSO $O-C$ values is $\sim$0.0002 days.

In all, I can use the 135 eclipse times 1969--2021 (Table 6) to construct an $O-C$ curve (Fig. 7).  I have also plotted the Court et al. (2019) sinusoidal model (thin black curve).  The Court sinusoidal model is denied by the addition of the {\it TESS} times, which show a fast decreasing $P$ instead of the predicted fast increasing $P$, and is denied by the many 1969 times of Mumford and most of the 1972--1973 times of Warner, which were ignored in Court's fig. 9.  So the Z Cha $O-C$ curve is not sinusoidal.  Importantly, the $O-C$ does not have any sort of parabolic shape.  That is, a parabola cannot provide a meaningful description of the $O-C$ curve in particular, or the general run of evolutionary $P$ changes, so I have no measure of $\dot{P}$.

Some unknown $P$-change mechanism is dominating over the expected steady $\dot{P}$.  This $O-C$ curve can be well-described by a doubly-broken line (see the magenta line segments).  The breaks around 1976 and 2000 have $\Delta P$/$P$ of $+$1.1 and $-$0.6 parts-per-million, with these being small compared to the usual period changes for nova eruptions.  So, empirically, Z Cha is suffering positive and negative period changes that occurred over a small number of years, and for which there is no known model to explain the changes.

Previous papers have claimed to use subsets of the above data to find sinusoidal models, with periods 28$\pm$2, 32.57, 37.5$\pm$0.5, and 43.5$\pm$0.8 years.  The basic iteration is for the workers to add a few times to the $O-C$ curve, describe the curve as a sinusoidal shape, postulate that this shape is actually a periodic signal, attribute the sinusoid to a third body in orbit around the inner binary, and claim the discovery of a new planet or brown dwarf.  We have now gone through four sets of iterations for Z Cha.  Such claims are ill-advised because the claims all have no more than 1.2 cycles in the $O-C$ curve, and no periodicity with only 1.2 cycles of poor data can ever provide useful confidence for the existence of a periodicity.  All four claims are only looking at a single dip and a single peak in the $O-C$ curve, and such can always be attributed to a periodic signal.  The four Z Cha sinusoid claims have all been refuted successively by the addition of subsequent times, with the latest iteration from my Fig. 7 where the Court et al. (2019) prediction of a fast increasing $P$ after 2018 is strongly denied by the new {\it TESS} times.  Unfortunately, such ill-advised claims for sinusoidal $O-C$ curves are made for the {\it majority} of those CVs in this paper, with all of them having already been refuted.

\begin{deluxetable}{llrl}
\tablenum{6}
\tablecaption{135 eclipse times for Z Cha (full table in machine readable formation in the electronic article)}
\tablehead{
\colhead{Eclipse Time (BJD)} & \colhead{Year} & \colhead{$O-C$} &
\colhead{Source}  }
\startdata
2440264.6831	$\pm$	0.0001	&	1969.12	&	0.0007	&	Mumford	\\
...			&	...	&	...	&	...	\\
2454252.072863	$\pm$	0.000062	&	2007.41	&	0.0002	&	AAVSO 2007	\\
2455117.009369	$\pm$	0.000038	&	2009.78	&	0.0000	&	AAVSO 2009	\\
2456630.015345	$\pm$	0.000088	&	2013.92	&	-0.0001	&	AAVSO 2013	\\
2456730.067769	$\pm$	0.000043	&	2014.20	&	-0.0002	&	AAVSO 2014	\\
2458083.048500	$\pm$	0.000064	&	2017.90	&	-0.0010	&	AAVSO 2017	\\
2458138.029209	$\pm$	0.000058	&	2018.05	&	-0.0008	&	AAVSO 2018	\\
2458389.017191	$\pm$	0.000060	&	2018.74	&	-0.0009	&	TESS 3	\\
2458476.181123	$\pm$	0.000017	&	2018.98	&	-0.0011	&	TESS 6	\\
2458580.405386	$\pm$	0.000014	&	2019.26	&	-0.0014	&	TESS 10	\\
2458609.534705	$\pm$	0.000013	&	2019.34	&	-0.0013	&	TESS 11	\\
2458632.555036	$\pm$	0.000060	&	2019.41	&	-0.0012	&	TESS 12	\\
2458670.102599	$\pm$	0.000014	&	2019.51	&	-0.0013	&	TESS 13	\\
2459048.335766	$\pm$	0.000052	&	2020.54	&	-0.0010	&	TESS 27	\\
2459125.069555	$\pm$	0.000015	&	2020.75	&	-0.0015	&	TESS 30	\\
2459218.938631	$\pm$	0.000017	&	2021.01	&	-0.0015	&	TESS 33	\\
2459318.842191	$\pm$	0.000013	&	2021.28	&	-0.0015	&	TESS 37	\\
2459346.183357	$\pm$	0.000018	&	2021.36	&	-0.0016	&	TESS 38	\\
2459371.587897	$\pm$	0.000051	&	2021.43	&	-0.0013	&	TESS 39	\\
\enddata	
\tablecomments{{\bf (1)}~~The $O-C$ values, in days, with the linear ephemeris of Court et al. (2019) for epoch BJD 2440000.06088 and period 0.0744992878 days.  {\bf (2)}~~Sources in the full table include references Bailey (1979), Baptista et al. (2002), Cook (1985b), Dai et al. (2009), Honey et al. (1988), Mumford (1971), Nucita et al. (2011), Pilarcik et al. (2018), Robinson et al. (1995), van Amerongen et al. (1990), Warner (1974), Wood et al. (1986).}
\end{deluxetable}

\begin{figure}
\epsscale{1.16}
\plotone{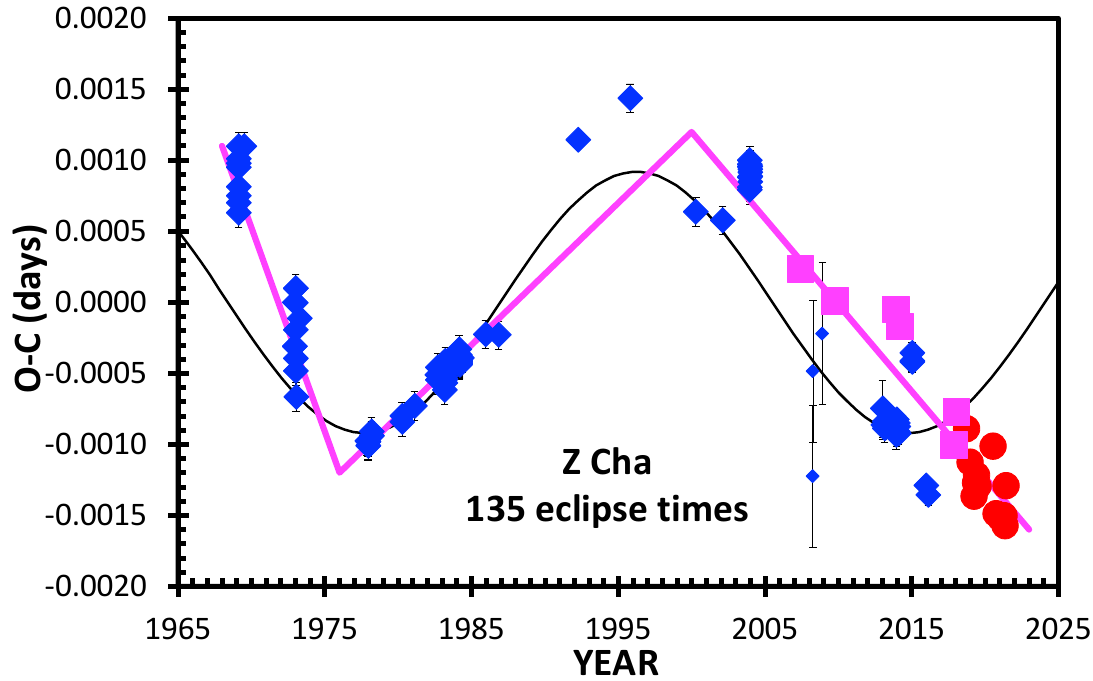}
\caption{$O-C$ curve for Z Cha.  This plot has 135 eclipse times, coming from the literature (blue diamonds), from {\it TESS} (red circles), and the AAVSO light curves (magenta squares).  The importance of this plot is that it shows a well-measured case where the $O-C$ curve is not a parabola in shape, nor does it have any sinewave superposed on top.  The large and significant period changes can be described as a three-segment broken line, with a fairly abrupt period increase around 1976 and a fairly abrupt period decrease around 2000.  This case of period change is in accord with no known theoretical mechanism.  The vertical scatter is likely due to the difficulties of defining and measuring the eclipse times, as well as intrinsic variability in the eclipse profiles. }
\end{figure}

\subsection{DV UMa; Dwarf Nova with $P$=0.0859 days}

{\bf DV UMa} is a faint dwarf nova of the SU UMa class (i.e., with superoutbursts and superhumps), discovered in 1982 on the Harvard plates.  In 1987, the system was discovered to be an eclipsing binary with period 0.0859 days.  The eclipse profile shows four fast steps for the ingress/egress of the WD and of the hotspot in the accretion disc.

The literature on DV UMa contains relatively few useable eclipse times.  One trouble is that most of the published times are for when DV UMa is in eruption, and phases of such measured eclipse are known to make for systematic errors in the O-C curve that vary substantially over each eruption, and so in-eruption times should never be used for $O-C$ curves of the orbital period.  Another problem is that  the papers report variously the times of minimum photometric brightness (`mid-eclipse'), or the time halfway from the WD ingress to egress (i.e., the time of conjunction), or both.  Published times that report only the mid-eclipse were converted to times of conjunction with the offset of -0.00067 days, as measured by Patterson et al. (2000).  All the eclipse times in Table 7 are consistently for the WD conjunction, and all converted to BJD.  I have not used the times published by Han et al. (2017b)\footnote{The Han times and analysis have many and deep problems:  (1) The tabulation in Han et al. (2017b) mixes up eclipse times from the two definitions, and this creates a step in their $O-C$ curve as an erroneous artifact.  (2) Their tabulation contains many times when DV UMa is certainly in eruption, with this including at least two of their own timings, all the early eclipse times from Patterson et al. (2000), all the eclipse times derived by Han from the AAVSO light curves, and all of the eclipse times in Nogami et al. (2001).  With these mistakes, Han created the two minima in their $O-C$ curve.  Between the first two mistakes, Han has constructed the claimed sinewave modulation as an artifact of errors.  (3) Even given their DV UMa times as stated, Han was unwise to claim a sinewave component in the $O-C$ curve with only $\sim$1 cycle of data, followed by their injudicious four-digit precision in the period and their requiring a high eccentricity for their newly `discovered' brown dwarf companion.  The coauthors Han and Qian have a long history of papers making ill-advised claims for sinusoidal $O-C$ curves, for stars discussed in this paper including EM Cyg, Z Cha, OY Car, V2051 Oph, AC Cnc, and DV UMa, with all these poor conjectures having already been refuted by latter data.  (4) Coauthors Han and Qian also have a long track record of publishing CV eclipse times that have substantial systematic errors.  This was proven by Pilar\u{c}ik et al. (2018) for OY Car, and I had independently realized in this paper that their data have substantial and systematic errors for Z Cha, V2051 Oph, EX Dra, OY Car, AC Cnc, EM Cyg, and DV UMa.  In light of the unreliability of the data and analyses by Han, I cannot use any of the quoted times from Han et al. (2017b).}.  The error bars are usually not quoted, so I have adopted the RMS of the seasonal $O-C$ values for each observer as the one-sigma total error. 

The DV UMa $O-C$ curve can be extended to 2023 by using the ZTF light curves from 2018--2023.  After eliminating the magnitudes taken while DV UMa is in eruption, we are left with 495 $zg$ and 842 $zr$ magnitudes.  The $zr$ magnitudes are made commensurate with the $zg$ light curve by an offset of 0.164 mag.  When folded on the orbital period, the light curve closely follows a single history, which is to say that flickering and secular variations are small, roughly at the 0.10 mag level.  The folded light curve clearly shows the photometric steps of the WD eclipse.  The ZTF light curve contains an adequate number of in-eclipse magnitudes for the time intervals 2018.2--2019.5, 2019.7--2020.5, and 2020.7--2022.4.  The time of the WD conjunction was derived by fitting the folded light curve between phases $-$0.06 to $+$0.06 to a template eclipse profile taken from Patterson et al. (2000).  The sparseness of the ZTF light curve matters little for the timing accuracy because the flickering and secular variations are small compared to the eclipse depth.

I have constructed an $O-C$ diagram for DV UMa (see Fig. 8) with the 37 eclipse times in Table 7.  A parabola looks to be a reasonable description of the shape of the curve, although small noise might be added on top.  My best-fitting parabola has $\dot{P}$=($+$2.80$\pm$0.8)$\times$10$^{-13}$.

\begin{deluxetable}{llrl}
\tablenum{7}
\tablecaption{37 eclipse times for DV UMa (full table in machine readable formation in the electronic article)}
\tablehead{
\colhead{Eclipse Time (BJD)} & \colhead{Year} & \colhead{$O-C$} &
\colhead{Source}  }
\startdata
2446854.74850	$\pm$	0.00090	&	1987.16	&	0.000304	&	Howell et al.	\\
...			&	...	&	...	&	...	\\
2458485.03532	$\pm$	0.00007	&	2019.00	&	0.000051	&	ZTF	\\
2458913.01100	$\pm$	0.00006	&	2020.17	&	0.000256	&	ZTF	\\
2459391.03885	$\pm$	0.00007	&	2021.48	&	0.000544	&	ZTF	\\
\enddata	
\tablecomments{{\bf (1)}~~The $O-C$ values, in days, uses a linear ephemeris with epoch BJD 2452780.46995 and period 0.0858526521 days.  {\bf (2)}~~Sources in the full table include references Feline et al. (2004), Howell et al. (1988), and Krajci (2007).}
\end{deluxetable}

\begin{figure}
\epsscale{1.16}
\plotone{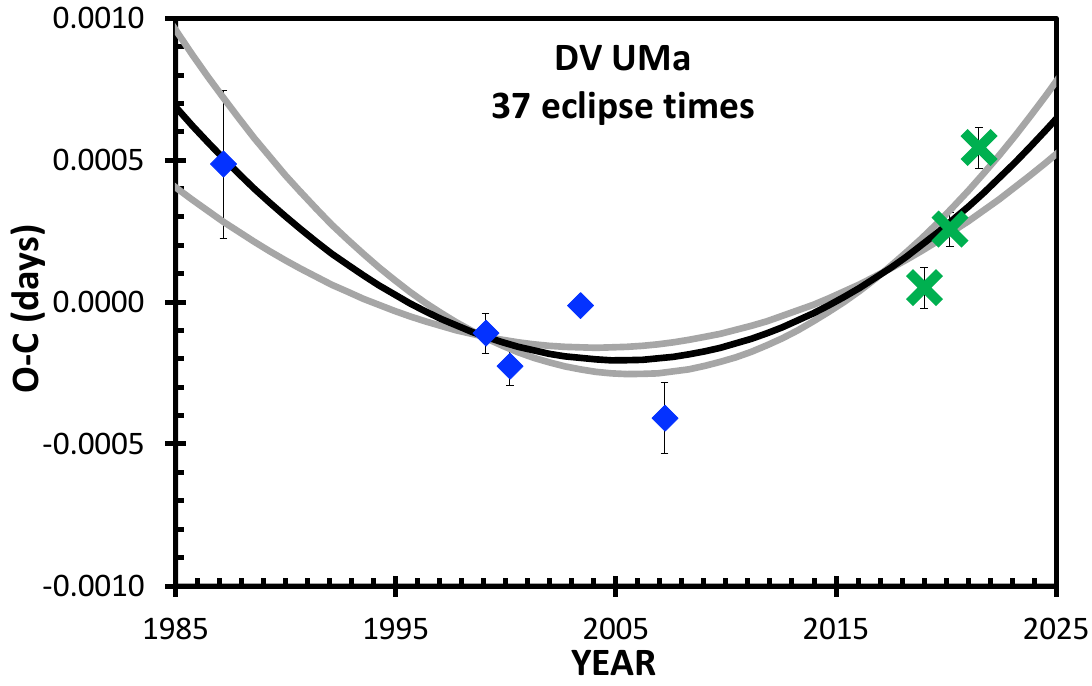}
\caption{$O-C$ curve for DV UMa.  My $O-C$ curve has 37 eclipse times in quiescence from 1987--2021, with 34 from the literature (blue diamonds) and 3 from ZTF light curves (green $\times$).   For display purposes only, I am plotting the seasonal average $O-C$ values for each observer.  My best-fitting parabola, with $\dot{P}$=($+$2.80$\pm$0.8)$\times$10$^{-13}$, is shown as the thick black curve, while the one-sigma extremes are shown as the thick gray curves.  }
\end{figure}

\subsection{IP Peg; Dwarf Nova with $P$=0.158 days}

{\bf IP Peg} is a normal dwarf nova of the U Gem class, with its usual quiescent $V$ of 15.0 mag.  The star was discovered as a variable in 1981, and as an eclipsing binary (with a $P$=0.158 days) in 1985.

The three papers with big studies of eclipse times have all made the unfortunate definition of the eclipse times as being the time of the WD egress, represented by a photometric step rise just after the photometric minimum.  This definition is unfortunate because the egress changes substantially, with durations varying from 10--300 seconds, and the edges depend greatly on the highly variable hotspot component, so the egress time is a poor phase marker for $O-C$ purposes.  This definition is also poor because the egress step is small, often very small, so it is hard to define and recognize in practice, even with high-speed high-SNR light curves.  The egress is completely invisible in all the other reported light curves, including those of {\it TESS} and ZTF.  We are left with the question of how to better define the eclipse times for $O-C$ purposes.  The clear answer is to use the time of photometric minimum light as the one measurable phase marker for the binary.  With this, the reported egress times will be offset to the minimum times before inclusion in the overall $O-C$ curve, with Wood et al. (1989) quoting the WD middle-egress as being placed at 0.04315 phase (for a correction to minimum of $-$0.00683 days).  But this is not a perfect solution, because shifts in the position of the hotspot will cause small shifts in the phases of minimum light.  This is just a version of the ubiquitous problem for all CVs where flickering and hotspot movements make for an irreducible noise in each eclipse time.  A further problem is that the eclipse bottom is not flat, but rather has a steep ingress followed by a slowly rising minimum, and this makes it difficult to measure the time of minimum.  The usual bisector analysis fails greatly due to the large asymmetries.  Parabola fits to the minimum have the derived time dependent upon the range of phases used in the fit.  This means that the many reported times of minimum light will have some measurement error causing noise in the $O-C$ curve.

The timings in Copperwheat et al. (2010) have fundamental problems.  One problem is that their reported times of egress have $O-C$ values with an RMS scatter of 1.7 seconds, which seems impossible given the intrinsic variations in the egress width and their integration time of 1.5 seconds.  A second problem is that their $O-C$ values are all far outliers from the general run of the $O-C$ curve, as well as from the specific contemporaneous measures by many observers.  The tabulated times are roughly 0.0063 days earlier than everyone else.  A third fundamental problem is that their tabulated times of mid-egress are all 0.0074 days earlier than in their quoted ephemeris.

Many eclipse timings have been published in the literature, plus 114 minimum times 1985--2020 collected in the Czech $O-C$ Gateway and in Bob Nelson's Database.  I have not used any measures of the shallow secondary eclipse, and I have tossed out three far outliers.  The one-sigma error bars are taken to be the RMS scatter in the $O-C$ for appropriate groups of eclipses.  All the times of minimum light were converted to BJD and tabulated in Table 8. 

{\it TESS} observed IP Peg in just one Sector so far (Sector 56 in 2022) with 120 s time resolution.  ZTF observed IP Peg with 1338 magnitudes sparsely covering 2018.7--2023.3 in three passbands ($zg$, $zr$, and $zi$), all with 30-second exposures.  To derive the times of minimum light, I have performed a chi-square minimization of the light curves versus a periodic template constructed from the $R$-band light curve of Wood et al. (1989).  The eclipse depth varies greatly with the band, so I have freely scaled the intensity of the template to fit the observed light curve.  The ZTF light curves were fitted for each band separately, covering one or two seasons as needed to get an adequate number of measures inside the eclipses.  I found that all the {\it TESS} and ZTF light curves deviate from the older light curves by having a fainter segment after the WD egress, so I have only fitted for phases from $-$0.06 to $+$0.04, over which the template closely matches the data.  With this, my measures of  the times of zero phase are consistent with the times of Wood, as well as the other reported times of photometric minimum.

I have constructed an $O-C$ curve (Fig. 9) with 193 eclipse times from 1984--2022.  We see a good parabola shape, with $\dot{P}$=($-$2.42$\pm$0.18)$\times$10$^{-11}$.  The $O-C$ curve shows some nominally significant deviations from the parabola, but I expect that this noise arises from ordinary measurement errors associated with defining the time of minimum.

\begin{deluxetable}{llrl}
\tablenum{8}
\tablecaption{196 eclipse times for IP Peg (full table in machine readable formation in the electronic article)}
\tablehead{
\colhead{Eclipse Time (BJD)} & \colhead{Year} & \colhead{$O-C$} &
\colhead{Source}  }
\startdata
2445933.41473	$\pm$	0.00200	&	1984.64	&	0.00307	&	Goranskij \\
...			&		&		&		\\
2458679.12452	$\pm$	0.00015	&	2019.53	&	0.00027	&	ZTF zr 	\\
2458722.15599	$\pm$	0.00040	&	2019.65	&	-0.00031	&	ZTF zg	\\
2459107.07058	$\pm$	0.00021	&	2020.71	&	-0.00104	&	ZTF zr 	\\
2459141.08437	$\pm$	0.00040	&	2020.80	&	-0.00156	&	ZTF zg	\\
2459310.04897	$\pm$	0.00014	&	2021.26	&	-0.00102	&	ZTF zi	\\
2459649.08507	$\pm$	0.00037	&	2022.19	&	-0.00049	&	ZTF zr 	\\
2459720.11895	$\pm$	0.00060	&	2022.38	&	-0.00113	&	ZTF zg	\\
2459838.14180	$\pm$	0.00002	&	2022.71	&	0.00000	&	TESS 56	\\
\enddata	
\tablecomments{{\bf (1)}~~The $O-C$ values, in days, uses a linear ephemeris with epoch BJD 2459520.002336 and period 0.30047730700 days.  {\bf (2)}~~Sources in the full table include references Goranskij et al. (1985), Wolf et al. (1993), and Wood et al. (1989).}
\end{deluxetable}

\begin{figure}
\epsscale{1.16}
\plotone{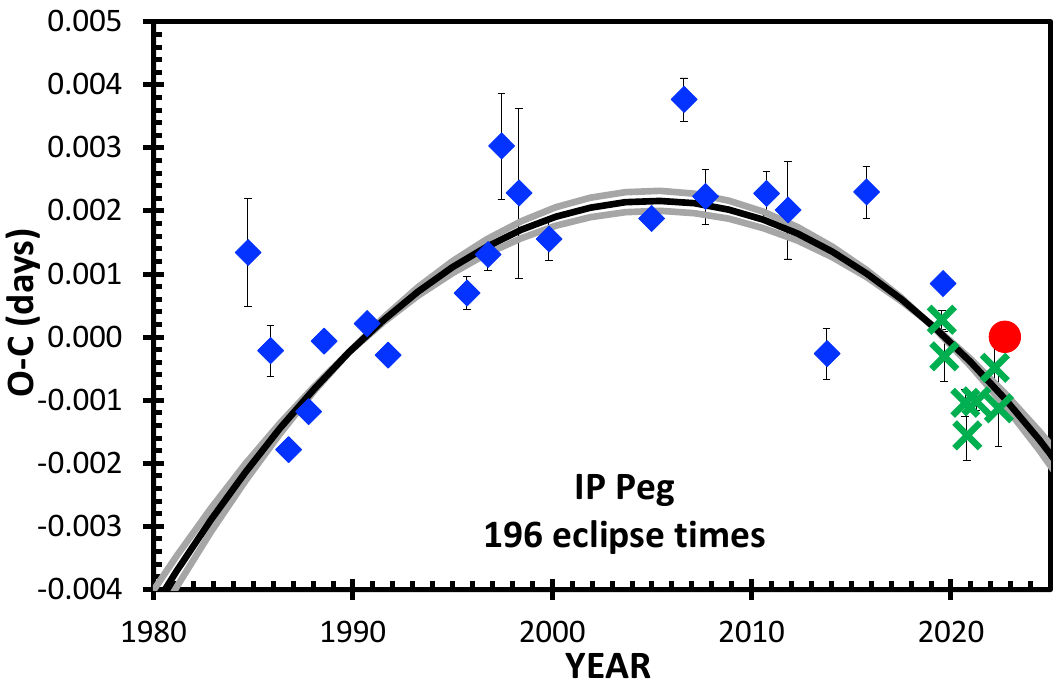}
\caption{$O-C$ curve for IP Peg.  I have 188 eclipse times from the literature (blue diamonds), 7 times from the ZTF light curves (green $\times$ symbols), plus 1 high-accuracy time from {\it TESS} (red circle).  For display purposes only, the literature times have been combined into one-year bins.  We see a strong downward curvature, with $\dot{P}$=($-$2.42$\pm$0.18)$\times$10$^{-11}$, as a measure of the evolutionary decay of the orbit. }
\end{figure}

\subsection{LX Ser; Novalike with $P$=0.158 days}

{\bf LX Ser} was discovered as a CV in 1979, and has been called ``Stepanyan's star''.  Eclipses were discovered in 1980, with amplitude 2.0--2.5 mag, duration 40 minutes, and orbital period 0.158 days.

A few papers in the literature reported eclipse times of minimum from 1980 to 2016, but only with modest coverage.  The great advance came with Boyd (2023) reporting 82 eclipse times well-spread-out from 2007--2022.  The $O-C$ curve constructed from all the eclipse times collected by Boyd (see the blue diamonds in Fig. 10) shows a concave-up parabola with substantial bumps.

I have no corrections or improvements on Boyd's data, but I have tried to extend his results.  The most important try was to measure the $B$ brightness of LX Ser on 204 Harvard plates 1892--1951, but the eclipse signal turned out to be of inadequate significance to get a useful eclipse time, primarily because the eclipses are shorter than most of the plate exposure times.  The only addition I can offer is times from {\it TESS} Sectors 24 and 51, for which I get average times of BJD 2458970.211311$\pm$0.000046 and 2459707.239359$\pm$0.000053.  With the two Sectors having 141 and 68 eclipses, the nominal timing accuracy (around 4 seconds) is by far the best of all.  But this timing accuracy is of little utility, because the scatter in the $O-C$ curve for groups of well-timed eclipses (around 30 seconds), as caused by intrinsic variations in the CV, so it matters little whether an eclipse is measured with 30-s, 4-s, or 0.001-s accuracy.  For $O-C$ purposes, this means that one {\it TESS} Sector has comparable utility as one eclipse time by Boyd. {\it TESS} has only two snapshots separated by two years, while Boyd has 82 times spread uniformly throughout 16 years.  So my addition of two {\it TESS} eclipse times can only make for confirmation and marginal improvement on Boyd's published $O-C$ curve. 

With all of the 82 new and 82 collected times in Boyd (2023), plus my 2 {\it TESS} times, I have constructed an $O-C$ curve for LX Ser (see Fig. 10).  The basic shape is a period-increasing parabola, with $\dot{P}$=($+$5.2$\pm$0.3)$\times$10$^{-12}$, plus some bumps.    The amplitude of the bumps is up to $\sim$40 seconds, which is comparable to bumps for other CVs that have been superposed on parabolas.

\begin{figure}
\epsscale{1.16}
\plotone{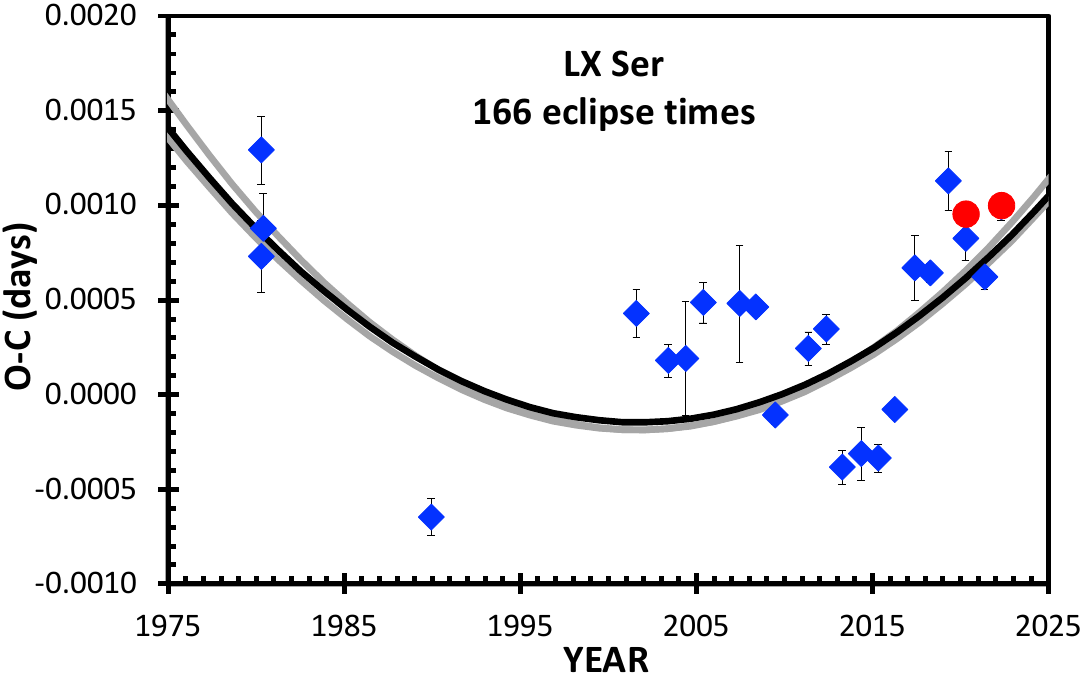}
\caption{$O-C$ curve for LX Ser.  This curve contains 82 new measures and 82 collected times from Boyd (1923), as shown with blue diamonds, with these being binned into yearly and observer bins.  My two new times from {\it TESS} are shown with red circles, but these points mainly serve to confirm what Boyd has better measured.  (Indeed, the second {\it TESS} points almost completely covers up the 2022 average value from Boyd.)  The $O-C$ is calculated for a fiducial ephemeris with an epoch of BJD 2452777.87598 and a period of 0.158432503 days.  The $O-C$ curve appears as a concave-up parabola to represent the evolutionary trend over 42 years.  The best-fitting parabola, with $\dot{P}$=($+$5.2$\pm$0.3)$\times$10$^{-12}$, is shown as a thick black curve, while the formal 1-sigma extremes are shown as thick gray curves.  This $O-C$ is not a perfect parabola, but instead has multiple small bumps with amplitude up to 40 seconds or so. }
\end{figure}

\subsection{UU Aqr; Novalike with $P$=0.164 days}

{\bf UU Aqr} was discovered as a variable star in 1924, and with little attention was considered to be some sort of a semi-regular variable.  This changed in 1984--1986 with the discovery of a strong UV excess, deep eclipses with a fast period at 0.164 days, and substantial flickering.  With only four eclipses in 1985, the system was largely unstudied until 1994 (Baptista, Steiner, \& Cieslinski 1994).  Since then, few eclipse times have been published, chiefly by Boyd (2023) and Bruch (2023).

Bruch (2023) made a full study of the $O-C$ curve.  He measured a highly accurate minimum time for {\it TESS} Sector 42 in August/September 2021, and he measured 99 eclipse times from AAVSO light curves from 2000-2019.  These AAVSO light curves were mostly from D. Boyd, who had previously published 15 of the minimum times (Boyd 2012).  Bruch also collected the original eclipse time from 1985, plus 39 times 1988--1992 as measured by Baptista et al. (1994).  Bruch constructed an $O-C$ curve with 140 times from 1985 to 2021.  These show a nice concave-down parabola.  Bruch derived $\dot{P}$=($-$2.240$\pm$0.004)$\times$10$^{-11}$, but the quoted uncertainty is much too small to be realistic.

Boyd (2023) also made a full study of the UU Aqr $O-C$ curve.  He has the big advantage using his own database of 53 good eclipse times 2007-2022, which forms the anchor for half the $O-C$ curve.  Boyd supplemented this with 53 eclipse time from the literature, to form an $O-C$ curve with 106 eclipses well-spread between 1985 and 2022.  The shape of the curve is a good and distinct concave-down parabola, for which Boyd derived $\dot{P}$=($-$2.08$\pm$0.11)$\times$10$^{-11}$.

\subsection{U Gem; Prototype Dwarf Nova with $P$=0.177 days}

{\bf U Gem} is famous as the prototype dwarf nova, the first discovered (in 1855), the brightest, and the best observed.  U Gem is a deep totally eclipsing binary, where a M4.5 main sequence star is going around a WD with a period of 0.177 days.

{\it TESS} has recorded U Gem with 20-s time-resolution for three consecutive Sectors in 2021, each with nearly-continuous coverage of near 135 orbits each.  The third Sector records a DN outburst, so it was not used for measuring eclipse minima.  The measurement errors for the minimum times are determined to an accuracy of 0.000012 days (1.0 seconds), with the usual timing jitter from flickering damped down by a factor $\sqrt{135}$.  The {\it TESS} eclipse times in BJD are around 40$\times$ more accurate than all other measures, so these two times form the solid and reliable anchor for the entire $O-C$ curve.

From 1999--2017, I have used the light curves in the AAVSO International Database to derive 71 times of mid-eclipse.  From the scatter about any smooth curve in the $O-C$, the RMS is 0.00096 days (83 seconds), which I take as the timing jitter from flickering.

The astronomical literature has 173 eclipse times, mostly from 1961--2010, as collected in Nelson's database.  I have not used the visual timings (because their accuracy is no longer adequate to be helpful), I have not used times during eruptions (to avoid any possible systematic offsets), and I have tossed out 2 far-outliers.  The RMS scatter around the $O-C$ curve is 0.00075 days (65 seconds).

I have exhaustively looked through the Harvard plates, and measured the B band brightness of U Gem, as based on the local comparison stars calibrated with the APASS sequence.  This has been supplemented with the DASCH light curve, where I have found that both DASCH and my by-eye measures have near-zero average difference, and an RMS difference that is close to the quoted error bars.  I have magnitudes going back to 1896, but the only time interval with adequate coverage to pull out the orbital modulation is from 1930--1940.  A substantial problem is that many of the plates have exposure times longer than one hour, and these have a coverage of $>$0.24 in phase, which is smearing out the light curve.  So I have restricted analysis to the 56 plates with 60 minutes or shorter exposures.  Most of these plates are from 45 to 60 minutes in exposure, with an average of 53.7 minutes.  To find the time of eclipse, I have used a modern B-band template (with an exactly timed mid-eclipse) convolved with the 53.7 minute exposure.  The convolved template has the eclipse with low amplitude (because the eclipse duration is greatly shorter than the typical exposure time), so the observed light curve is primarily being phased-up with the prominent hump nearly a quarter in phase before the eclipse.  This template is shifted in time until the chi-square is minimized, with the one-sigma range being that over which the chi-square is within 1.0 of the minimum.  The particular minimum in this optimized template is chosen to be one near the average time for the 1930--1940 plates.  With this, I get an eclipse time in HJD, which I then convert to BJD 2428429.0977$\pm$0.0028.

Bailey (1975) has collected 219 visual magnitudes from 1857--1892.  The equivalent of a Fourier 
transform returns a highly significant peak at exactly the modern $P$, and his folded light curve returns the typical V-band shape for U Gem.  So he is picking out the eclipse time, and this has a phase of mid-eclipse at 0.03$\pm$0.03 on a stated ephemeris with units of HJD.  The average date of the light curve is stated (1876 March 20), so we can calculate the HJD of mid-eclipse for near that average time.  The difference between HJD and Ephemeris Time in 1876 is $-$3.6 seconds.  With this, we have a confident eclipse time in 1876 of BJD 2406334.0514$\pm$0.0071.

In all, I have 244 eclipse times from 1876--2021 (see Table 9), with these being plotted as an $O-C$ curve in Fig. 11.  To avoid the visual crowding, I have combined the $O-C$ values into bins, mostly of 1 year width.  All but two of the times are from 1961--2021, and these show a highly significant concave-up parabola shape, with no significant deviations over the 60 years.  This best-fitting parabola happens to pass close to the 1876 measure.  When all 244 times are fitted together to a parabola, the curvature is nearly unchanged, because almost all the statistical weight in the fit is from 1961--2021.  The best parabola has $\dot{P}$=($+$4.8$\pm$0.7)$\times$10$^{-12}$.

\begin{deluxetable}{llrl}
\tablenum{9}
\tablecaption{244 eclipse times for U Gem (full table in machine readable formation in the electronic article)}
\tablehead{
\colhead{Eclipse Time (BJD)} & \colhead{Year} & \colhead{$O-C$} &
\colhead{Source}  }
\startdata
2406334.0514	$\pm$	0.0071	&	1876.25	&	0.01885	&	Bailey	\\
2428429.0977	$\pm$	0.0028	&	1936.71	&	0.00907	&	HCO 1930--1940	\\
2437638.827335	$\pm$	0.00075	&	1961.93	&	0.00095	&	Krzeminski	\\
...			&	...	&	...	&	...	\\
2458168.440130	$\pm$	0.00096	&	2018.13	&	0.00084	&	AAVSO (BPO)	\\
2458171.624530	$\pm$	0.00096	&	2018.14	&	0.00093	&	AAVSO (LMA)	\\
2458172.509330	$\pm$	0.00096	&	2018.15	&	0.00120	&	AAVSO (LMA)	\\
2459512.044677	$\pm$	0.000012	&	2021.81	&	0.00265	&	TESS 44	\\
2459538.050172	$\pm$	0.000020	&	2021.88	&	0.00294	&	TESS 45	\\
\enddata	
\tablecomments{{\bf (1)}~~The $O-C$ values, in days, use calculated times for a linear ephemeris with epoch BJD 2447516.03131 and period 0.1769062190 days.  {\bf (2)}~~Sources in the full table include references Arnold et al. (1976a), Bailey (1975), Beuermann \& Pakull (1984), Biro et al. (2006), Dai \& Qian (2009), Diethelm (2005), Eason et al. (1983), Krzeminski (1965), Lampens et al. (2017), Marsh et al. (1990), Mumford (1969; 1970; 1974; 1975; 1976a; 1977), Nakajima (2007), Paczynski (1985), Patrick (1984), Simon (2003), Warner (1973), and Zhang \& Robinson (1987).}
\end{deluxetable}

\begin{figure}
\epsscale{1.16}
\plotone{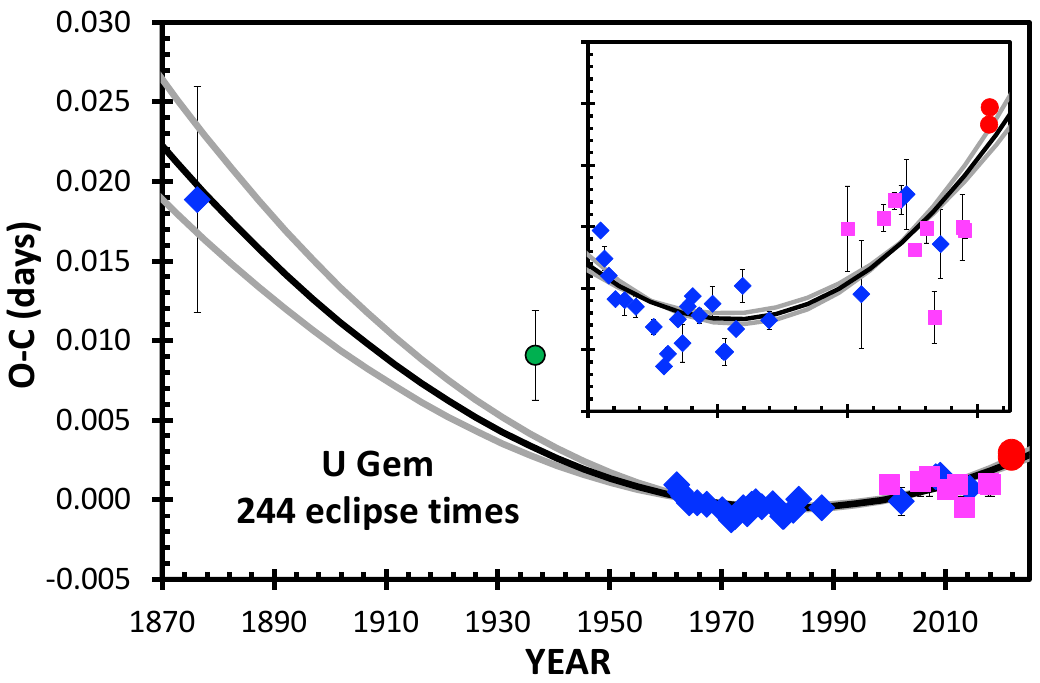}
\caption{$O-C$ curve for U Gem from 1876 to 2021.  The eclipse times are shown as red circles for {\it TESS}, as magenta squares for the AAVSO times, as blue diamonds for the literature measures, and as a green circle for the Harvard plates time.  The times from 1961 to 2021 (shown as a close-up in the inset) define a good concave-up parabola (the black curve) with $\dot{P}$=($+$4.8$\pm$0.7)$\times$10$^{-12}$. The 1-sigma changes are shown with the gray curves.  The 1961--2021 parabola passes nicely through the 1876 datum, so we see a smooth evolutionary effect over 145 years.}
\end{figure}

\subsection{UX UMa; Prototype Novalike with $P$=0.197 days}

{\bf UX UMa} is a prototype for novalike CVs.  It was discovered as an eclipsing binary in 1933, with deep eclipses every 0.197 days.   The star is relatively bright, typically around 13 mag (with substantial variations on all time-scales), so a large number of eclipse timings are available since 1933.  

I have collected and compiled these many times, converted the times in HJD to BJD, and displayed these in Table 10.  I have not used any visual timings, as their accuracy is always poorer than the many contemporaneous times from other methods.  I have also thrown out several $>$3-sigma outliers.  From the literature, I have 555 eclipse times covering 1914--2015.

I have also collected CCD time series from the AAVSO International Database, and then derived minimum times from parabola fits to the eclipse minima.  These give 16 mid-eclipse times in JD, which I then convert to HJD and then to BJD.  These are only for the eclipse times from the AAVSO database that have not already been published by G. Samolyk in a series of 19 papers from 2008--2022 that appears in the {\it Journal of the AAVSO}.

I have also collected the sparse light curves of ZTF, with 1578 magnitudes from 2018--2023.  The $zr$ magnitudes were offset by 0.052 mag so as to be effectively commensurate with the $zg$ magnitudes.  The light curves beautifully define the eclipses in each of the six observing seasons with coverage.  These seasonal light curves were then fitted with a periodic template, where the time of minimum is taken for one of the template minima near the average time of the data.  This results in 6 times of mid-eclipse for the 2018--2023 observing seasons.

{\it TESS} has observed UX UMa in four Sectors to date.  Schaefer (2021) has already reported on 243 individual eclipse times from two Sectors in 2019, with these being explicitly listed now in Table 10.  These times were used to prove that there exists a substantial and irreducible jitter in eclipse times arising from the ordinary flickering suffered by all CVs, and in the case of UX UMa this inescapable timing noise is 24 seconds (0.00027 days).  Further, the {\it TESS} data set provided a nice opportunity to test many ways of defining the eclipse times, and it was found that the most consistent measures all involve only the lowest part of the eclipsing light curve (e.g., by a parabola fit close around the minimum, or by bisectors from low in the eclipse), while measures involving ingress and egress times are always poor.  For the two {\it TESS} Sectors in 2020, Bruch (2023) has provided a representative time for each.

In all, I have 822 eclipse times, all converted to BJD, in Table 10.  These form an $O-C$ curve in Fig. 12.  We see a noisy parabola that is concave-down (the thick black curve in the figure), for which I formally get $\dot{P}$=($-$2.62$\pm$0.20)$\times$10$^{-12}$.  This is not a good parabola, with the deviations being comparable to the sagitta of the parabola.  Much of the deviations are likely intrinsic noise from UX UMa, and this can be seen by the 0.002 day scatter for {\it TESS} and other times.  Still, from 1970 to 2000, the eclipse times appear to be systematically early from the best-fitting parabola by up to 0.002 days (170 seconds).  This could be considered as a decade-long bump, perhaps where the hot spot has moved to one side for much of that time.  The formal parabola fit was forced to be above the 1970--2000 values because of the slope from 2010--2023 as measured with high statistical weight, so a flatter parabola can well-fit all the data except for the post-2010 slope.  A reasonable fit that covers all the $O-C$ curve is shown as the thin black line (with $\dot{P}$ equal to $-$1$\times$10$^{-12}$), for which the post-2010 slope is just part of a small bump, and the 1970--2000 deviations seem to be of low significance.  It is not clear how to characterize the best parabola fit.  Nevertheless, we can place reasonable approximate values on the real long-term evolution of UX UMa over the last 109 years.  With substantial error bars, we know that $\dot{P}$ is something like ($-$2$\pm$1)$\times$10$^{-12}$.

\begin{deluxetable}{llrl}
\tablenum{10}
\tablecaption{822 eclipse times for UX UMa (full table in machine readable formation in the electronic article)}
\tablehead{
\colhead{Eclipse Time (BJD)} & \colhead{Year} & \colhead{$O-C$} &
\colhead{Source}  }
\startdata
2420238.2452	$\pm$	0.0030	&	1914.29	&	0.00168	&	Kukarkin	\\
2427307.3952	$\pm$	0.0005	&	1933.64	&	-0.00103	&	Johnson et al.	\\
...			&	...	&	...	&	...	\\
2458762.2129	$\pm$	0.0002	&	2019.76	&	-0.00210	&	TESS	\\
2458762.4093	$\pm$	0.0002	&	2019.76	&	-0.00240	&	TESS	\\
2458762.6061	$\pm$	0.0002	&	2019.76	&	-0.00228	&	TESS	\\
2458762.8029	$\pm$	0.0002	&	2019.76	&	-0.00209	&	TESS	\\
2458762.9999	$\pm$	0.0002	&	2019.76	&	-0.00179	&	TESS	\\
2458914.0431	$\pm$	0.0001	&	2020.18	&	-0.00214	&	Bruch	\\
2458916.2064	$\pm$	0.0002	&	2020.18	&	-0.00222	&	ZTF	\\
2458944.1338	$\pm$	0.0001	&	2020.26	&	-0.00214	&	Bruch	\\
2459326.0696	$\pm$	0.0001	&	2021.30	&	-0.00195	&	ZTF	\\
2459710.1682	$\pm$	0.0002	&	2022.36	&	-0.00240	&	ZTF	\\
2459946.1750	$\pm$	0.0007	&	2023.00	&	-0.00119	&	ZTF	\\
\enddata	
\tablecomments{ {\bf (1)}~~The $O-C$ values, in days, use calculated times for a linear ephemeris with epoch BJD 2451319.78020 and period 0.1966712860 days.   {\bf (2)}~~Sources in the full table include references Africano \& Wilson (1976), Agerer \& Hubscher (2000), Arnold et al. (1976b), Baptista et al. (2003), Brat et al. (2007), Bruch (2023), de Miguel et al. (2018), Diethelm (2007; 2011; 2012b), Johnson et al. (1954), Kjurkchieva \& Marchev (1994), Krzeminski \& Walker (1963), Kukarkin et al. (1969), Mumford (1976b), Nather \& Robinson (1974), Nelson (2004), Panek \& Zink (1978), Paschke (2009), Quigley \& Africano (1978), Rubenstein et al. (1991), Rutten et al. (1992), Safar \& Zejda (2002), Samolyk (2008; 2009; 2010; 2011; 2012; 2013a; 2013b; 2014; 2015; 2016a; 2016b; 2017; 2018; 2019; 2020; 2021; 2022), Schaefer (2021), Todoran (1967), Walker (1968), Walker \& Herbig (1954), and Zejda et al. (2006).}
\end{deluxetable}

\begin{figure}
\epsscale{1.16}
\plotone{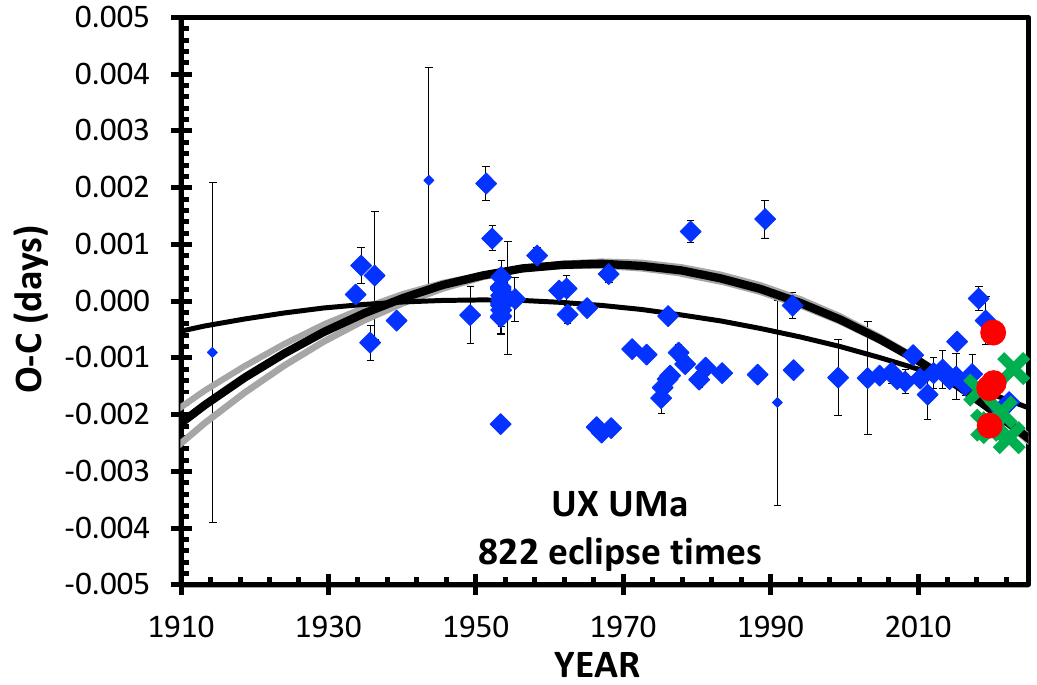}
\caption{$O-C$ curve for UX UMa.  The 822 eclipse times from Table 10 are averaged into time bins from each source, with literature times in blue diamonds, {\it TESS} times in red circles, and ZTF times in green $\times$ symbols.  The formal fit is shown with the thick black parabola (flanked by 1-sigma ranges with gray curves).  This formal parabola was forced to lie above many of the 1970--2000 eclipse times because the post-2010 data have a moderate slope of high statistical weight.  If the post-2010 slope is taken to be just some ordinary bump on top of the evolutionary parabola, then the pre-2010 data have a much better fit, as shown by the thin black curve.  It is not clear what is the best representation of the evolution, but it appears that $\dot{P}$ is negative and small, something like $\dot{P}$=($-$2$\pm$1)$\times$10$^{-12}$.}
\end{figure}

\subsection{EX Dra; Dwarf Nova with $P$=0.210 days}

{\bf EX Dra} is an ordinary dwarf nova that was discovered in a quasar survey in 1989.  EX Dra has deep eclipses with a period of 0.210 days.  Many eclipse times have been reported since 1991, with the residuals in the $O-C$ fits always being touted as one or two sinewaves, all on the basis of roughly one `cycle'.  These many sinewaves are all mutually inconsistent, and denied by subsequent timings.

The reported times of minimum have substantial confusion.  Han et al. (2017a) apparently defined their times as halfway between ingress and egress, with this being a poor measure of the times of conjunction due to the variable asymmetry of the eclipses.  Fiedler et al. (1997), Baptista et al. (2000) and Shafter \& Holland (2003) reported only the times of the WD egress, to which must be subtracted 0.0114 days to get the times of minimum.  Pilar\u{c}ik et al. (2012) measured the times of the WD egress, then claimed to subtract 0.0228 days to get what they call times of minimum light, but the $O-C$ curve shows that they actually subtracted only 0.0114 days.  The few eclipse times in Ho\u{n}kov\'{a} et al. (2013), Hubscher (2014; 2015) and Kruspe, Schuh, \& Traulsen (2007) have no information about the method for measuring the times of minimum, but these are all systematically late by 0.001 days, so these apparently have some systematic problem, and I cannot use these measures.  To be explicit, the times I collect in Table 11 are all consistently BJD times of photometric minimum. 

Following prior practice, I have not used any eclipse times from when EX Dra is having a dwarf nova eruption.  The shape and depth of the eclipse light curves vary substantially with color and from eclipse-to-eclipse.  Further, ordinary flickering makes for inescapable variations in the phase of minimum brightness.  For sparse light curves, these variations make for substantial variations in minimum times by any method.  In particular, the sparse light curves of ZTF and the Harvard plates cannot produce any useful $O-C$ measures.

EX Dra was measured by {\it TESS} in 23 Sectors, from Sector 14 (July 2019) to Sector 59 (November 2022).  I have measured the representative time of minimum light for each of 9 Sectors by fitting a parabola to the folded light curve for orbital phases from $-$0.05 to $+$0.05.  The light curve in this range is closely a symmetric parabola, and the wonderful {\it TESS} light curve has fast time resolution (20 and 120 seconds) and good photometric accuracy (2.0 and 1.6 per cent for a single flux measure, respectively) all for roughly 60 eclipses in each Sector averaged together, so the true time of minimum light can be measured with high accuracy.

In all, I have 136 times of eclipse minimum (see Table 11), with the 9 {\it TESS} times representing $\sim$540 eclipses.  These are made into an $O-C$ diagram in Fig. 13.  We see a structure that looks like a poor concave-up parabola.  I can fit a parabola to this (the black curve), with a $\dot{P}$=($+$3.05$\pm$0.32)$\times$10$^{-11}$.  This value might express the long-term average evolution of the system.  But there are substantial deviations from a perfect parabola, with the bumps deviating up to 0.001 days (86 seconds).  These deviations are comparable in size to the bumps seen in other CVs, where the up-bumps and down-bumps are caused by measurement problems of various types and by intrinsic changes like from hot spot movements.  In such an expected case, the bumps will average to zero, and the underlying parabola would represent the evolution.

\begin{deluxetable}{llrl}
\tablenum{11}
\tablecaption{136 eclipse times for EX Dra (full table in machine readable formation in the electronic article)}
\tablehead{
\colhead{Eclipse Time (BJD)} & \colhead{Year} & \colhead{$O-C$} &
\colhead{Source}  }
\startdata
2448398.4528	$\pm$	0.00010	&	1991.39	&	Fiedler et al.	&	0.003239	\\
...	&      ...      &	...&	...    \\
2458694.202758	$\pm$	0.000048	&	2019.57	&	TESS 14	&	0.003313	\\
2458809.038245	$\pm$	0.000059	&	2019.89	&	TESS 18	&	0.003042	\\
2458880.207063	$\pm$	0.000088	&	2020.08	&	TESS 21	&	0.003082	\\
2458950.116245	$\pm$	0.000143	&	2020.27	&	TESS 23	&	0.003110	\\
2459020.025303	$\pm$	0.000052	&	2020.47	&	TESS 26	&	0.003014	\\
2459403.160938	$\pm$	0.000060	&	2021.52	&	TESS 40	&	0.002896	\\
2459594.203530	$\pm$	0.000043	&	2022.04	&	TESS 47	&	0.002455	\\
2459759.214567	$\pm$	0.000083	&	2022.49	&	TESS 53	&	0.002696	\\
2459924.015152	$\pm$	0.000077	&	2022.94	&	TESS 59	&	0.002423	\\
\enddata	
\tablecomments{{\bf (1)}~~The calculated times are for a linear ephemeris with epoch BJD 2452474.8040 and period 0.2099373990 days.  {\bf (2)}~~Sources in the full table include references Fiedler et al. (1997), Baptista et al. (2000), Shafter \& Holland (2003), Pilarcik et al. (2012), Han et al. (2017).}
\end{deluxetable}

\begin{figure}
\epsscale{1.16}
\plotone{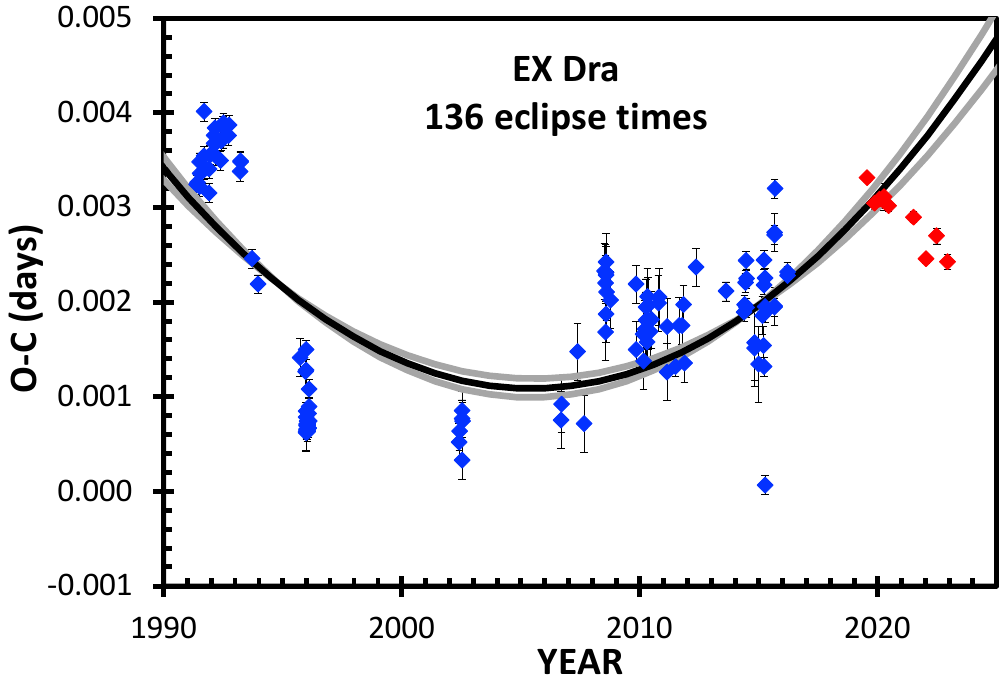}
\caption{$O-C$ curve for EX Dra.  This curve contains 136 eclipse times from Table 11, with the blue diamonds for measures taken from the literature, and with the red circles for the {\it TESS} average times for each Sector.  The general run of the $O-C$ curve is like an upturned parabola, as shown by the black curve with $\dot{P}$=($+$3.05$\pm$0.32)$\times$10$^{-11}$.  But the deviations from the parabola are a substantial fraction of the parabola's sagitta, with deviations of up to 86 seconds (0.001 days).  The $O-C$ curve might be best characterized as a `parabola $+$ bumps'.}
\end{figure}

\subsection{RW Tri; Novalike with $P$=0.232 days}

{\bf RW Tri} is a novalike CV that was discovered as an eclipsing binary in 1937, with a nearly-triangular primary eclipse that is 2.5 mag deep recurring every 0.232 days.  At $V$=12.5 outside of eclipse, the system is relatively easy to observe, so there are many eclipse times reported in the literature from 1937 to the present.

RW Tri has nearly 300 times of eclipse reported in the literature that have been collected into Nelson's database.  I can add 149  further published times from the literature, as cited in Table 12.  I have eliminated the duplicates, averaged the independent times for the same eclipse, and eliminated several $>$5-sigma outliers.  The times reported by Smak (1995) are systematically late from all other measures by $\sim$0.005 days, this being as if they had the wrong sign for the heliocentric correction, their quoted HJD times of eclipses are equal to their plotted JD times of eclipses so there is an internal inconsistency in their applied corrections, so I cannot use these times with any confidence.  We are left with 449 eclipse times from 1937 to 2020, and I have converted them to BJD.

I have extracted 5 time series from the AAVSO database that were not used by Bruch (2023), calculated the times of minimum light, and converted these times to BJD.  These are included in Fig. 14 as part of the yearly averages of the above literature values.

RW Tri was observed in {\it TESS} Sectors 18 (2019) and 58 (2022), all with 120-second time resolution.  I have fitted a symmetric nearly-triangular eclipse shape to the folded light curves for each Sector, producing a minimum time in the middle of each Sector, with this representing the average over nearly 99 orbits each.  The individual times of all the eclipses in Sector 18 were already reported in Schaefer (2021), and these show an eclipse-to-eclipse timing jitter of 0.0081 days due to ordinary flickering in the light curve.  This random error is irreducible for any one eclipse, no matter how well observed.  Fits to the $O-C$ curve must add 0.00081 days in quadrature to the reported measurement errors so as to get the total error for use in calculating the chi-square.  However, for the {\it TESS} times reported here, they are averaged over a hundred particular cases of flickering, so the flickering jitter will be reduced by close to a factor of 10$\times$ to near 0.00008 days.  My two {\it TESS} averaged eclipse times and their measurement errors are listed in Table 12.

I have searched by hand all the Harvard plates from 1889--1937, and I find 191 plates with useful magnitudes.  Five plates show RW Tri significantly fainter than its out-of-eclipse brightness.  Each of these measures can be converted into a time of minimum light, by means of a modern B-band template convolved with the exposure times.  So I have five eclipse times, converted to BJD, going back to 1891.  Two of the plates have exposure times of close to one hour, so their resultant minimum times are of little utility.  The three useful plates require a substantial concave-down curvature.

I have used all 461 eclipse times to create an $O-C$ curve (Fig. 14), and then fit a parabola.  I derive $\dot{P}$=($-$4.6$\pm$0.8)$\times$10$^{-12}$ with a chi-square fit.  The shape is a good parabola, although there are likely real bumps, downward in 2004 and upward in 2017, with amplitudes near 230 seconds.

\begin{deluxetable}{llrl}
\tablenum{12}
\tablecaption{461 eclipse times for RW Tri (full table in machine readable formation in the electronic article)}
\tablehead{
\colhead{Eclipse Time (BJD)} & \colhead{Year} & \colhead{$O-C$} &
\colhead{Source}  }
\startdata
2412043.7721	$\pm$	0.0036	&	1891.85	&	-0.0071	&	HCO (I 4746)	\\
2413148.7061	$\pm$	0.0048	&	1894.88	&	0.0030	&	HCO (I 11898)	\\
2415721.6818	$\pm$	0.0100	&	1901.92	&	0.0017	&	HCO (I 28036)	\\
2423004.6731	$\pm$	0.0012	&	1921.86	&	0.0024	&	HCO (I 40697)	\\
2426624.8440	$\pm$	0.0100	&	1931.77	&	0.0112	&	HCO (RH 3469)	\\
...			&	...	&	...	&	...	\\
2458428.78161	$\pm$	0.00080	&	2018.85	&	-0.0048	&	AAVSO, DERA	\\
2458479.33242	$\pm$	0.00080	&	2018.99	&	-0.0045	&	AAVSO, BPO	\\
2458479.56483	$\pm$	0.00080	&	2018.99	&	-0.0040	&	AAVSO, BPO	\\
2458749.93977	$\pm$	0.00080	&	2019.73	&	-0.0050	&	AAVSO, DERA	\\
2458803.040764	$\pm$	0.000051	&	2019.87	&	-0.0052	&	TESS 18	\\
2458839.67762	$\pm$	0.00080	&	2019.97	&	-0.0059	&	AAVSO, GKA	\\
2459897.064011	$\pm$	0.000047	&	2022.87	&	-0.0074	&	TESS 58	\\
\enddata	
\tablecomments{ {\bf (1)}~~The $O-C$, in days, is calculated for a fiducial ephemeris with an epoch of BJD 2441129.36487 and a period of 0.23188330 days.  {\bf (2)}~~Sources in the full table include references Africano et al. (1978), Boyd (2012; 2023), Bruch (2023), Mandel (1965), Polsgrove et al. (2006), Protitch (1958), Robinson et al. (1991), Rutten et al. (1992), Schoembs et al. (1987), Schoembs \& Hartmann (1983), Smak (2019), Subebekova et al. (2020), Walker (1963), Wonkler (1977).}
\end{deluxetable}

\begin{figure}
\epsscale{1.16}
\plotone{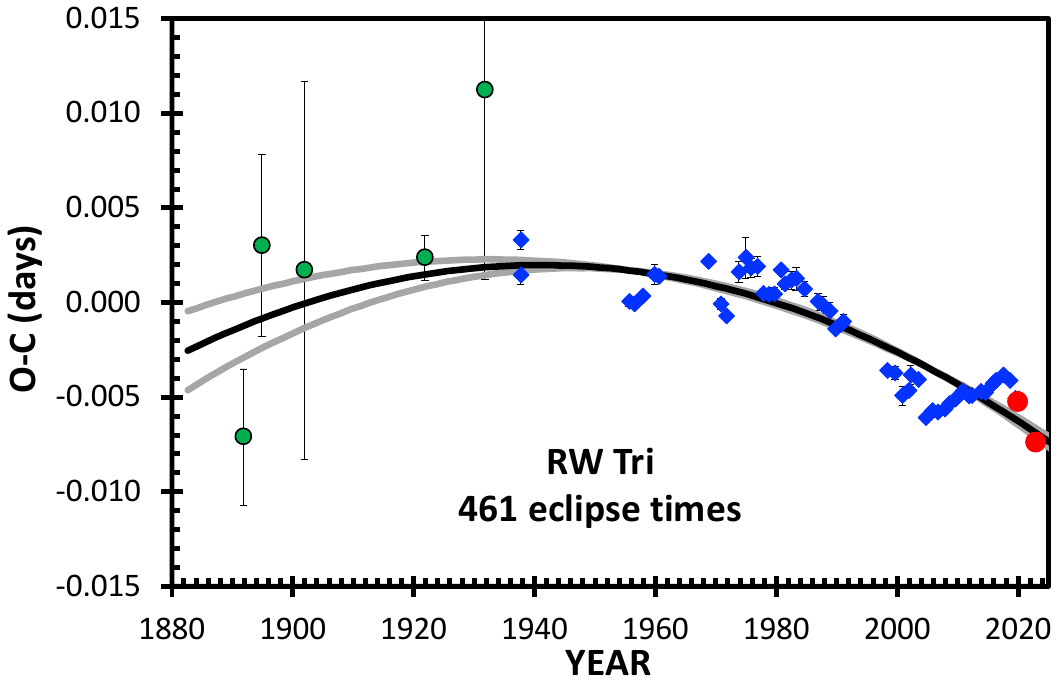}
\caption{$O-C$ curve for RW Tri.  This curve contains 5 eclipse times from the Harvard plates (green circles), 2 times from {\it TESS} with very high accuracy (red circles), plus 454 times from the AAVSO and the literature binned into yearly averages (blue diamonds).   The general run of the $O-C$ curve is shown with a black parabola, with $\dot{P}$=($-$4.6$\pm$0.8)$\times$10$^{-12}$, that quantifies the real evolution over the last 132 years.}
\end{figure}

\subsection{EM Cyg; Novalike with $P$=0.291 days}

{\bf EM Cyg} is an ordinary dwarf nova, with $V$ at minimum near 13.6, and frequent eruptions brightening to near $V$=12.0.  The system was discovered as a variable star in 1928, recognized as a CV in 1953, and eclipses (with $P$=0.291 days) were detected in 1957.  The eclipses are variable, broad, and ill-defined, with amplitudes of $\sim$0.1 mag that are superposed on ellipsoidal modulations of similar amplitude.  Frequent fast flickering usually distorts the shape of the eclipse light curve, making for a large jitter in eclipse times.

Liu et al. (2021) collect 58 minimum times from the literature 1962--2016, and report three times from their own photometry.  I add 10 times from Brat et al. (2011) and Beuermann \& Pakull (1984).

{\it TESS} has observed EM Cyg in Sectors 14, 40, 41, and 54, each of which has the star in quiescence only about half the time.  Liu et al. (2021) have measured minimum times for 90 eclipses in Sector 14.  The $O-C$ curve for these 90 times has a large RMS scatter of 0.0043 days.  A substantial amount of this scatter is due to one far-outlier and due to their inclusion of eclipse times in eruption and quiescence.  The eclipse times during the DN eruption are systematically 0.004 days {\it early} when compared to those in quiescence, likely due to the shifting position of the center-of-light in the binary from the DN eruption.  The best procedure is to exclusively use eclipse times from quiescence, as these will be the most stable and reproducible.  For {\it TESS} times, I have used my own measures from four Sectors, with eruptions excluded (Table 13).

The AAVSO International Database contains many long runs of CCD time series photometry through V-band filters, with some from 2005--2009 and many in 2021, 2022, and 2023.  The runs were selected only if they had durations $>$0.15 days (to avoid problems with sparse light curves) and had the average $V$-magnitude fainter than 13.1 (to avoid DN eruptions).  All these 
long runs were by observers R. Tomlin (AAVSO observer code TRE) and K. Graham (GKA). 

ZTF has 738 useable magnitudes in the $zg$, $zr$, and $zi$ filters from 2018--2022.  Many of these were eliminated due to the star being brighter than its quiescent level.  And the run of $zg$ magnitudes is too short for normalization.  We are left with 290 magnitudes, where the $zi$ magnitudes were offset by 0.20 mag so as to be commensurate with the $zr$ magnitudes.  The ZTF light curve is sparse, with most magnitudes far in time from any other measure, and this creates substantial additional noise.  The brightness in quiescence varies by up to 0.5 mag, so a folded light curve of sparse magnitudes will have a scatter up to 0.5 mag, as compared to the eclipse depth of $\sim$0.1 mag.  An eclipse profile can only be seen by many sparse points phase-averaged together.  With this situation, the only way I can get a signal with adequate timing accuracy is to fold all 290 magnitudes from 2018--2022.  The result is a representative eclipse time near the average observing date, as given in Table 13.

With these 80 eclipse times (1962--2023), I have constructed an $O-C$ diagram (Fig. 15).  This $O-C$  curve is nearly flat with no obvious structure.  The scatter is likely random and intrinsic to the star, due to the timing jitter caused by the large flickering in EM Cyg.  I have fit a parabola, which has $\dot{P}$=($+$6.6$\pm$3.4)$\times$10$^{-12}$.

It might be instructive to compare this result (a simple parabola with $+$6.6$\pm$3.4 in units of 10$^{-12}$) to prior fits of the $O-C$ curve for this one DN.  For comparison, prior published $\dot{P}$ measures are $-$208.4, $-$25$\pm$3, and $+$45.5$\pm$9.7, all in the same units of 10$^{-12}$ (Pringle 1975; Dai \& Qian 2010; Liu et al. 2021).  This situation is typical for CVs, where old values are often orders-of-magnitude or wrong-sign different from each other and from the latest value.  The problem is that the older results did not use all the then-available data, and they could not use later timings that greatly extend the years of observation.  Further frequent problems, for EM Cyg and many other CVs, are when data and analysis mistakes make for deviations in the $O-C$ that will dominate the curves.  EM Cyg also has two claimed periodicities in its $O-C$ curve, for periods of 17.74 and 26.14 years, as published with four-digit precision (Dai \& Qian 2010; Liu et al. 2021).  Both claims are certainly not significant because the RMS scatters of the times about their fitted sinewaves are 1$\times$ and 3$\times$ the claimed semi-amplitude, and both claims are certainly not significant because the $O-C$ coverages are only 2.6$\times$ and 1.9$\times$ the claimed cycle length.  Further, both periodicity claims are denied when further data is added in (see Fig. 15).  This situation is the same for most of the CVs in this paper, where the many claimed periodicities are certainly insignificant and most have been denied by later timings.

\begin{deluxetable}{llrl}
\tablenum{13}
\tablecaption{New eclipse times for EM Cyg}
\tablehead{
\colhead{Eclipse Time (BJD)} & \colhead{Year} & \colhead{$O-C$} &
\colhead{Source}  }
\startdata
2454172.0273	$\pm$	0.0005	&	2007.19	&	0.0017	&	AAVSO 2005--2009	\\
2458691.0138	$\pm$	0.0008	&	2019.57	&	0.0053	&	TESS 14	\\
2458985.1193	$\pm$	0.0016	&	2020.37	&	0.0026	&	ZTF 2018--2022	\\
2459405.1912	$\pm$	0.0014	&	2021.52	&	0.0008	&	TESS 40	\\
2459439.2300	$\pm$	0.0016	&	2021.61	&	0.0033	&	TESS 41	\\
2459451.1566	$\pm$	0.0003	&	2021.65	&	0.0002	&	AAVSO 2021	\\
2459772.0282	$\pm$	0.0006	&	2022.53	&	0.0023	&	TESS 54	\\
2459799.0828	$\pm$	0.0002	&	2022.60	&	0.0014	&	AAVSO 2022	\\
2460111.2292	$\pm$	0.0003	&	2023.45	&	0.0013	&	AAVSO 2023	\\
\enddata	
\tablecomments{The $O-C$ values, in days, use calculated times for a linear ephemeris with epoch BJD 2455408.38986 and period 0.29090913 days.}
\end{deluxetable}

\begin{figure}
\epsscale{1.16}
\plotone{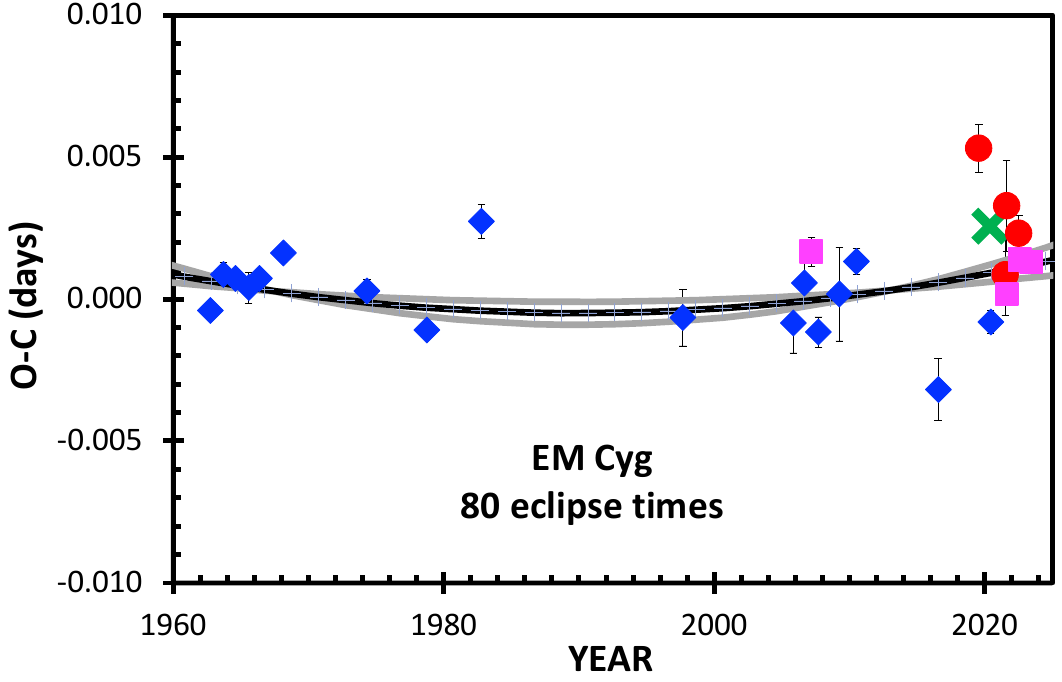}
\caption{$O-C$ curve for EM Cyg.  The red circles are for the four {\it TESS} Sectors, the magenta squares are times from AAVSO long-run $V$-band CCD photometry, the green $\times$ symbols shows the average time for ZTF from 2018--2022, and the blue diamonds are from times reported in the literature as combined into bins of width 1- or 2-years.  The $O-C$ curve is nearly flat, with the scatter due to the large flickering suffered by EM Cyg.  The best-fitting parabola is shown with the black curve, while the one-sigma range in $\dot{P}$ is shown with the gray curves.}
\end{figure}

\subsection{AC Cnc; Dwarf Nova with $P$=0.300 days}

{\bf AC Cnc} is a faint dwarf nova of the Z Cam class, which means that it has occasional standstills in brightness.  AC Cnc is also an eclipsing binary with period 0.300 days, with a simple symmetric eclipse profile of amplitude  $\sim$1 mag.  

This binary was well-covered by two campaigns for K2 (Campaign 5 in 2015 and Campaign 18 in 2018) with 58 second time resolution for 75 and 51 days of nearly-gap-free coverage (Schlegel \& Honeycutt 2019).  AC Cnc was also well-covered by {\it TESS} during three Sectors (44--46 in 2021), with 20 second time resolution, each with 24--27 days of nearly-gap-free coverage (Bruch 2022).  I have used the {\it TESS} 20-second light curves for Sectors 44--46 to find eclipse times from parabola fits to the minima between phases $-$0.05 to $+$0.05, averaged over each Sector.  I find that the average $O-C$ varies over each {\it TESS} orbit (i.e., half a Sector) by 0.00015 days (13 seconds), apparently varying on time-scales of nearly 10 days, and this can only be due to intrinsic changes in AC Cnc.  For $O-C$ purposes, the best set of eclipse times is actually by D. Boyd, observing with a 0.25-m and 0.35-m telescopes near Oxford, reporting 49 eclipse times 2007--2023 (Boyd 2023).  This data set is great because it has good coverage over the last 16 years with a single set of instruments and methods.  The real accuracy of the Boyd eclipse times for a single eclipse is somewhat better than those from K2 and {\it TESS}, while the intrinsic variations of the AC Cnc timing jitter is what provides the ultimate accuracy on the $O-C$ curve.  So what really matter is the number of independent epochs, which is 49 for Boyd, 2 for K2, and 1 for {\it TESS}.  The changes in the $O-C$ curve are well-measured from 2007--2023 by Boyd, while K2 and {\it TESS} only provide three confirming snapshots.

The literature contains 19 photographic eclipse times from 1952--1979 (Kurochkin \& Shugarov 1980), with this forming the critical extension back in time that makes for good accuracy in measuring the $\dot{P}$.  The timing accuracy derived from these archival plates is substantially worse than possible with modern high-speed photometry.  But this accuracy is more than adequate to show the shape of the $O-C$ curve.  The eclipse times and the derived ephemeris are give with time-stamps in Julian Days, so I have had to correct each to HJD and then BJD.  (Qian et al. 2007 failed to make this required correction, with the result that their $O-C$ has a systematic jump of 0.0054 days after the archival data, and this mistake is what creates their claimed sinusoidal $O-C$ curve.)  

The literature also contains 30 isolated eclipse times 1982--2015, with these nearly equally from amateur and professional sources.  I have not included any visual timings, nor any timings based on the shallow secondary eclipse.  These times do not have quoted error bars, so I adopt the RMS scatter for the groups of observers.

All 103 eclipse times in BJD are tabulated in Table 14, along with the source citations.  The $O-C$ curve is given in Fig. 16.  The plotted $O-C$ values have had some time averaging over 1--4 year time bins for each observer.  I have fitted a parabola, for which I get $\dot{P}$=($-$2.29$\pm$0.34)$\times$10$^{-11}$.   

The $O-C$ shape is that of a parabola, with small superposed bumps.  The upward bump centered on 2019 is highly significant, from Boyd's many epochs, with K2 and {\it TESS} only providing confirmation at three epoch.  This bump has a small amplitude of 0.001 day (86 seconds).  Other bumps might be around 2001 and 1980.

\begin{deluxetable}{llrl}
\tablenum{14}
\tablecaption{103 eclipse times for AC Cnc (full table in machine readable formation in the electronic article)}
\tablehead{
\colhead{Eclipse Time (BJD)} & \colhead{Year} & \colhead{$O-C$} &
\colhead{Source}  }
\startdata
2434059.35378	$\pm$	0.00500	&	1952.13	&	-0.00442	&	Kurochkin et al.	\\
...			&	...	&	...	&	...	\\
2459520.00234	$\pm$	0.00007	&	2021.84	&	0.00000	&	TESS 44	\\
2459538.03095	$\pm$	0.00007	&	2021.89	&	-0.00003	&	TESS 45	\\
2459566.27581	$\pm$	0.00002	&	2021.96	&	-0.00003	&	TESS 46	\\
\enddata	
\tablecomments{{\bf (1)}~~The $O-C$ values, in days, uses a linear ephemeris with epoch BJD 2459520.002336 and period 0.30047730700 days.  {\bf (2)}~~Sources in the full table include references Bruch (2022), Boyd (2023), Diethelm (2012a), Dvorak (2005), Hubscher (2007; 2016), Hubscher et al. (2009, 2012), Karjci (2006), Kurochkin \& Shugarov (1980), Qian et al. (2007), Thoroughgood et al. (2004), Yamasaki et al. (1983), Zejda et al. (2006), and Zhang (1987).}
\end{deluxetable}

\begin{figure}
\epsscale{1.16}
\plotone{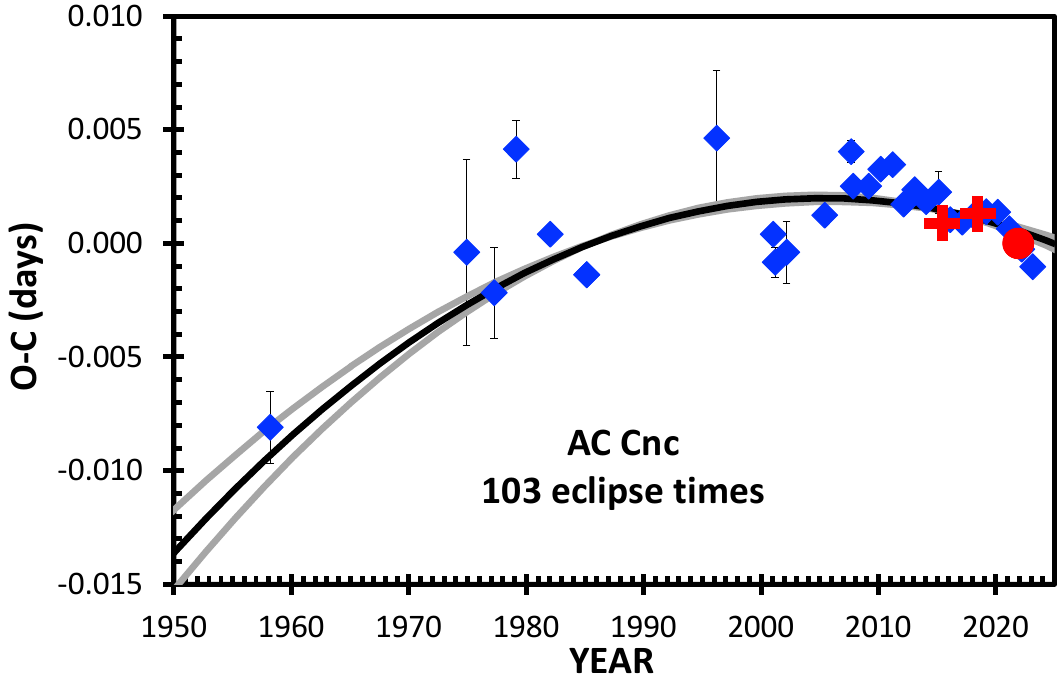}
\caption{$O-C$ curve for AC Cnc.  My $O-C$ curve has 98 eclipse times from the literature (blue diamonds), 2 averaged eclipse times from the K2 light curves (the red plus-signs), plus my 3 averaged eclipse times from the three {\it TESS} sectors (the overlapping red circles).  For display purposes only, I show time-binned values for separate observers.  The general run of the $P$ evolution over the last 71 years has $\dot{P}$=($-$2.29$\pm$0.34)$\times$10$^{-11}$. }
\end{figure}

\subsection{V Sge; the unique V Sge star with $P$=0.514 days}

{\bf V Sge} is a unique CV with by-far the highest known accretion rate at 2.3$\times$10$^{-5}$ M$_{\odot}$ yr$^{-1}$.  The star has been steadily brightening at the rate of $-$0.84$\pm$0.10 mag/century for the last 134 years.   The system is a 3.3 M$_{\odot}$ main sequence star accreting through its Roche lobe on to a 0.85 M$_{\odot}$ carbon/oxygen WD in a 0.514 day orbital period.  Critically, this is the only known interacting binary with RLOF that has a mass ratio ($q=M_{\rm comp}/M_{\rm WD}$) that is much larger than 1, with $q=3.9$.  With this $q$, V Sge necessarily has exponentially-increasing runaway accretion and is in-spiraling fast.

To measure the $\dot{P}$, I have collected and reduced a light curve with 68250 Johnson $V$ magnitudes from 1904--2020 and 2095 Johnson $B$ magnitudes from 1890--2017.  The older parts of the light curve are 689 magnitudes from Harvard for 1890--1989 plus 514 magnitudes from Sonneberg Observatory for 1956--1996.  The AAVSO database has 21392 Johnson $V$ mags for 1999--2019 and 46270 visual magnitudes converted to Johnson $V$ for 1904--2020.  I have also extracted times of minima from {\it TESS} and collected many eclipse times from the literature.  In the end, I have 196 eclipse times from 1903--2019.  These form an $O-C$ curve that shows a perfect and deep concave-down parabola.  I fit the parabola to get $\dot{P}$=($-$5.06$\pm$0.06)$\times$10$^{-10}$.

\subsection{Five AM CVn Binaries}

AM Canum Venaticorum (AM CVn) stars are CVs with orbital periods shorter than $P_{\rm min}$ that show no hydrogen in their spectra.  For $P$ from 5 to 65 minutes, the donor star can only be some version of a hydrogen-poor degenerate star, like a helium WD. 


{\bf HM Cnc} has the shortest-period known for any binary, at 321.6 seconds.  Munday et al. (2023) reports on 143 optical times close to the maximum flux from 2000--2022, making for a decreasing $P$ over time, with $\dot{P}$=($-$3.657$\pm$0.001)$\times$10$^{-11}$.


{\bf V407 Vul} (RX J1914+24) is a double-degenerate binary with a period of 569.5 s.  Ramsay et al. (2005) collect 8 times (from 1993 to 2003) of when the X-ray light curve has a sharp rise to maximum flux, and they form an $O-C$ curve with a significant curvature, so the orbit is speeding up at a rate of ($+$3.17$\pm$0.10)$\times$10$^{-12}$.  

{\bf ES Cet} has a period of 620.2 seconds.  de Miguel et al. (2018) report on timings of minimum light from 452 hours over 145 nights from 2003 to 2017, with their $O-C$ curve showing a nice concave-up parabola, with $\dot{P}$=($+$3.2$\pm$0.1)$\times$10$^{-12}$.

{\bf AM CVn} is the class prototype, with its 0.0119 day orbital period for a helium WD donor star.  There is a long and confused history for the unravelling of the positive and negative superhump periodicities, harmonics, sidebands, cycle count problems, and possible period-doubling, where the Fourier transforms show a bewildering number of sharp peaks that are ever changing.  After 1999 or so, the orbital period is agreed to be the stable 1028.7322 second signal, even though this modulation only has a full amplitude of 0.005 mag.  Once identified, the orbit can be marked going back in time to 1992.  Patterson et al. (2019) covers 40 years and $\sim$1400 nights of observation from 1978--2019, although only the 1992--2019 measures are useable for $O-C$ purposes.  They derive $\dot{P}$=($-$2.7$\pm$0.3)$\times$10$^{-11}$.

{\bf YZ LMi} (SDSS J0926+3624) was the first AM CVn star to have eclipses, these with a 1699 s period.  Schlindwein \& Baptista (2018) collect 8 eclipse times (from 2006, 2009, and 2012) to construct an $O-C$ digram with a significant concave-up parabola shape with $\dot{P}$=($+$3.2$\pm$0.4)$\times$10$^{-13}$.  This fit can be tested with the highly-accurate time of mid-eclipse reported by Anderson et al. (2005), where the eclipse epoch is BJD\footnote{Anderson writes that this epoch is `HJD((TT)', which is an eclipse time that is far outside the ephemeris from Schlindwein \& Baptista, however if there is an easy one-letter typo changing the epoch from BJD to HJD is corrected, then the reported epoch become a plausible fit to the ephemeris.} 2453473.725393$\pm$0.000003, as taken from many eclipses.  The 2018--2022 ZTF folded light curve shows the eclipse, but even taken all together, the eclipse timing accuracy is greatly to poor to be useful.  With the one added time from Anderson, I confirm the previous period change, with $\dot{P}$=($+$3.9$\pm$0.5)$\times$10$^{-13}$, as shown in Fig. 17.

\begin{figure}
\epsscale{1.16}
\plotone{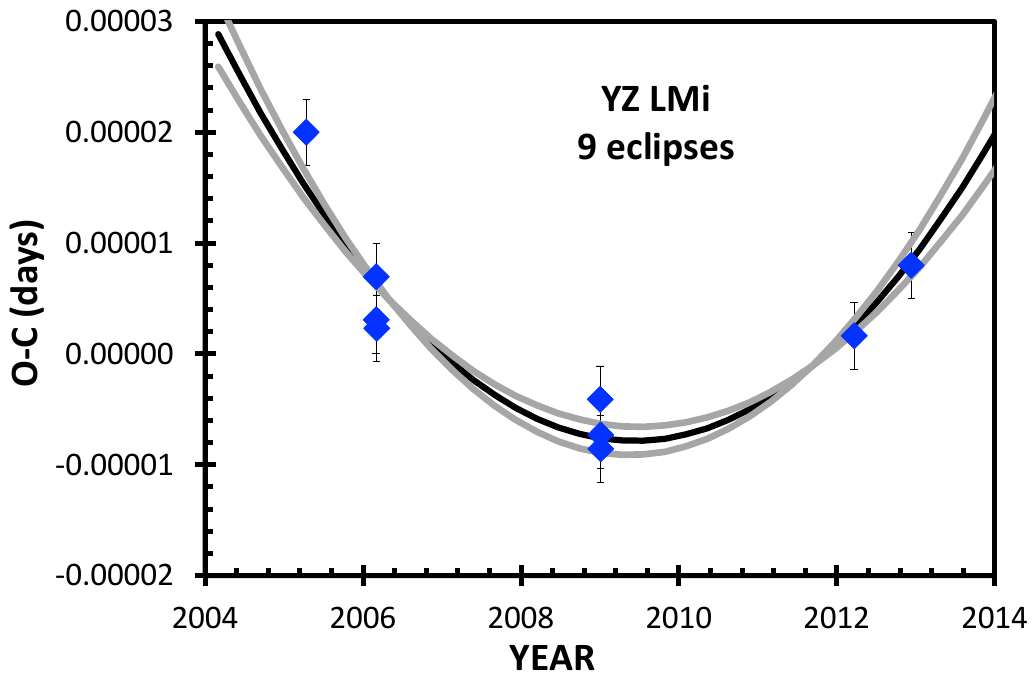}
\caption{$O-C$ curve for YZ LMi.  Only 9 eclipse times are available, but they form a good parabola (the black curve) with $\dot{P}$=($+$3.9$\pm$0.5)$\times$10$^{-13}$.  The one-sigma extremes of $\dot{P}$ are depicted with the two gray curves. }
\end{figure}

\subsection{Three Compact Binary Supersoft Sources}

The CBSS stars are CVs shining at near the Eddington luminosity and are bright sources of soft (15--80 eV) X-ray photons.  The X-rays can only be emitted from hydrostatic nuclear burning on the WD, with this requiring an accretion rate $\gtrsim$10$^{-7}$ M$_{\odot}$ yr$^{-1}$.

Some CBSS stars have been included in a group named as the `V Sge stars' (Steiner \& Diaz 1998).  The basis for this is just the high luminosity and similarities in a few high-energy emission lines.  However, other than V Sge itself with mass ratio $q$=3.9, all the other members of this class have $q$$\le$0.5 (Oliveira \& Steiner 2007).  This is a fundamental difference, where V Sge has a completely different system structure, evolutionary past, and evolutionary future than all the rest of the so-called V Sge stars.  So it is best to simply adopt a terminology where V Sge is the unique member of the V Sge group and the other stars are labelled CBSS.

{\bf V617 Sgr} is a CBSS with primary and secondary eclipses for a period of 0.207 days.  Zang et al. (2023b) report 149 new eclipse times, mostly with AAVSO data.  They have also collected eclipse times from Steiner et al. (1999, 2006) and Shi, Qian, \& Fern\'{a}ndez-Laj\'{u}s (2014), while about half of the eclipse times in Patterson et al. (2022) are included due to the same light curves in the AAVSO database.  I have not been able to make any useful additions, most importantly, the Harvard plates show no significant signal at the orbital period.  So I can do little better than simply use the Zang et al. fits for 188 eclipse times from 1987--2022.  They report that $\dot{P}$=($+$4.60$\pm$0.08)$\times$10$^{-10}$.  They note apparently-sinusoidal variations in the $O-C$ curve with low amplitude (0.0032 days) at a 26 year period superposed on the parabola.  However the significance is not high, only a little more than one orbit is covered, and Shi et al. used largely the same data to propose a 4.5 year periodicity, so I regard this as dubious.  In all cases, the possible sinewave amplitude is so low as to have negligible effect on the measured $\dot{P}$.

{\bf QR And} was discovered as a supersoft X-ray source in 1990, then the optical counterpart was recognized as a high-luminosity CV with high-excitation lines, so the system is a CBSS.  The optical light curves show an eclipsing system with an orbital period of 0.6605 days, where the primary eclipse has amplitude near 0.4 mag, with substantial flickering superposed.  Patterson et al. (2022) measured 23 times of mid-eclipse, and collected 15 mid-eclipse times from the literature, for an $O-C$ curve from 1956--2019, to get $\dot{P}$=($+$1.06$\pm$0.06)$\times$10$^{-9}$.  I can improve this measure by adding mid-eclipse times from 60 Harvard plates 1937--1950 with the middle time in 1941, from each year of ZTF data in 2018--2022, plus from {\it TESS} Sector 57 in 2022 (see Table 15).  I have also added 8 times from Nelson's database, plus 3 times from Beuermann et al. (1995).  With all these 56 times (1941--2022) and the resultant $O-C$ plot (Fig. 18), I derive $\dot{P}$=($+$1.13$\pm$0.04)$\times$10$^{-9}$.

{\bf WX Cen} is a CBSS with a 0.417 day period, showing a near-sinewave modulation (with the minimum being fairly sharp) of full-amplitude 0.32 mag.  Zang et al. (2023a) measured 218 minimum times from 2003--2021 based on light curves from the AAVSO and {\it TESS} databases.  With 8 more times from Oliveira \& Steiner (2004) and 4 more times from Qian et al. (2013),  Zang et al. constructed an $O-C$ curve with 229 times of minimum from 2000--2021 plus one time in 1991.  They then fit a parabola to derive $\dot{P}$=($-$1.2$\pm$0.1)$\times$10$^{-9}$.

\begin{deluxetable}{llrl}
\tablenum{15}
\tablecaption{56 eclipse times for QR And (full table in machine readable formation in the electronic article)}
\tablehead{
\colhead{Eclipse Time (BJD)} & \colhead{Year} & \colhead{$O-C$} &
\colhead{Source}  }
\startdata
2430153.1917	$\pm$	0.0150	&	1941.43	&	0.2316	&	HCO 1937--50	\\
...			&	...	&	...	&	...	\\
2458372.4816	$\pm$	0.0070	&	2018.69	&	-0.0298	&	ZTF, 2018	\\
2458726.4922	$\pm$	0.0060	&	2019.66	&	-0.0268	&	ZTF, 2019	\\
2459073.2356	$\pm$	0.0037	&	2020.61	&	-0.0258	&	ZTF, 2020	\\
2459405.4560	$\pm$	0.0030	&	2021.52	&	-0.0177	&	ZTF, 2021	\\
2459834.1128	$\pm$	0.0044	&	2022.69	&	-0.0006	&	ZTF, 2022	\\
2459868.4574	$\pm$	0.0002	&	2022.79	&	0.0000	&	TESS 57	\\
\enddata	
\tablecomments{{\bf (1)}~~The calculated $O-C$ values (in days) are for a linear ephemeris with epoch HJD 2459868.45741 and period 0.66046180 days.  {\bf (2)}~~Sources in the full table include references Beuermann et al. (1995) and Patterson et al. (2022).}
\end{deluxetable}

\begin{figure}
\epsscale{1.16}
\plotone{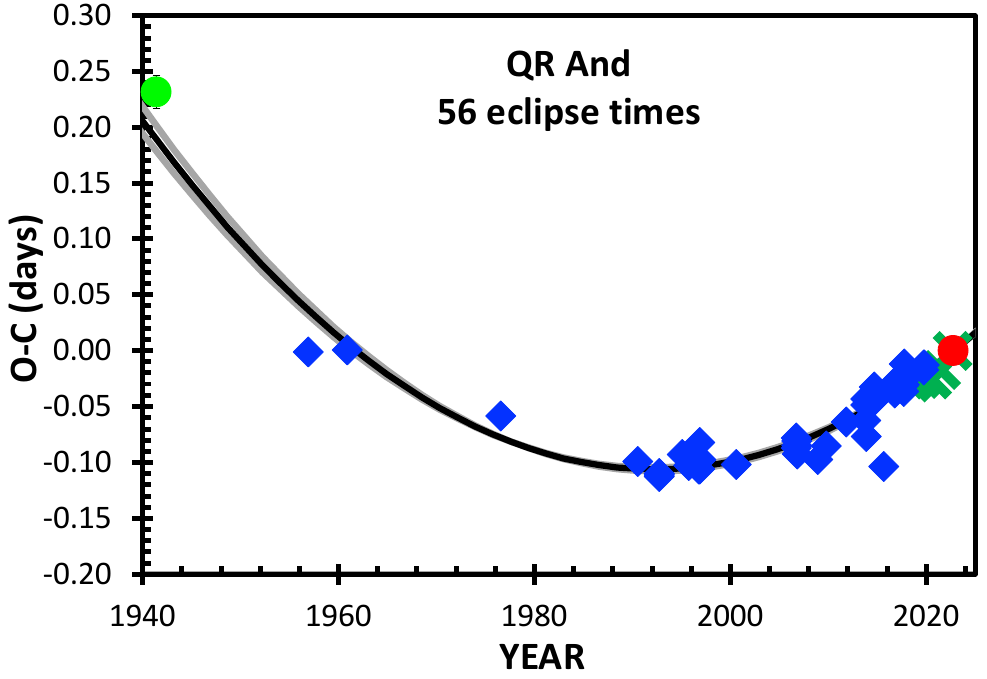}
\caption{$O-C$ curve for QR And.  The measures from the literature are shown in blue diamonds, the measure from the Harvard plates is the one green circle in the upper left, the one highly-accurate measure from {\it TESS} is shown by the red circle in the far right, and the five seasonal averages from ZTF are green $\times$ symbols overlapping on the right side.  The best-fitting parabola is the black curve, which is closely flanked by the gray curves showing the one-sigma deviations in $\dot{P}$.  We see that the evolution of $P$ has been a steady increase (i.e., a good concave-up parabola) for the last 81 years, all with $\dot{P}$=($+$1.13$\pm$0.04)$\times$10$^{-9}$. }
\end{figure}

\subsection{Thirteen CVs of the SW Sex Class}

{\bf V482 Cam} (HS 0728+6738), {\bf SW Sex}, {\bf DW UMa}, {\bf TT Tri} (HS 0129+2933, {\bf V1315 Aql}, {\bf PX And}, {\bf V1024 Cep} (HS 0455+8315), {\bf HS 0220+0603} (HS 0220), {\bf BP Lyn}, {\bf BH Lyn}, {\bf V1776 Cyg}, {\bf 1RXS J064434.5+334451} (1RXS J0644), and {\bf V363 Aur} are CVs of the SW Sex class.  All of these have eclipse timings with many per year from 2007--2023 measured by Boyd (2023).  This vast amount of telescope work was exhausting and done with precision and accuracy.  Boyd's large database of 17 years of well covered eclipse times covers large gaps left by the rest of the world, and provides the strong anchor for the $O-C$ curves.  Boyd (2023) also collected all the historical eclipse times reported in the literature, forming these into nice $O-C$ curves.  Boyd then fitted parabolas to the curves, resulting in $\dot{P}$ measures for each system, along with their one-sigma error bars.  I have not been able to find any errors or problems with the Boyd data or analysis, while I can make no significant additions to the timings.  So my conclusion for these 13 novalike systems is to simply adopt the $\dot{P}$ values as quoted in Boyd (2023), see Table 1.

Boyd (2023) also provides many eclipse times for five other SW Sex stars, as individually discussed elsewhere in this Section for LX Ser, UU Aqr, AC Cnc, RW Tri, and BT Mon.  This impressive work for 18 CVs of one class makes a clear and independent case for the distribution of CV $\dot{P}$ values.  We see 5 systems with strong positive $\dot{P}$, 2 systems with $\dot{P}$$\sim$0 parabolas, 10 systems with strong negative $\dot{P}$, and one system with a wiggly curve for the $O-C$.  Five of the systems have {\it positive}-$\dot{P}$ which is impossible for MBM, the {\it negative}-$\dot{P}$ systems have values an average of 40$\times$ larger than required by MBM, and the wiggly-curve $O-C$ curve for SW Sex is impossible within the MBM.  In all, these CVs provide a complete refutation of the core MBM predictions.

\subsection{Eight Classical Novae}

{\bf V1500 Cyg}, {\bf RR Pic}, {\bf DQ Her}, {\bf T Aur}, {\bf HR Del}, {\bf BT Mon}, {\bf QZ Aur}, and {\bf V1017 Sgr} are classical novae for which Schaefer (2023) has already made the measures of the post-nova $\dot{P}$ (see Table 1).  Just with these 8 CNe, we already see the refutation of MBM because half the systems have {\it positive} $\dot{P}$, which is impossible with MBM.

The original goal of this work has been was to measure the sudden orbital period change ($\Delta P$) across the nova event, for which I have good measures for six CNe (plus four RNe).  These measures show that 5-out-of-6 measures of CN $\Delta P$ are {\it negative}, which is impossible by standard theory and not represented in the MBM.  This is particularly bad for the MBM because the long-term period change is the sum of the steady $\dot{P}$ between eruptions plus the sudden period change across each nova event ($\Delta P$) as averaged over the recurrence time-scale ($\tau_{\rm rec}$), so the real evolutionary period change is $\dot{P}$+$\Delta P$/$\tau_{\rm rec}$.  For 3-out-of-6 CNe, the $\Delta P$/$\tau_{\rm rec}$ effect dominates greatly over the observed $\dot{P}$.  For example of QZ Aur (with $\Delta P$ measured at $-$1.04$\times$10$^{-4}$ days and $\tau_{\rm rec}$ estimated at 420 years), $\Delta P$/$\tau_{\rm rec}$ is 17$\times$ larger (more negative) than the observed $\dot{P}$, so the slow-and-steady effects assumed by MBM are negligibly small and hence wrong.  For example of RR Pic (with $\Delta P$ measured at $-$2.91$\times$10$^{-4}$ days and $\tau_{\rm rec}$ estimated at 1000 years), $\Delta P$/$\tau_{\rm rec}$ is much larger and of the opposite sign as the observed $\dot{P}$, so the MBM neglect of the effect makes for the long-term period change prediction to have the wrong-sign.  These stark failures of the MBM predictions are also for HR Del and most of the RNe.  So, for the majority of nova systems, MBM fails to account for the real $\Delta P$ and hence gives predictions with large errors.

\subsection{Six Recurrent Novae}

{\bf T Pyx}, {\bf IM Nor}, {\bf CI Aql}, {\bf U Sco}, {\bf V394 CrA}, and {\bf T CrB} are recurrent novae for which Schaefer (2023) has presented $\dot{P}$ measures.  Schaefer (2023) is only the last in a long series of papers over the previous three decades reporting on eclipse times and period changes in RNe.  I have $\dot{P}$ measures independently for more than one inter-eruption interval for U Sco (4 inter-eruption intervals), T CrB (2 intervals), T Pyx (2 intervals), and CI Aql (2 intervals).  The first intervals for U Sco, T CrB, and CI Aql have a sufficiently large uncertainty in $\dot{P}$ so that changes of $\dot{P}$ are not of high significance.  Importantly, for U Sco and T Pyx, I measure that the $\dot{P}$ value underwent large and highly-significant changes at the times of the nova events.  The RN U Sco had measured $\dot{P}$ values (in dimensionless units of 10$^{-12}$) of $-$1100$\pm$1100 before the 2010 eruption, $-$21100$\pm$3200 between the 2010 and 2016 eruptions, and $-$8800$\pm$2900 between the 2016 and 2022 eruptions.  The RN T Pyx has measured $\dot{P}$ values (in the same units) of $+$649$\pm$7 before the 2011 eruption, and $+$367$\pm$27 after that eruption.

These sudden changes in $\dot{P}$ are impossible by standard theory and by the MBM.  That is, the $\dot{P}$ values for each system are governed by the binary and stellar properties, with these unchanged by any nova eruption.  So we have a refutation of the MBM for these systems.

For the higher question of what is really going on, the mystery only deepens.  The problem is that the U Sco and T Pyx eruptions involved are accurately measured to have identical photometric and spectroscopic properties (Schaefer 2010; Schaefer et al. 2013).  With each eruption being apparently identical with the previous eruption, some hidden feature of the nova must change the $\dot{P}$ for the years and decades to come.  The eruption events must set some sort of an adjustable switch that controls and changes the AML law.

\section{TEST THE CRITICAL PREDICTION OF MBM}

The MBM makes an explicit and unique prediction of $\dot{P}$ as a function of $P$.  (See Fig. 1 for the blue curve from K2011, and the purple curve for $\dot{P}_{\rm MBM}$.)  Within the MBM, this prediction is required and has no exceptions or any variations.  This MBM prediction of $\dot{P}$ is its most fundamental requirement.  Now, after decades of neglect, this core prediction can be used to test the MBM.  I find that all the MBM $\dot{P}$ predictions fail by orders-of-magnitude in a variety of ways:

{\bf (1.)}  MBM makes the strong prediction that no CV can have {\it positive} $\dot{P}$.  This is directly refuted for the CVs with main-sequence or sub-giant companions, where 43 per cent of the systems have positive $\dot{P}$.   Such is impossible for MBM.

{\bf (2.)}  MBM predicts that the negative-$\dot{P}$ systems must be close to the unique track shown in Fig. 1.  Contrarily, we see that all 25 systems lie below the predictions.  Fig. 1 has a logarithmic vertical axis, and we see that most of the systems are 1--3 orders-of-magnitude in deviation from the MBM prediction.  For these negative-$\dot{P}$ systems, the average MBM error in $\dot{P}$ (and hence $\dot{J}$) is a factor of 110$\times$.  This means that even for ignoring the positive-$\dot{P}$ violators, MBM is operating with an AML law that averages $\sim$1 per cent of the actual AML law. 

{\bf (3.)}  MBM does not allow for any mechanism that has the long-term $O-C$ deviate from a simple parabola.  (Bumps in the $O-C$ caused by flickering noise, hot spot movement, and the various types of measurement error do not affect the long-term parabola shapes.)  Contrarily, SW Sex, OY Car, and Z Cha have $O-C$ curve shapes that vary greatly up-and-down (see fig. 5 of Boyd 2023, Fig. 4, and Fig. 7).  These weird $O-C$ shapes are highly significant with no chance for measurement problems that matter, they have both positive and negative period changes, and their size is larger than possible for intrinsic noise.  So these three stars provide yet more proof that there is an unknown period-change mechanisms that dominate in at least 6 per cent of CVs.  Such period changes are impossible in the MBM.

{\bf (4.)}  MBM does not include the effects of $\Delta P$ from nova events.  That is, the MBM takes the long-term period change to be $\dot{P}$, whereas it is actually $\dot{P}$+$\Delta P$/$\tau_{\rm rec}$ for those CVs that have nova eruptions.  For the majority of the measured cases, MBM either gets the wrong-sign for the real evolutionary changes, or gets them wrong by over one order-of-magnitude.  So, MBM mis-calculates the evolutionary $\dot{P}$ because it fails to include the large negative-$\Delta P$ values for the majority of systems with novae.

{\bf (5.)}  MBM does not allow for a nova eruption to change the steady $\dot{P}$ in quiescence between eruptions.  But for all the cases with good measures (twice for U Sco and once for T Pyx), the observed $\dot{P}$ changed significantly, even up to one order-of-magnitude change.  Such is impossible for MBM.

\section{TEST THE INDIRECT MBM PREDICTIONS}

The single distinct and critical assumption of the MBM is its AML `recipe', so the one fundamental test of MBM is to check its predictions of $\dot{J}$ and $\dot{P}$.  For this one most-important test, MBM has failed badly in its predictions for 52 CVs.  But this strict refutation is new, while most workers will remember the long-past successful predictions of the MBM.  Historically, back in the 1980s, the MBM became the consensus model because it provided a quantitative explanation for the range of the Period Gap and the minimum $P$.  These claimed-successful predictions are only {\it indirect} tests of the MBM.  With the failure of the one {\it direct} test (for $\dot{P}$ measures), it is useful to give a modern evaluation of the indirect tests:


\subsection{Failure of $P_{\rm min}$ prediction}

The MBM makes a strong prediction as to the minimum orbital period ($P_{\rm min}$) that will be achieved by systems with nondegenerate companion stars.  Below the Period Gap, the magnetic braking mechanism does {\it not} operate (because the stars are fully convective and have no significant magnetic field), so the only AML mechanism is from gravitational radiation (GR).  With the binary evolution driven only by GR, the model requires that the binary orbit slowly contracts until reaching a minimum $P$, after which the period `bounces' and starts increasing again.  By the MBM, the $P_{\rm min}$ must apply to all binaries (with nondegenerate companions) with a unique solution.  The basic model predicts that $P_{\rm min}$ equals 65 minutes, although with corrections for irradiation effects and more, the limit becomes 72 minutes for the standard-MBM (K2011).

In a large survey of CV periods made with the Sloan Digital Sky Survey, G\"{a}nsicke et al. (2009) found a sharp spike in the period distribution which gives the minimum period of 82.4$\pm$0.7 minutes.  The difference with the standard-MBM prediction of 72 minutes is stark.  That is, there is no possibility that a GR-only model can result in $P_{\rm min}$ as large as 82.4 minutes, and there is no chance that the spike in the period distribution can actually be as low as 72 minutes. 

Undaunted, K2011 tried to repair the standard-MBM by arbitrarily increasing the AML below the Period Gap\footnote{This conjectured source of additional AML is with no idea as to the physical mechanism, where the postulated functional dependency is assumed to scale like the GR contribution.  Alternative conjectures are possible, for example the eCAML model of Schreiber, Zorotovic, \& Wijnen (2016).  In this model, the postulated extra AML is {\it consequential}, where the extra $\dot{J}$ is proportional to the accretion rate.  The eCAML rate is also an assumption with no physics and no mechanism.  When the proportionality constant is adjusted arbitrarily, eCAML reproduces the observed $P_{\rm min}$ (Belloni et al. 2018).  Such un-evidenced conjectures are attempts to repair the failures of the standard-MBM.}.  They formalized this by setting the AML rate to be some factor ($f_{\rm GR}$) times the GR rate.  The standard-MBM (with only GR below the Period Gap) has $f_{\rm GR}$=1.  K2011 find the best-fitting value to match the observed $P_{\rm min}$ results from $f_{\rm GR}$=2.47$\pm$0.22.  This addition of an unknown AML that dominates over GR is one of the founding principles of the revised-MBM of K2011.  That is, in the revised-MBM, the observed $P_{\rm min}$ is returned only if there is some unknown and unrecognized AML mechanism at work below the Period Gap that is 1.47$\pm$0.22 times that of the standard model.  Thus, the standard model has completely missed the dominant AML mechanism below the Period Gap.

The MBM prediction of $P_{\rm min}$ is commonly thought to be a success, and a reason to believe in the MBM.  However, the revised-MBM prediction of $P_{\rm min}$ matches the observed value only because it was put in by hand, so it is not a prediction.  The standard-MBM prediction is a failure, because the predicted $P_{\rm min}$=72 minutes is in strong contradiction to the observed $P_{\rm min}$=82.4$\pm$0.7 minutes.  So, depending on which version of MBM is being tested, the prediction is actually either a non-prediction (revised-MBM) or a failure (standard-MBM).

\subsection{Failure of $P_{\rm gap}$ predictions}

The lower and upper edges of the Period Gap ($P_{\rm gap-}$ and $P_{\rm gap+}$) are touted as a successful prediction of MBM.  Knigge (2006) analyzed a set of superhump CVs to determine that the observed Gap is from 2.15$\pm$0.03 to 3.18$\pm$0.04 hours.  K2011 used the standard-MBM model to determine that the Gap should extend from 2.24 to 3.52 hours, with poor agreement for the upper edge position.  They also freely adjust the AML strength to create their revised-MBM, where the Gap should extend from 2.24 to 3.24 hours.  This revised-MBM did not make a successful prediction for the Gap, rather, the $P_{\rm gap+}$ was put in by hand.  So like for $P_{\rm min}$, depending on the version of MBM being tested, the prediction is actually either a non-prediction (revised-MBM) or a failure (standard-MBM).

As a test of the MBM predictions for the Gap position, I made a large effort to collect, validate, and update all nova $P$ values, including discovery of 49 new nova $P$, for a total of 159 reliable nova $P$, so we suddenly have a large enough sample to well-measure the Gap for novae (Schaefer 2022).  I found that the nova Gap extends from 1.70--2.66 hours.  This is greatly different from the MBM predictions.  

The situation for MBM predictions of the Gap is actually much worse.  The reason is that the MBM predicts that all CVs will follow one evolutionary track and all CVs will have the identical period-range for the Period Gap.  That is, MBM predicts that the Gap edges are universal for all CV types.  This can be tested by collecting measured Gap edges for five well-defined classes of CVs.  The Gap is 1.70--2.66 hours for novae, 1.94--3.14 hours for nova-likes, and 2.22--3.38 hours for dwarf novae (Schaefer 2022).  K2011 reports the Gap is 2.15--3.18 hours for CVs with superhumps.  Further, Schreiber, Belloni, \& Schwope (2024) measured the Gap to be from 2.45--3.18 hours for non-magnetic systems.  These systems are from the large Sloan Sky Survey volume-limited sample, and so are dominated by the lowest accretion rate systems.  Further, they demonstrated that polars do {\it not} have a Period Gap for any range.  (Belloni et al. (2020) calculate that the white dwarf magnetic fields will have zero effect on the magnetic braking mechanism for all intermediate polars and most polars, while a small fraction of polars will have only small reduction in the magnetic braking effect.  This means that the MBM predicts that polars should have the same Period Gap as the non-magnetic CVs.)  These are displayed in Fig. 19 as a number-line showing the ranges of the Gap.  It is highly significant that the Gap edges change greatly depending on the CV type.  That is, $P_{\rm gap-}$ varies from 1.70 to 2.45 hours, while $P_{\rm gap+}$ varies from 2.66--3.38 hours.  The Gap for novae has little overlap with the Gap for polars.  (I am thinking that the progression of the Gap edges to longer-and-longer periods is caused by the progression to lower-and-lower average accretion rate, $\langle \dot{M} \rangle$, for the four classes as we move from the highest average accretion rate for the novae to the lowest average accretion rate for the Sloan volume-limited sample.)  And the failure of the MBM prediction for the  Gap position is compounded for the complete lack of any Gap for the polars, with this being another failure of the MBM predictions.  So the MBM prediction of a single required evolutionary track and a universal Gap has failed.

The cause of this large change in the Gap position is not of importance for recognizing that the Gap does change greatly with CV class.  The only way for the Gap position to change greatly is for CVs to have many greatly-different evolutionary tracks.  Based on the calculations of K2011 (see the bottom panel of their fig. 14), the range of $P_{\rm gap+}$ from 2.66 hours for novae to 3.38 hours for dwarf novae requires that the group average AML varies by a factor of 4.4$\times$, just in this comparison between two CV groups that intermix.  In all cases, the bottom line is that the MBM prediction for a single unique Gap has failed completely.

\begin{figure}
\epsscale{1.15}
\plotone{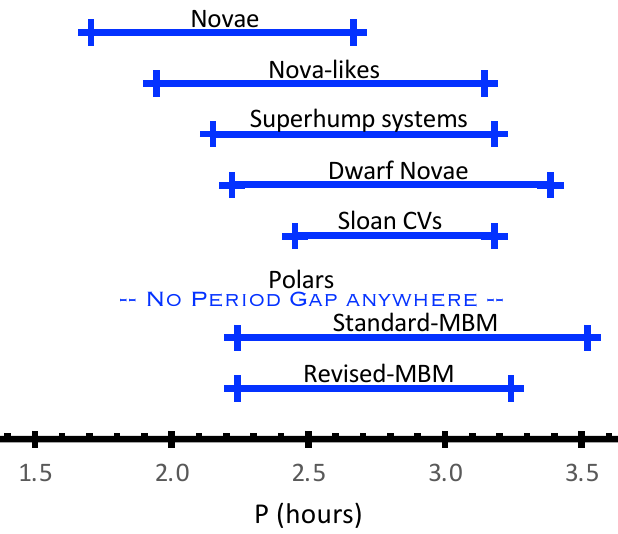}
\caption{Period Gap varies with CV class.  The period range of the Period Gap (from $P_{\rm gap-}$ to $P_{\rm gap+}$) should be constant for all CVs, according to the MBM.  But in reality, the range of the observed Gap varies significantly and substantially with the CV class.  This constitutes a refutation of the MBM prediction that all CVs follow along the same unique evolutionary track. }
\end{figure}

\subsection{Failure of $\dot{M}$ predictions}

The MBM also makes specific and unique predictions for the {\it secular} accretion rate of all interacting binaries, for example, see fig. 11 of K2011 for $P$$<$6 hours.  According to the MBM, all systems with the same $P$ must have the same $\dot{M}$.  Contrarily, it is widely known since the 1980s that interacting binaries with the same $P$ have a very wide range of $\dot{M}$.  (See K2011 for historical and scientific details.)  Warner (1987) shows this wide distribution, with systems just above the Gap spread over a range of 300$\times$, while systems just below the Gap spread over a range of 1000$\times$.  For my systems in Table 1, those just above the Gap have $\log \dot{M}$ (in M$_{\odot}$ yr$^{-1}$) varying from -11.1 for DQ Her to -7.0 for KV UMa.  For the nova systems in Shara et al. (2018) just above the Period Gap, the range of $\log \dot{M}$ is -11.0 to -7.3, with an average of -8.9 and an RMS of 0.8.  For comparison, just above the Period Gap, the MBM prediction is that all systems will have $\log \dot{M}$ near to -9.0.  The point is that real systems have a huge dispersion in $\dot{M}$, contrary to the MBM prediction.

This severe problem has been known for decades.  The proposed solution is that ``The observed scatter in $\dot{M}_2$ was therefore quickly interpreted as evidence for mass-transfer-rate fluctuations on unobservably long time-scales, but still satisfying $\tau_{\rm var} \ll \tau_{\rm ev}$" (K2011).  This picture is that the accretion rate changes greatly on time-scales of $\tau_{\rm var}$, consisting of `episodes' of relatively constant accretion.  Modern astronomers have watched the brightnesses of many individual CVs with no great changes over intervals from many decades to over a century, so $\tau_{\rm var}$$\gg$100 years.  All we see is the brightness and $\dot{M}$ for the current episode, and we see that the many systems have accretion rates that are spread over 2--3 orders-of-magnitude.  Further, the scatter over a range of two orders-of-magnitude in $\dot{M}$ are still seen when the measures are averaged over 10$^3$--10$^5$ years, as derived from the white dwarf surface temperatures (Pala et al. 2022).  On some sufficiently long time-scale, for each individual star, the time-average accretion rate, $\langle \dot{M} \rangle$, is conjectured to closely approximate the MBM predicted accretion rate, $\dot{M}_{\rm MBM}$.  Here, the `sufficiently long time-scale' must be substantially shorter that the evolutionary time-scale of the system, $\tau_{\rm ev}$, which K2011 calculate to be of order 100 million years.  So the picture is that MBM needs episodes lasting $\gg$10000 years or so to average out to $\dot{M}_{\rm MBM}$ in $\ll$100 million years or so.

Two classes of physical mechanisms have been identified to make the extreme variability of $\dot{M}$ for superposition on the slow {\it secular} evolution that is postulated to average out to $\dot{M}_{\rm MBM}$ (K2011).  The first class of mechanism has the nova events triggering one full cycle where the accretion goes from some high value, down to near-zero, and then back up to the high value.  This mechanism was first invented as a means to solve the MBM $\dot{M}$ problem (MacDonald 1986)  The established version of this cycle is now called the `Hibernation' model (Shara et al. 1986).  Within this Hibernation case, there is no expectation that the very-long-term average will be anywhere near $\dot{M}_{\rm MBM}$.  But this does not matter, because Hibernation has been completely refuted (Schaefer 2023).  The second class of mechanism has a long-term cycle being driven by irradiation of the companion star.  This mechanism was also first invented as a means to solve the MBM $\dot{M}$ problem (King et al. 1995).  There has been no useful observational evidence to support the existence of these irradiation-induced cycles.  There is no reason to expect that the accretion rates averaged over the cycle will be anywhere near $\dot{M}_{\rm MBM}$.  K2011 calculates that irradiation-induced cycles have durations of 10$^6$ to 10$^9$ years, which is comparable or longer than $\tau_{\rm ev}$, in which case the existence of a {\it cycle} becomes meaningless.

This attempt at a solution is an invention of a speculative mechanism designed solely to solve the $\dot{M}$ problem, and for which there is no observational evidence for existence of the mechanism.  Further, this attempted explanation requires an additional unevidenced presumption that the very-long-term average $\langle \dot{M} \rangle$ will smooth out to $\dot{M}_{\rm MBM}$ in all systems.  It is a poor argument to invent a new unobserved mechanism to explain one dataset, as such a procedure is just assuming the desired answer.

The attempted explanation requires that the $\dot{M}$ be spread over a wide range of 2--3 orders-of-magnitude for each $P$, with this effect on binary evolution being greatly stronger than the predicted effect of MBM.  With the MBM effect always being much smaller than the postulated extreme-variability mechanism, it matters little what the MBM predicts.  So by postulating a large and unknown mechanism, this is an admission that MBM is adding only negligible effects.

In all cases, independent of whether any speculative ad hoc model can provide an explanation, the MBM model has greatly failed its prediction that $\dot{M}$ follows a single unique evolutionary track as a function of $P$.

\subsection{Failure of period-distribution predictions}

The number-distribution of orbital periods (d$N$/d$P$) for close binaries depends on the rate of AML ($\dot{J}$) as a function of $P$.  That is, for any initial population of binary orbits at any $P$, the number of binaries with a lower period will be comparable if $\dot{J}$ is a slow function, while the number of binaries with this same lower period will be greatly smaller if $\dot{J}$ is a strongly rising function of period.  For the case of detached binaries (with constant $M_{\rm comp}$ and $R_{\rm comp}$), we have $\dot{J}_{\rm MBM}$$\propto$$P^{-3}$, which is a strong function of period.  El-Badry et al. (2022) derive that detached binaries should have a distribution with d$N$/d$P$$\propto$$P^{7/3}$ for systems dominated by magnetic braking.  The population of low-mass detached main-sequence eclipsing binaries is just such a set where the MBM AML rates should dominate, and their distribution can be well-measured due to their large numbers known for short periods.  Further, with the lack of mass transfer, these systems provide a simple test of MBM without complications.  These binaries provide a test of the magnetic braking model for stars closely like CV companions.  (Indeed, it was just this population that was critical in the early days of MBM, for which originally only four such systems were known, see Patterson 1984.)  So MBM makes a strong prediction that the number distribution should go as the $7/3$ power of period, which is to say that the relative numbers of short-period binaries must be small.

El-Badry et al. (2022) have constructed an impressive sample of 3879 eclipsing binaries from the ZTF data base with periods under 10 days.  These are all detached eclipsing binaries involving only main-sequence stars, with all contact systems excluded.  The minimum $P$ varies from 0.19 days for the low-mass stars (near 0.1 M$_{\odot}$) up to 0.42 days for the high-mass stars (near 0.9 M$_{\odot}$).  The stars were separately analyzed for ranges of absolute magnitude, over a wide range.  Their strong result is that the number distribution is closely flat for all periods under 2 days, i.e., d$N$/d$P$$\propto$$P^0$.  That is, for stars just like those in CVs, the real AML effects make for $\dot{J}_{\rm MBM}$$\propto$$P^{-1}$.  This is a stark failure of the MBM prediction.

El-Badry et al. work backwards from their observed number-distributions to derive that these detached systems have $\dot{J}_{\rm MBM}$$\propto$$P^{-1}$.  This is not merely some `tweak' or modification of MBM, rather this new AML law would produce a completely different evolution model.  The expected values for $P_{\rm min}$, $P_{\rm gap-}$, and $P_{\rm gap+}$would all change greatly, making for big disagreements with the observed values.  Critically, by going from an AML law scaling as $P^{-3}$ to $P^{-1}$ makes for gross changes in all studies of binary evolution, demographics and population synthesis.  Going to $\dot{J}_{\rm MBM}$$\propto$$P^{-1}$ is such a big change from the MBM that the model is too greatly different to go by the same name.

In the end, the MBM has failed the critical direct test of its AML recipe (for $\dot{P}$) and all the historically-important indirect tests of its AML recipe (for $P_{\rm min}$, $P_{\rm gap-}$, $P_{\rm gap+}$, $\dot{M}$, and d$N$/d$P$).

\section{THE $\dot{P}$ MUST BE DOMINATED BY SOME UNKNOWN PHYSICAL MECHANISM}

In my many long discussions with founders and leaders of the MBM, a wide variety of possible methods have been mentioned as speculative ways to explain the deviations from MBM $\dot{P}$ predictions for individual systems.  Most of these have been examined in detail and quantitatively in Schaefer (2023), with there now being no chance that these speculations have any possibility of reconciling the MBM.  Here, I will examine the conclusion that no known physical mechanism can explain the measured $\dot{P}$ values for more than a few systems.  To do this, I must go through the known physical mechanisms for period changes, asking the question of whether each mechanism can account for the wide diversity of my 52 measured CVs?

\subsection{Magnetic Braking}

The MBM titular physical mechanism for AML is magnetic braking.  This mechanism postulates that the companion star emits a substantial stellar wind, the wind particles are forced into co-rotation with a postulated significant magnetic field, thus gaining angular momentum, which is then carried out of the binary system.  This slows the rotation of the companion, which is tidally locked into synchronous rotation, so the slowing of the companion's spin is transferred to angular momentum loss of the orbit, hence making the $P$ decrease over time.  This mechanism certainly is operational in CVs, at least at some level.

A deep problem is that the magnetic braking mechanism requires that each and every CV companion star have a large stellar wind and a large surface magnetic field, yet no one has ever reported evidence for any stellar wind or any surface magnetic field for any CV companion star.  (The exception is for the few CVs with red giant companions, like T CrB, which apparently have substantial stellar winds.)  It is poor to advance a model where the two critical components are both merely conjectural.

A further deep problem arising from the magnetic braking mechanism is that it is implausible to think that all the CV companions for any given $P$ will have the same stellar wind strengths and surface magnetic fields so as to produce the same $\dot{J}_{\rm MBM}$ and $\dot{P}_{\rm MBM}$.  Yet, the MBM requires that all stars of a given $P$ have the same AML rate and hence the same stellar winds and surface magnetic fields.

Historically, the MBM founders knew that the CV binaries must be evolving to shorter-and-shorter $P$, so some AML mechanism must be operational, and the magnetic braking mechanism was the only available idea (Patterson 1984, see Section {\rm VI}).  The existence of some sort of AML mechanism was known for young stars, by watching the progressive slow-down in the spin periods of isolated stars, averaged over age from T Tauri stars up to old stars, all roughly one solar-mass.  The variations in the spin-down rate as a function of age is dominated by the fast changes of the T Tauri stars and the young cluster stars.  This crudely-observed rotational-slowing was extrapolated by two orders-of-magnitude in equatorial velocity, and then applied to old low-mass CV companion stars.  As far as I am aware, this rotational-slowing is the {\it only} observational calibration of the AML loss rate for CVs.  That is, the only evidential basis for the AML of CV companion stars is dominated by a far-extrapolation of the AML of isolated one-solar-mass T Tauri stars.  This is not a satisfactory basis for the foundation of the AML that drives CV evolution and demographics.

Historically, the theoretical basis for the magnetic braking mechanism is filled with a myriad of contradictory scaling relations, where the power-law indices for each quantity vary over ranges of up to 4, which is to say that there is no accepted or plausible model for the physics of the magnetic braking on CV companions.  This is characterized by K2011, where they consider angular momentum loss rates scaling as the companion's radius raise to a power $\gamma$, with $\gamma$ running from 0--4 and with the proportionality constant varying over a range of 10$^5$.  K2011 selected candidate physics models that the ``particular MB prescriptions we do consider were selected mainly on the basis that they have proven themselves to be popular".  This emphasizes that no one has any idea as to the strength of the magnetic braking mechanism, nor how it depends on the stellar parameters, nor on the detailed physics.  Again, this is not a satisfactory basis for the foundation of the AML that drives CV evolution and demographics.

A further deep problem is that the poorly-observed young-star AML relation does not have any established physical mechanism.  That is, young-stars are observed to slow down with age, but there is no evidence pointing to any particular mechanism.  The assignment of the magnetic braking mechanism to this young-star AML has no evidential basis.  Historically, the magnetic braking mechanism was selected as the only known mechanism that had an imagined plausibility.  That is, while the magnetic braking has never been seen in any star system, theory calculations show that the effect must be operational at some non-zero level.  So it is only speculation that the young-star AML arises from the magnetic braking mechanism, and only speculation that the young-star AML mechanism is dominant for the old low-mass companion stars in CVs.  That is, the MBM name is poor because there is no evidence that magnetic braking is involved at any significant level.  The `MB' part of MBM is mis-named, as there is no evidence or need for the involvement of the magnetic braking physical mechanism.  As the MBM is implemented from its origins up to K2011, a better name might be `Assumed-Power-Law-Braking-Model'.

So how well can the magnetic braking mechanism explain the observed $\dot{P}$ values?  Well, the real magnetic braking mechanism can only {\it lose} angular momentum from the system, so it can only contribute {\it negative} $\dot{P}$.  This means that all of the {\it positive}-$\dot{P}$ systems cannot be explained, and that is roughly half of my systems.  Further, the magnetic braking mechanism cannot explain any period changes on time-scales of less than $\sim$10,000 years for the tidal synchronization (Paterson 1984, eq. 46).  This means that the magnetic braking mechanism cannot explain the complex and fast $\dot{P}$ changes seen in OY Car, Z Cha, and SW Sex.  This also means that magnetic braking cannot explain the sudden large changes in $\dot{P}$ for U Sco, T Pyx, and T CrB.   This means that some other physical mechanism is required for the majority of the CVs, and that this other mechanism must dominate over any conceivable effects of magnetic braking. 

\subsection{Gravitational Radiation}

A known physical mechanism for AML is gravitational radiation (GR).  This mechanism is certainly operational for CVs, and is likely one of the dominant effects below the Period Gap.  The GR contribution is already included in the MBM predictions for $\dot{P}$ in K2011.  In all cases, the GR AML contribution to $\dot{P}$ is very small, negative, and invariant.  It is impossible that the GR effect on $P$ can account for the large deviations between the observed period changes and the MBM predictions.

\subsection{Mass Transfer in the Binary}

A substantial physical mechanism for $P$ changes in CVs is the transfer of mass between the stars, arising from the accretion process.  This $\dot{P}_{\rm mt}$ from mass transfer is included in the predictions for the MBM (K2011).  All CVs have the mass ratio $q$$\lesssim$1, for which the $\dot{P}_{\rm mt}$ value must be positive.  (The one exception is V Sge, with $q$=3.9, for which $\dot{P}_{\rm mt}$ must be negative.)  To set the scale, for a CV with $P$=4 hours, the MBM required case (companion star mass of 0.33 $M_{\odot}$, mass ratio of $q$=0.44, and accretion rate of $\dot{M}$=1.5$\times$10$^{-9}$ $M_{\odot}$ year$^{-1}$) has the resultant $\dot{P}_{\rm mt}$=3.5 in units of 10$^{-12}$.

Can the physical mechanism of mass transfer reconcile the failed MBM predictions with the 58 observed $\dot{P}$ measures?  Within the MBM, the answer is certainly `no'.  This is because the MBM predictions allow for no variations in $\dot{P}_{\rm mt}$, while the mass transfer effects are already included in the K2011 predictions.

Can the MBM predictions of $\dot{M}$ be violated such that the mass-transfer plus assumed-magnetic-braking combination result in $\dot{P}$ values as observed (instead of as values near $\dot{P}_{\rm MBM}$)?  The answer is still `no'.  That is, for plausible variations in $\dot{M}$ and hence $\dot{P}_{\rm mt}$, the revised predictions cannot match the observations.  As exceptions for this conclusion, three of my 52 stars can have reasonable $\dot{P}_{\rm mt}$ values that make for a match between prediction and observation, with these being discussed later in this subsection.

The basic problem with attempts to reconcile the model is that $\dot{P}_{\rm mt}$ is usually small compared to the deviations from $\dot{P}_{\rm MBM}$, and mass transfer effect only make for positive period changes.  The CVs with negative $\dot{P}$ are (almost) all far below the MBM curve (see Fig. 1), so changing the $\dot{M}$ will not change the predicted $\dot{P}$ by enough to reconcile with observations.  That is, even dropping the accretion rate to zero will not lower the model prediction by enough to matter.   To take a specific example, for the MBM-idealized CV with $P$=4 hours, arbitrarily lowering the accretion rate to zero will only decrease the model $\dot{P}$ by 3.5 in units of 10$^{-12}$, to a combined value of near $-$4 in those units.  Most of the CVs with periods from 3--5 hours violate even this limit, while most of these systems are nova-like CVs, which necessarily have accretion rates $>$1.5$\times$10$^{-9}$ $M_{\odot}$ year$^{-1}$ (Dubus et al. 2018).  So in practice, changing the mass transfer does not bring the model predictions into agreement with observations.

In an attempt to reconcile the CVs with positive $\dot{P}$, we can try increasing the $\dot{M}$ value (from that required by MBM) so that the predicted total period change might become positive enough.  For (almost) all of the CVs with positive $\dot{P}$, the mass transfer rate cannot be changed to any acceptable value (even in violation of the basic MBM requirement) so as to match the observed $\dot{P}$.

Can the mass-transfer mechanism explain the large changes seen in the $O-C$ curves for OY Car, Z Cha, and SW Sex?  If so, then we should see large changes in the accretion rate synchronized with the $O-C$ period changes.  Such $\dot{M}$ changes would be visible as brightness changes in the AAVSO light curves.  {\bf OY Car} has its $O-C$ curve (see Fig. 4) concave-down from 1979 to 1997 and concave-up from 1997 to 2023, so there might be a sharp change in $\dot{M}$ and $V$ around the year 1997.  But the AAVSO light curve shows the out-of-eruption brightness to be closely constant from 1978 to 2023.  So the period changes of OY Car cannot be explained by changes in the mass transfer.  {\bf Z Cha} has its $O-C$ curve (see Fig. 7) showing fairly sharp period changes around the years 1976 and 2000.  If these period changes arise from variations in the mass-transfer mechanism, then we would expect sudden dimming and brightening episodes in those years.  The out-of-eruption and out-of-eclipse AAVSO light curve is constant from 1978--2023, and I have found no useful coverage before 1978.  At first sight, the light curve appears to show a 0.7 mag brightening as a plateau from 2001 to 2007, but this is entirely due to one prolific observer, while contradicted by many quasi-simultaneous measures and limits from another observer and from the early CCD photometry, so I conclude that this one observer simply was using a poor sequence of comparison stars.  So we have no indication that variations in the mass-transfer mechanism created the sudden period change in the year 2000.  {\bf SW Sex} has its $O-C$ curve (see Boyd 2023) showing concave-up parabolas from 1981--2002 and 2019--2023, and a concave-down parabola from 2002--2019.  So if the SW Sex period changes are governed by mass-transfer variations, then we would expect relatively high $\dot{M}$ from 1981--2002 and 2019--2023, and low accretion from 2002--2019.  That is, in the light curve, SW Sex would have a distinct low-state from 2002--2019.  The AAVSO light curve mainly covers 2007--2022, mostly from CCD measures of D. Boyd (Boyd 2023), while the ZTF light curves have excellent coverage of the out-of-eclipse light for the 2018--2023 seasons.  These light curves show that SW Sex was consistently {\it bright} by $\sim$0.6 mag from 2012--2020.  This behavior is exactly opposite that predicted for the mass-transfer mechanism.  This shows that mass transfer variations are not governing the period changes of SW Sex.  In the end for these three stars, their period changes cannot be governed by the mass-transfer mechanism.

The observed $\dot{P}$ values deviate by orders of magnitude from the MBM prediction, both positively and negatively.  If these deviations are driven by the mass-transfer mechanism, then $\dot{P}$ should be strongly correlated with $\dot{M}$ for my 52 CVs.  That is, the CVs with high accretion rates should systematically have large-positive $\dot{P}_{\rm mt}$, and hence large-positive $\dot{P}$.  CVs with low $\dot{M}$ should have near-zero $\dot{P}_{\rm mt}$, and hence have the most-negative $\dot{P}$ values.  I have tested this by collecting measured $\dot{M}$ values for all my CVs, and seeking the predicted correlation between $\dot{P}$ and $\dot{M}$.  The plot shows huge scatter over a large range, running positive-to-negative $\dot{P}$, with the range getting larger for larger $\dot{M}$.  This demonstrates that the predicted correlation does not exist, so the deviations from the MBM predictions are not determined by variations in the mass transfer mechanism.  Indeed, half of the CVs look to show the opposite correlation.  That is, the majority of the CVs with the highest accretion rates ($\sim$10$^{-7}$ $M_{\odot}$ year$^{-1}$), the RNe and the CBSS systems, have the largest-negative $\dot{P}$ values, contrary to the predicted correlation.  

In all, from the above analyses, we see that the MBM deviations are not explained by the mass-transfer mechanism.

Nevertheless, three CV systems have their large deviations from MBM explained by their very-large accretion rate.  {\bf T Pyx} has a near-maximal accretion rate, with this being in complete violation of MBM for its short $P$ below the Gap.  The reason for this violation of MBM is because T Pyx is coming out of an extreme high-state kicked off by an ordinary classical nova eruption around the year 1860 (Schaefer, Pagnotta, \& Shara 2010).  When the realistic $\dot{P}_{\rm mt}$ is included, we expect a large-positive $\dot{P}$, just as we observe.  {\bf IM Nor} is a sister-RN of T Pyx, and so likely has its large-positive-$\dot{P}$ explained simply by an expected huge $\dot{M}$.  {\bf V Sge} has the largest $\dot{M}$ out of all CVs by far, with this runaway accretion driven by the positive feedback associated with its unique mass ratio of $q$=3.9.  With this $q$$\gg$1, the mass transfer will make a large-{\it negative} $\dot{P}_{\rm mt}$, and that is exactly what we see.  The observed $\dot{P}$ is greatly more-negative than allowed by MBM, but when a realistic $\dot{P}_{\rm mt}$ is added in, the model is reconciled with the data.  These three systems are in highly exotic states, so it is fair to wonder whether MBM applies to them.  Still, these three CVs have their large deviations from MBM adequately explained by the mass-transfer mechanism.  But none of the other CV $\dot{P}$ measures can be reconciled by the mass-transfer mechanism.

\subsection{Stellar Activity Cycles}

Another proposed idea is to explain some of the $\dot{P}$ values as due to the hypothetical effect of stellar activity cycles on the companion star.  Specifically, the `Applegate mechanism' (Applegate 1992) postulates that starspot cycles will make the companion star change its {\it shape} substantially, so that an equatorial expansion will slow down the stellar rotation, with the angular momentum then being transferred to the binary orbit for a $P$ change.  This mechanism is imagined to result in sinusoidal variations in the $O-C$ curve, with a period that of the star's starspot cycle.  If the $O-C$ catches only a fraction of the sinewave, then we could be fooled into reporting an evolutionary parabola.  This idea has severe problems, both theoretical and practical:  

{\bf First}, the CV companion stars are claimed to be remarkably close in structure to ordinary isolated main sequence stars, while all such have cycle periodicities from 4.6--14.9 years (Garg et al. 2019), so we would be fooled only for $O-C$ curves with durations ($\Delta Y$) of $<$7 year or so.  With this, zero of my CVs with main sequence or sub-giant companions could possibly fool us.  

{\bf Second}, the Applegate mechanism has never been seen (independently from a $\sim$1 cycle up-and-down in an $O-C$ curve), as reported in the big survey of Richman, Applegate, \& Patterson (1994).  In this current paper, I hav demonstrated that there are no significant sinewave $O-C$ curves for my 52 CVs.  

{\bf Third}, the Applegate mechanism should be working on our own Sun with an 11-year cycle, yet solar oblateness changes in our Sun are observed to be $<$6$\times$10$^{-6}$ over our Sun's magnetic cycle (Fivian et al. 2008).  If our Sun does not have any effect of even minuscule size, then the CV companion stars have no chance of displaying a measurable effect.  

{\bf Fourth}, the Applegate mechanism cannot work effectively on time-scales shorter than the spin-orbit tidal synchronization time-scale, with this being from 110 to 3000 years for CV companions.  So any variations driven by activity cycles with 4.6--14.9 year periods will have the hypothetical spin-ups and spin-downs that alternate on fast time-scales, much faster than the orbit can respond.  That is, even if the star shape changes with a period equal to the activity cycle, the $P$ value cannot change so fast as to reflect the activity cycle.  So the $O-C$ curve cannot show any sinewave on the activity cycle.  Therefore, the observed parabolas cannot be caused by the Applegate mechanism.  

{\bf Fifth}, theory models of the Applegate mechanism have never justified or modeled (either in detail or conceptually) how the postulated magnetic field changes can actually make shape changes.  So the most critical part of this idea is missing the physics.  

We are left with five strong reasons for knowing that the stellar activity cycles are not creating the reported $\dot{P}$ values, much less having any measurable effect for any star.

\subsection{Third-body Effects}

Yet another possibility to explain my $\dot{P}$ measures is that the $O-C$ parabola might just be a short snapshot from sinusoidal $O-C$ variations caused by a brown-dwarf or planet in orbit around the inner binary.  The idea is that a long-period third-body orbiting the inner binary will make the CV binary move back-and-forth with the observed eclipse times varying sinusoidally due to light travel time delays.  If the observed $O-C$ curve $\Delta Y$ covers only a third or a half of the outer-orbit centered roughly on the time of conjunction, then Earth observers will see an $O-C$ shape that can be mistaken as part of a parabola.  However, this explanation can work for few, if any, of my 52 CVs:  

{\bf First}, Shara et al. (2021) has used speckle-imaging to search for the third-body around the six nearest CVs, with zero-out-of-six having third bodies with periods from 3--1000 years and masses $>$0.25 M$_{\odot}$.  This proves that the needed triples are uncommon at best.  

{\bf Second}, the third-body explanation for the parabolic $O-C$ curves cannot work for T Pyx, U Sco, and T CrB, where the $\dot{P}$ is observed to change suddenly across a nova event.  Further, Schaefer (2023) tells how to place an upper limit on the possible $\dot{P}$ from a third body, and this upper limit is violated by 12 of my 52 CVs.  And SW Sex, OY Car, and Z Cha have the $O-C$ variations far from any sinewave shape or any periodicity, so we are not being fooled by these cases.  And five of the CVs have the measured $\dot{P}$ so small that there is no effect for the third body to excuse.  In all, 20 out of my 52 CVs cannot have a third body fooling us into seeing a parabolic curve of significant sagitta.  

{\bf Third}, out of the remaining 32 CVs, the circumstances required to fool us with a third body are rare, so few the expected number of instances is close to zero.  The size of the parameter space for one CV to fool us with a third body orbit is small, requiring that a massive third body exists (under $\frac{1}{6}$ of the cases), requiring inclination $\gg$0$\degr$ ($\sim$$\frac{3}{4}$ of the cases), requiring the third body period to be from roughly 2$\Delta Y$ to 4$\Delta Y$ (under 10 per cent of the cases), and requiring that the middle of the $O-C$ observing interval be near one of the conjunctions (under 50 per cent of the cases).  To fool us, these conditions all need to be met, and such will happen for $<$0.6 per cent of the CVs.  The expectation value for the number of my remaining 32 CVs for which a third body will fool us to see a parabolic $O-C$ curve is $<$0.006$\times$32 which equals $<$0.2 false alarms.  That is, the 32 remaining CVs are expected to have zero cases where we are being fooled by a third body.  In all, near-zero of my 52 CVs can have us being fooled by a third body.  So MBM cannot be excused by the third-body speculation.

\subsection{Magnetic White Dwarfs}

Could it be that a highly magnetized white dwarf will change the AML for individual systems to explain my measured $\dot{P}$ values as a specialized exception to the MBM model?  The idea is alluring because highly-magnetic white dwarfs might lie undiscovered in many CVs, and we can imagine their B-fields amplifying or reversing their companion's magnetic braking.  Fortunately, we have real physics models that point out how and when the white dwarf fields can change the magnetic braking (Li, Wu, \& Wickramasinge 1994; Belloni et al. 2020).

The white dwarf magnetic field falls off as the inverse-cube of the distance from the white dwarf.  The surface field strength of the white dwarf, $B_{WD}$, varies from small values up to 10$^8$ Gauss for extreme polars.  A binary near the Period Gap (with a binary separation of near 1 R$_{\odot}$) around a white dwarf of radius 0.01 R$_{\odot}$ will have the field near the companion star reduced from $B_{WD}$ by a factor of near 10$^9$.  For an extreme polar (with $B_{WD}$=10$^8$ Gauss) the field strength will be 0.1 Gauss around the companion star.  This is to be compared to the median field strength of 3000 Gauss for {\it isolated} M dwarfs (Han et al. 2023).  Even an extreme white dwarf field can only have negligible effects around the companion star.  But far from the companion, the white dwarf field can be comparable to the companion field.  In this case, some fraction of the companion's field lines will connect to the white dwarf, making for closed loops.  A further fraction of the companion's field lines will connect back onto the companion star itself, creating yet more closed loops.  These closed loops mean that any stellar wind ejected from their footprints will necessarily move along the field lines and be captured by one of the stars.  This creates a `dead zone', where some fraction of the wind does not escape the binary and does not carry away any angular momentum.  In the case of a low-field white dwarf, the dead zone is only created by the closed loops in the companion's own field, so the magnetic braking goes as in the MBM.  In the case of a high-field white dwarf, the dead zone extends to cover the entire companion, such that all of its stellar wind never escapes the binary to carry away any angular momentum, and magnetic braking is completely shut down.  So here is a physical mechanism by which the white dwarf's field can reduce or turn-off the magnetic braking, but only for the most extreme polars.

Li et al. (1994) calculate for the binary parameters of AM Her (with stellar masses of 0.7 and 0.4 M$_{\odot}$ in a 5 hour orbit) there is a sharp cut-off at $B_{WD}$=7$\times$10$^7$ Gauss.  That is, for white dwarf field strengths of less than this limit, there is zero effect on the magnetic braking strength.  Belloni et al. (2020) calculate that for magnetic field strengths $>$100$\times$ larger, the magnetic braking in the system is completely turned-off.  An important point is that in all cases, the white dwarf magnetic field can only {\it reduce} the magnetic braking effect.  Under no conditions will the magnetic braking be amplified or reversed.

The fraction of CVs with $B_{WD}$$>$7$\times$10$^7$ Gauss is small.  For isolated white dwarfs, 11--15\% have surface fields $>$1000 Gauss (Jordan et al. 2007).  For a volume limited sample of isolated white dwarfs, 95 per cent complete out to 20 pc from our Sun, Bagnulo \& Landstreet (2021) found 12 out of 152 (8 per cent) have $B_{WD}$$>$10$^7$ Gauss, that 8 out of 152 systems (5 per cent) have surface fields larger than the threshold of 7$\times$10$^7$ Gauss, and the highest field is 3$\times$10$^8$ Gauss.  For CVs, Pala et al. (2020) have constructed a volume-limited sample out to 150 pc with 42 systems, 12 of which are polars, for 29 per cent.  Further, of their 12 polars, only 2 have magnetic fields larger than 70 MG, for 17 per cent with small-number statistics.  Out of a population for 69 polars with measured field strengths (Ferrario, de Martino, \& Gansicke 2015), 5 have $B_{WD}$$>$7$\times$10$^7$ Gauss (for 7 per cent), while the maximum is 1.6$\times$10$^8$ Gauss.   So for CVs, roughly 29 per cent are polars and roughly 7 per cent, or perhaps a bit more, of these have fields with non-zero effect on magnetic braking, for $\sim$2 per cent of all CVs.  With 2 per cent of CVs having $B_{WD}$$>$7$\times$10$^7$ Gauss, the odds are that $\sim$1 of my 52 CVs has a magnetic white dwarf extreme enough to make for non-zero effects on magnetic braking.  Any system with $>$70 MG field would be easily recognized among my well-observed sample of 52 CVs with $\dot{P}$ measures, and none are seen, so the final answer is that zero of my CVs can have anything other than zero effects from $B_{WD}$.  The conclusion is that white dwarf magnetic fields cannot explain any violations of the MBM predictions for even any one CV.

In all cases, the magnetic white dwarfs cannot account for any of the discrepancies between my observed $\dot{P}$ measures and the MBM predictions.  The reason is that the magnetic effect can only reduce the magnetic braking.  The magnetic effects cannot amplify or reverse the magnetic braking.  So there is no way to get {\it positive} $\dot{P}$, so the magnetic effect has no chance of accounting for roughly-half of the MBM counter-examples.  Further, the remaining counter-examples from MBM are almost all with the period changes being {\it greatly more negative} that required by MBM, and such is the opposite of what is possible with the reduction of magnetic braking.  In no case would a special-pleading exception to MBM make for agreement with the observations.

In summary, an extreme $B_{WD}$ can only reduce the magnetic braking effect, making the size of $\dot{P}_{\rm MBM}$ closer to zero, but such an extreme case can have non-zero effects for only 0.6 per cent of CVs.  Even if some one of my 52 CVs had such an extreme $B_{WD}$, the effect cannot produce a greatly more-negative deviation or a positive-$\dot{P}$.  So there is no possibility that magnetic white dwarfs can excuse even one of my MBM violators.

\subsection{Various Mechanisms}

With only the results from Schaefer (2023) covering CNe and RNe, in ordinary discussions with one of the leaders in MBM research, they advanced a speculation that something about the nova event made for a special-$\dot{P}$-state for a century or so, after which the anomalous transient stopped and the system returned to $\dot{P}_{\rm MBM}$ for the remaining eons until the next eruption.  This wishful means to save the MBM is part of the reason why I tested in this paper many CV types with no proximate nova event.  That is, the DN and novalike systems might have had a nova event, but not recently, so the excuse about nova transient effects cannot apply to them all.  And the CBSS and V Sge systems do not have nova eruptions.  The distributions of $\dot{P}$ as a function of $P$ is the same for all these no-nova systems.  So this conjectural mechanism does not work for reconciling the MBM prediction.

Another possible worry is that somehow some unspecified selection effect might be creating the $\dot{P}$ distribution in Fig. 1.  But the RNe, CNe, AM CVn stars, SW Sex stars, CBSS stars, and V Sge were included in this paper because all class members were exhaustively tested for whether a long-term $O-C$ curve could be constructed.  Further, the DN and Novalike variables were selected for inclusion in this paper on the basis of the number and year-range of possible eclipse/minimum times, long before any shape of the $O-C$ curve was known.  Another form of selection effect arises from the numbers of systems in the Milky Way, where systems with large $|\dot{P}|$ must have a short lifetime as a CV, and hence be undercounted.  But this effect works in the wrong direction to excuse the MBM, because the effect selectively misses the systems have the largest $|\dot{P}|$, thus preferentially eliminating only the systems with the largest deviations.  So no selection effect can excuse the MBM prediction failures.

Early in my work on $\dot{P}$ measures, when I had only a few measures, various workers (including myself) worried that I was dealing with small-number-statistics.  Plus, we could always point to a few individual cases (like T CrB and V Sge) as somehow being special such that MBM might not apply.  These worries are a large part of the reason for the extensions to many CVs as reported in this paper.  Now, I have 52 CVs of all types, with many ordinary CVs in a large sample.

\subsection{Some New Mechanism is Needed}

This Section has evaluated all known physical mechanisms that can produce $\dot{P}$ in CVs.  None of the known mechanisms can individually or collectively produce the observed deviations from the MBM.  In this case, my observed $\dot{P}$ measures must be dominated by some unknown physical mechanism.  The forced existence of an unknown physical mechanism that dominates CV evolution is one way of expressing the main results from this paper.

The discovery and testing of ideas for this unknown mechanism is now the biggest issue in CV research.  Indeed, with the AML rates dominating the evolution of all close binaries (including CVs, X-ray binaries, supernova progenitors, and all close detached binaries), the discovery of the true AML mechanism is required before any calculations of binary evolution and demographics can be used with any confidence.

\section{AVERAGING $\dot{P}$ OVER VERY LONG TIMES}

In my long discussions with the founders, textbook writers, and primary-practitioners of MBM, all of the potential issues in the prior section were raised in efforts to reconcile the failed MBM predictions with observations.  These attempts are good science, because the robustness of the MBM prediction failures must be critically tested, and all alternatives considered.  All of the above reconciliation attempts have failed.  Nevertheless, there is one final idea mentioned and highlighted by all my consultants.  The final idea is that there exists some unknown mechanism that makes real CVs vary their actual $\dot{P}$ over huge ranges from large-negative to large-positive values with time-scales of $\gg$100 years, while the very-long time average for each individual system somehow averages out closely to the MBM prediction.  That is, the MBM only requires that the $\dot{P}$ value averaged over $\sim$10 million years be close to its predicted value.  Humans have only watched the period changing on time-scales of a century or so, hence we can only see one episode for each star, and this will have the observed $\dot{P}$ being roughly constant, at any value inside a huge negative-to-positive range.  When hundreds-to-thousands of episodes for each star are averaged together, the conjecture is that the overall average $\dot{P}$ must come out close to the $\dot{P}_{\rm MBM}$ value.  The existence of some unknown physical mechanism that dominates over MBM is already assured (see the previous Section), so the primary open question is whether the very-long-time average happens to be sufficiently close to $\dot{P}_{\rm MBM}$.

Let us look at some numbers related to this idea.  Not including the case of T CrB, the observed $\dot{P}$ measures range from $-$21100 to $+$9500, in units of 10$^{-12}$.  For the MBM core range of $P$ from 3--5 hours, the observed $\dot{P}$ measures range from $-$60 to $+$460, in units of 10$^{-12}$.  The time-averaging-to-$\dot{P}_{\rm MBM}$ idea would have every individual CV fluctuate over this entire range.  For any given period in the range of 3--5 hours, the required $\dot{P}_{\rm MBM}$ varies from $-$0.26 to $-$2.2, in units of 10$^{-12}$.  At any given $P$, MBM requires that the AML deviate from the model prescription by less than a factor of 2$\times$ so that the upper edge of the Period Gap is sufficiently sharp (K2011, fig. 14).  That is, close above the Period Gap, MBM requires that all CVs have their long-term $\langle \dot{P} \rangle$ within a range from $-$0.13 to $-$0.52, in units of 10$^{-12}$.  So the conjectural idea is that some unknown mechanism makes all CVs $\dot{P}$ vary over a range of width 520 in units of 10$^{-12}$, yet all somehow manage to average out to a specific value near $\dot{P}_{\rm MBM}$ inside a range of width 0.39 in the same units.  This hypothetical averaging process reduces the range by a factor of 1300.

The time-averaging-to-$\dot{P}_{\rm MBM}$ idea has many severe problems:  

{\bf First}, by making an unevidenced assumption that the unknown period-change mechanism will average out to $\dot{P}_{\rm MBM}$, the MBM advocate is just assuming the desired answer.  

{\bf Second}, by invoking a new physical mechanism that always dominates over MBM by an average of over 2 orders-of-magnitude, the MBM advocate is actually denying the MBM model.  With the MBM effect constituting $\lesssim$1 per cent of the period changes, it really does not matter whether the MBM AML model is included or nor.  
 
 {\bf Third}, part of the motivation for requiring a single invariant track for $\dot{P}$ versus $P$ is so it makes a sharp upper edge for the Period Gap.  However, CVs entering the Gap will have the edge determined by their current $\dot{P}$, not by some long-term $\langle \dot{P} \rangle$.  So as each CV approaches the Gap, its operational $P_{\rm gap+}$ will be determined by the currently operational $\dot{P}$, which ranges from $-$60 to $+$460, in units of 10$^{-12}$.  So each CV will have a greatly different $P_{\rm gap+}$, and the resultant Period Gap cannot have the sharp observed upper edge.  Thus, an MBM advocate who invokes extreme-$\dot{P}$-variability is requiring the MBM to predict a very broad edge for the Gap, as strongly contradicted by observations.

{\bf Fourth}, the episodes with not-large changes in $\dot{P}$ must be greatly longer than a century (or else I would see large changes in my $O-C$ curves), and Schaefer (2023) calculates that there cannot be enough episodes in the $<$10 million year evolution time so as to average out the wild fluctuations in $\dot{P}$. 

 {\bf Fifth}, a critical point is that the observed $\dot{P}$ values do {\it not} average out to $\dot{P}_{\rm MBM}$.  For CVs between the minimum $P$ and the lower edge of the Period Gap ($P$ from 0.055 to 0.083 days), the $\langle \dot{P} \rangle$ is $+$140, in units of 10$^{-12}$, while the MBM requires a value somewhat smaller than $-$0.06, in the same units.  For CVs just above the Period Gap, with periods from 3--5 hours, the $\langle \dot{P} \rangle$ is $+$25, in units of 10$^{-12}$, while the MBM requires a value near $-$0.6 in the same units.  For CVs with a period from 5--12 hours, $\langle \dot{P} \rangle$ is $-$140, in units of 10$^{-12}$, while the MBM requires a value near $-$3 in the same units.  So $\langle \dot{P} \rangle$ does {\it not} average out to $\dot{P}_{\rm MBM}$.  It does not matter why or how the CVs change their $\dot{P}$, because the time-averaging-to-$\dot{P}_{\rm MBM}$ idea fails because real CVs do {\it not} average out to $\dot{P}_{\rm MBM}$.

\section{CONCLUSIONS}

Here, I have collected 58 measures of $\dot{P}$ from 52 CVs of all types.  These were collected to test the required and exacting predictions of the MBM.

The MBM predictions fail completely:  First, nearly half of the CVs have {\it increasing} $P$, which is impossible in MBM. Second, those CVs with decreasing $P$ have their measured $\dot{P}$ deviating from the MBM predictions with an average deviation of 110$\times$, all in the same direction. Third, OY Car, Z Cha, and SW Sex have large chaotic period changes that are impossible for the MBM. Fourth, the MBM is missing the long-term effects on evolution caused by large sudden period decreases seen in almost all classical nova systems. Fifth, U Sco, T Pyx, and T CrB are observed to have suddenly and large changes in $\dot{P}$ across their nova events, with this being impossible in the MBM.

These many $\dot{P}$ measures cannot be accounted for by the MBM, nor by the conjectural magnetic braking mechanism, nor by any known physical mechanism.  But the CV period changes are fast and large, so there must be some {\it unknown} physical mechanism that is driving these CV systems.  And this unknown mechanism is dominating by typically two-orders of magnitude over the sum of all known mechanisms.  The discovery and testing of some new physical mechanism is now the single most important program for the fields of CVs and all close binaries in general.

Until this problem is solved, our community is faced with the dilemma that all the MBM calculations published in the past are made using the critical assumption of AML that is orders-of-magnitude in error.  So we can have no confidence in the validity of all prior MBM work and results.  This problem extends to most prior work on evolution, demographics, and population synthesis of all close binaries.

\begin{acknowledgments}
I am thankful for many detailed discussions with Joseph Patterson (Columbia), Juhan Frank (Louisiana State University), Tom Maccarone (Texas Tech), Saul Rappaport (MIT), Michael Shara (American Museum of Natural History), and Christian Knigge (University of Southampton). 
\end{acknowledgments}

%

\vspace{5mm}
\facilities{AAVSO, DASCH, TESS, ZTF, MAST}







{}


\end{document}